\documentclass[reprint,amsmath,amssymb]{revtex4-2}

\usepackage{graphicx,amsmath,amsfonts}
\usepackage{amssymb,ulem}
\usepackage{dcolumn}
\usepackage{mathrsfs}
\usepackage{xcolor}
\usepackage{calligra}
\usepackage{bm}
\usepackage{enumerate}
\usepackage{pgfplots}
\usepackage{tikz}
\usetikzlibrary{decorations.markings}
\usetikzlibrary{plotmarks}
\usepackage[export]{adjustbox}

\usepackage{stackengine} 
\newcommand\oast{\stackMath\mathbin{\stackinset{c}{0ex}{c}{0ex}{\ast}{\bigcirc}}}

\newcommand{\n}{\hat{\mathbf{n}}}
\newcommand{\eps}{\boldsymbol{\varepsilon}}

\newcommand{\tp}{\left( t \right) }

\newcommand{\hPib}{\hat{\boldsymbol{\Pi}}}

\newcommand{\Pib}{\boldsymbol{\Pi}}
\newcommand{\Pb}{\mathbf{P}}

\newcommand{\hQb}{\hat{\mathbf{Q}}_\nu}

\newcommand{\Eb}{\mathbf{E}}
\newcommand{\Bb}{\mathbf{B}}

\newcommand{\Lgr}{\mathcal{L}}

\newcommand{\Ua}{\mathbf{U}^a}
\newcommand{\Ub}{\mathbf{U}^{a'}}

\newcommand{\Wl}{\mathbf{U}^\parallel}

\newcommand{\Ab}{\mathbf{A}}

\newcommand{\hA}{\hat{\mathbf{A}}}

\newcommand{\hYb}{\hat{\mathbf{Y}}}

\newcommand{\I}{\tensor{\mathbf{I}}}

\newcommand{\A}{\mathbf{A}}

\newcommand{\rb}{\mathbf{r}}
\newcommand{\rp}{\left( \mathbf{r} \right)}
\newcommand{\rpp}{\left( \mathbf{r}' \right)}
\newcommand{\rpt}{\left( \mathbf{r};t \right)}
\newcommand{\dV}{\text{d}^3{\bf r}}
\newcommand{\dk}{\text{d}^3{\bf k}}
\newcommand{\dS}{\text{d}^2{\bf r}}
\newcommand{\ds}[1]{{\underline{#1}}}
\newcommand{\dss}[1]{\underline{\underline{#1}}}

\makeatletter
\newsavebox{\@brx}
\newcommand{\llangle}[1][]{\savebox{\@brx}{\(\m@th{#1\langle}\)}%
  \mathopen{\copy\@brx\kern-0.5\wd\@brx\usebox{\@brx}}}
\newcommand{\rrangle}[1][]{\savebox{\@brx}{\(\m@th{#1\rangle}\)}%
  \mathclose{\copy\@brx\kern-0.5\wd\@brx\usebox{\@brx}}}

\colorlet{LightGray}{gray!10}
\makeatother

\definecolor{matlab1}{HTML}{0072BD}
\definecolor{matlab2}{HTML}{D95319}
\definecolor{matlab3}{rgb}{0.4940,0.1840,0.5560}

\newlength{\mywidth}
\newlength{\myheight}
\newlength{\mydelta}
\newif\iffig
\figfalse

\begin{document}

\title{Operative Approach to Quantum Electrodynamics in Dispersive Dielectric Objects Based on a Polarization Modal Expansion}

\author{Carlo Forestiere}
\affiliation{Department of Electrical Engineering and Information Technology, Universit\`{a} degli Studi di Napoli Federico II, via Claudio 21,
 Napoli, 80125, Italy}
\author{Giovanni Miano}
\affiliation{ Department of Electrical Engineering and Information Technology, Universit\`{a} degli Studi di Napoli Federico II, via Claudio 21,Napoli, 80125, Italy}
\begin{abstract}
In this paper we deal with the macroscopic electromagnetic response of a finite size dispersive dielectric object, in unbounded space, in the framework of quantum electrodynamics, using the Heisenberg picture. We keep the polarization and the electromagnetic field distinct to enable the treatment of the polarization and electromagnetic fluctuations on equal footing in a self-consistent QED Hamiltonian. We apply a Hopfield type scheme to account for the dispersion and dissipation of the matter. We provide a general expression of the time evolution of the polarization density field observable as functions of the initial conditions of the matter field observables and of the electromagnetic field observables. It is a  integral operator whose kernel is a linear combination of the impulse responses of the dielectric object that we obtain within the framework of classical electrodynamics. 
The electric field observable is expressed in terms of the polarization density field observable by means of the full wave dyadic Green's function for the free space. The statistical functions of the observables of the problem can be expressed through integral operators of the statistics of the initial conditions of the matter field observables and of the electromagnetic field observables, whose kernels are linear or multilinear expressions of the impulse responses of the dielectric object.
 We expand the polarization density field observable in terms of the static longitudinal and transverse modes of the object to diagonalize the Coulomb and Ampere interaction energy terms of the Hamiltonian in the Coulomb gauge. Few static longitudinal and transverse modes are needed to calculate each element of the impulse response matrix for dielectric objects with sizes of the order up to $\min\limits_{\omega}\{c_0/[\omega \sqrt{|{\chi}(\omega)|}]\}$ where ${\chi}(\omega)$ is the susceptibility of the dielectric. We apply the proposed approach to different scenarios describing the dielectric susceptibility by the Drude-Lorentz model.

\end{abstract}
\maketitle


\section{Introduction}
In the last twenty years, there has been a large interest for macroscopic quantum electrodynamics in presence of metal and dielectric structures motivated by the prospect of using plasmonic and photonic devices for quantum optics and quantum technology applications (e.g., \cite{tame_quantum_2013}, \cite{flamini_photonic_2018}). While the problem of quantization of the macroscopic electromagnetic field in nondispersive and homogeneous
dielectrics has been successfully tackled since the
work of Jauch and Watson \cite{jauch_phenomenological_1948},
for dispersive and finite size dielectric objects in the unbounded space the problem has been significantly more difficult. 
 
Glauber and Lewenstein \cite{glauber_quantum_1991}  proposed two quantization schemes for the electromagnetic field in the presence of non-dispersive and non homogeneous dielectrics in the unbounded space, both based on the generalized Coulomb gauge $\nabla \cdot[\varepsilon(\mathbf{r}) \mathbf{A}]=0$. In the first scheme, they expand the electromagnetic field in terms of the full wave eigenmodes of the dielectric object, which is a continuum set of basis functions. In the second scheme, they expand the electromagnetic field in terms of a continuum set of basis functions based on plane waves that satisfy the generalized Coulomb gauge. They also discuss the relation between the two quantization schemes in the framework of electromagnetic scattering theory.
 
To deal with dispersive dielectrics there is the need to introduce dynamical variables that  represent the degrees of freedom of the matter. The established models are mainly based on either Hopfield type schemes or Langevin-noise schemes (e.g., \cite{hopfield_theory_1958, matloob_electromagnetic_1995, gruner_green-function_1996, scheel_macroscopic_2008}). 

Hopfield represented the polarization field of a homogeneous dielectric as a harmonic oscillating bosonic field linearly coupled to the electromagnetic field \cite{hopfield_theory_1958} and quantized the entire system by applying the Coulomb gauge. This model was introduced by Fano \cite{fano_atomic_1956}, who justified it in terms of an atomic medium. It can also be applied to the oscillations of free electrons in metals.

Huttner and Barnett \cite{huttner_quantization_1992} extended the Hopfield model to include the losses of the matter by coupling the polarization field to the electromagnetic field and to a continuum of reservoir bosonic fields. They use the Hamiltonian in the Coulomb gauge, apply the standard canonical quantization method to the entire system and, assuming the homogeneity of the medium, diagonalize the Hamiltonian in a closed form using the Fano method. Suttorp and Wubs \cite{suttorp_field_2004} have dealt with the response of an inhomogeneous dielectric in the Heisenberg picture using the classical dyadic Green's function for the electric field in the presence of the dielectric object. 

In the Huttner-Barnet model, the diagonalization of the matter Hamiltonian (polarization field + reservoir field) yields a set of dressed continuum fields that are coupled to the electromagnetic field.
This fact suggests that absorptive dielectrics can be equivalently described by a single continuum set of harmonic oscillating fields directly coupled to the electromagnetic field \cite{bhat_hamiltonian_2006, philbin_canonical_2010}. In \cite{bhat_hamiltonian_2006}, following \cite{glauber_quantum_1991}, the electromagnetic field is expressed in terms of the full wave eigenmodes of a non-dispersive reference dielectric object, then the Hamiltonian is quantized, and eventually it is diagonalized by the Fano method. In \cite{philbin_canonical_2010} the Hamiltonian based on the Coulomb gauge is quantized and diagonalized using the dyadic Green's function for the electric field in the presence of the dielectric object.

The Hopfield-type models have been applied in combination with the Power-Zienau-Wooley Lagrangian (e.g., \cite{gubbin_real-space_2016, dorier_canonical_2019}) and the Hamiltonian has been diagonalized by the Fano method. In these approaches, the diagonalization requires the solution of a classical electromagnetic scattering problem or the solution of a Lippmann-Schwinger type equation. 

The Hopfield model has been also used to quantize plasmons in metal particles in the full-retarded regime by expanding the current density field in terms of the electrostatic modes of the particle \cite{forestiere_quantum_2020}. A canonical
quantization scheme with numerical mode decomposition for diagonalizing the Hamiltonian has been recently proposed \cite{na_diagonalization_2021}. 

The Langevin-noise schemes are based on the introduction of phenomenological fluctuating currents to deal with the problem of dissipation and dispersion \cite{matloob_electromagnetic_1995, gruner_correlation_1995,vogel_quantum_2006}. The electromagnetic field operators are expressed in terms of the noise current operator by using the dyadic Green's function for the electric field in the presence of the dielectric object \cite{gruner_green-function_1996,dung_three-dimensional_1998}. These schemes are widely applied in many contexts (e.g., \cite{scheel_macroscopic_2008, franke_quantization_2019,hanson_langevin_2021}).

The Hopfield-type schemes and the Langevin noise schemes are equivalent if in the Langevin noise schemes the quantized photonic degrees of freedom associated with the fluctuating radiation field are added to the degrees of freedom of the material oscillators \cite{drezet_equivalence_2017,dorier_critical_2020}.  In both schemes the diagonalization of the Hamiltonian requires the full wave solution of a classical electromagnetic scattering problem in unbounded space: either the computation of the wave eigenmodes of the dielectric object, or the computation of the Green's function in the presence of the dielectric object, or the solution of three dimensional Lippmann-Schwinger type equations.

In this paper, we propose an ``operative" full wave approach to evaluate the macroscopic electromagnetic response of a dispersive dielectric of finite size in unbounded space that does not involve a complete diagonalization of the Hamiltonian. We use the Heisenberg picture to describe the time evolution of the observables of the matter and of the electromagnetic field. We keep the matter and the electromagnetic field distinct. We use a Hopfield type model to describe the matter and the coupling with the electromagnetic field. We expand the matter field observables through the electroquasistatic (longitudinal) modes \cite{fredkin_resonant_2003,mayergoyz_electrostatic_2005} and the magnetoquasistatic (transverse) modes \cite{forestiere_magnetoquasistatic_2020} of the object. These modes are size-independent, and do not depend on the material \cite{fredkin_resonant_2003,mayergoyz_electrostatic_2005,forestiere_magnetoquasistatic_2020}. They are the natural modes of the polarization field in the small size limit.
We apply the Coulomb gauge, and we use the transverse plane waves to represent the radiation field observables. The separation between matter and electromagnetic field allows us to include on an equal footing both electromagnetic field and matter fluctuations in a self-consistent QED Hamiltonian (e.g. \cite{drezet_quantizing_2017} and references therein). The expansion of the matter field observables in terms of the static longitudinal and transverse modes of the object allows to diagonalize the Coulomb and Ampere interaction energy terms of the Hamiltonian.
 Using this approach, we obtain a general expression for the time evolution of the polarization density field observable as functions of the initial conditions of the matter field observables and of the electromagnetic field observables. It is a linear integral operator whose kernel is a linear expression of the impulse responses of the dielectric object that we obtain within the framework of classical electrodynamics \cite{forestiere_time-domain_2021}. 
The electric field observable is expressed in terms of the polarization density field observable by means of the dyadic Green's function for the free space. The statistical functions of these observables are integral operators of the statistics of the initial conditions of the matter field observables and of the electromagnetic field observables. The kernels are linear or multilinear expressions of the impulse responses of the dielectric object. The use of the static modes significantly reduces the computational burden for the evaluation of the impulse responses of dielectric objects with sizes of the order up to $\min\limits_{\omega}\{c_0/[\omega \sqrt{|{\chi}(\omega)|}]\}$ where ${\chi}(\omega)$ is the susceptibility of the dielectric.

The paper is organized as follows. In Sec. \ref{sec:Formulation classical}, we introduce the classical Hamiltonian formulation, in the Coulomb gauge, of the electromagnetic response of dispersive dielectrics of finite size in unbounded space. In Sec. \ref{sec:bases}, we quantize the matter and the radiation fields. In Sec. \ref{sec:modal}, we represent the matter field observables in terms of the static longitudinal and transverse modes of the object, and the radiation field observables in terms of the transverse plane wave in free space. In Sec. \ref{sec:modalh}, we express the Hamiltonian observable of the system in terms of the coordinate and conjugate momentum operators of the matter and radiation field observables. In Sec. \ref{sec:heisenberg}, we first derive the Heisenberg equations of motion for the coordinate and conjugate momentum operators. We then reduce the full set of equations to a system of differential - integral equations of convolution type for the coordinate operators of the matter field. In Sec. \ref{sec:polarization}, we first obtain the equations governing the evolution of the coordinate operators of the polarization. We then give an expression of the polarization density field observable based on the impulse response of the dielectric object that we obtain within the framework of classical electrodynamics. In Sec. \ref{sec:Electric} we give the expression of the electric field observable in terms of the polarization density field operator. In Sec. \ref{sec:Num} we first summarize the principal steps of the numerical procedure to calculate the impulse responses, then we analyze the computational burden. In Sec. \ref{sec:Infinite} we apply the approach to an infinite homogeneous dielectric, a dielectric slab, and a dielectric sphere. We use the sphere to validate the numerical procedure for the computation of the impulse responses. In Sec. \ref{sec:disk} we consider a disk with rounded edges, which is very relevant in nano-photonic. We used the Drude-Lorentz model to describe the susceptibility of the material.  In Sec. \ref{sec:Conclusion} we conclude with a summary and a discussion of the main achievements. 

\section{Classical field equations}
\label{sec:Formulation classical}

\begin{figure}
    \centering
    \includegraphics[width=0.5\textwidth]{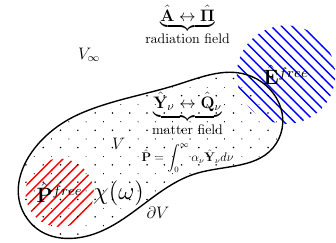}
    \caption{Region $V$ occupied by a dispersive dielectric with susceptibility $\chi \left(\omega\right)$ and boundary $\partial V$; the overall unbounded space is denoted as $V_\infty$. The dielectric is described by a Hopfield type model. The conjugate operators $\left\{ \hat{\mathbf{Y}}_\nu \left( \rb, t \right), \hat{\mathbf{Q}}_\nu \left( \rb, t \right) \right\}$ describe the matter field observables in the Heisenberg picture. The conjugate operators $\left\{ \hat{\mathbf{A}} \left( \rb, t \right), \hPib \left( \rb, t \right)\right\}$ describe the radiation field observables in the same picture. The polarization density field observable in the Heisenberg picture is described by the operator $\hat{\mathbf{P}}(\rb;t)$. It is related to the matter field operator $\hat{\mathbf{Y}}_\nu(\rb;t)$ through the coupling parameter $\alpha_\nu$ that characterizes the linear interaction between the matter and the electric field. The temporal evolution of the polarization density field operator is driven by the free electric field operator $\hat{\mathbf{E}}^{free}\rpt$ and the free polarization field operator $\hat{\mathbf{P}}^{free}\rpt$. They take into account the influence of the initial condition of the matter and radiation field operators, through which the initial quantum state of the system comes into play.}
    \label{fig:DielectricParticle}
\end{figure}

We consider a linear, isotropic, homogeneous, dispersive, and absorptive dielectric of finite size,  shown in Fig. \ref{fig:DielectricParticle}. We denote the space occupied by the dielectric with $V$, its boundary by $\partial V$, the (unit vector) normal to $\partial V$ that points outward by $\mathbf{n}$, the unbounded space by $V_\infty$, and the radius of the smallest sphere that contains $V$ by $a$.  The diameter  $2a$ of such sphere is equal to the largest linear dimension of the dielectric object.
Throughout this manuscript, we use the scalar product 
\begin{equation}
    \langle {\bf F}, {\bf G} \rangle_W = \int_W \dV\, {\bf F}^* \rp \cdot \mathbf{G} \rp,
\end{equation}
and the norm $\left\| {\bf F} \right\|_W = \sqrt{ \langle {\bf F}, {\bf F} \rangle_W }$. The scalar product is defined in $V$ if the domain is not explicitly indicated.

\subsection{Dielectric constitutive relation}

The electric polarization density field $\mathbf{P}\rpt$ describes the macroscopic state of the dielectric. Due to the linearity, isotropy, and homogeneity of the dielectric, the field $\mathbf{P}\rpt$ is solenoidal in $V$ but its normal component to $\partial V$ is different from zero. Therefore, a surface polarization charge lies on $\partial V$ with a surface density equal to $P_{n}=\mathbf{P} \cdot \mathbf{n}$. We denote by $\mathbf{E}\left(\mathbf{r};t \right)$ the macroscopic electric field and by $\mathbf{B}\left(\mathbf{r};t \right)$ the macroscopic magnetic field.
The macroscopic response of the dielectric for $t\ge0$ is described by
\begin{equation}
   \label{eq:constrel}
    \mathbf{P}\rpt = \left\{
    \begin{array}{cl}
        \varepsilon_0 \, \displaystyle \zeta(t)*\mathbf{E} \left(\mathbf{r};t \right)+ \mathbf{P}^{free}(\mathbf{r};t)& \text{in} \; V,  \\
          0 & \text{in} \; V_\infty \backslash V,
    \end{array}
    \right.
\end{equation}
where $*$ denotes the time convolution product, $\varepsilon_0$ is the vacuum permittivity, $\zeta ( t)$ is the inverse Fourier transform of the dielectric susceptibility $\chi(\omega)$,
\begin{equation}
\label{eq:InvFourier}
    \zeta (t) =\frac{1}{2\pi}\int_{-\infty}^{+\infty}d\omega\,\chi(\omega)e^{i\omega t}.
\end{equation}
$\mathbf{P}^{free}(\mathbf{r;t})$ takes into account the contribution of the initial state (at $t=0$) of the polarization.

The real part of susceptibility $\chi_r(\omega)$ is an even function of $\omega$, and the imaginary part $\chi_i(\omega)$ is an odd function, thus ${\chi}(-\omega)={\chi}^*(\omega)$. Since the dielectric is absorptive, ${\chi}_i(\omega)$ is negative for $\omega >0$. 
The causality implies that $ \zeta (t)=0$ for $t<0$, therefore, ${\chi} \left(\omega\right)$ obeys the Kramers–Kronig relations for $-\infty<\omega<+\infty$,
\begin{subequations}
\begin{eqnarray}
\label{eq:KK}
    {\chi}_r(\omega)&=\displaystyle\frac{2}{\pi}\mathcal{P}\displaystyle\int_0^\infty d\omega'\,\frac{\omega'{\chi}_i(\omega')}{\omega'\,^2-\omega\,^2}, \\
    {\chi}_i(\omega)&=-\displaystyle\frac{2\omega}{\pi} \mathcal{P}\displaystyle\int_0^\infty d\omega'\, \frac{ {\chi}_r(\omega')}{\omega'\,^2-\omega\,^2},
\end{eqnarray}
\end{subequations}
where $\mathcal{P}$ denotes the Cauchy principal value.

The electrodynamics of a dispersive and absorptive dielectric can be studied through a Hamiltonian formulation by modeling the medium as a continuum set of harmonic oscillators (e.g., \cite{bhat_hamiltonian_2006}, \cite{philbin_canonical_2010}). The harmonic oscillator field with natural frequency $\nu$, where $0\le\nu<\infty$, is described by the coordinate vector field $\mathbf{Y}_\nu\rpt$ defined in $V$. Throughout the paper, we indicate the continuum set $\{\mathbf{Y}_\nu\rpt\}$ as the ``matter fields". The polarization density field is expressed in terms of the matter fields as 
\begin{equation}
   \label{eq:polfield0}
    \mathbf{P}\rpt = \int_0^\infty d\nu\, \alpha_\nu \mathbf{Y}_\nu\rpt,
\end{equation}
where $\alpha_\nu$ is the coupling parameter characterizing the interaction between the matter field and the electric field. The choice
\begin{equation}
\label{eq:coupling}
    \alpha_\nu = \sqrt{\frac{2{\sigma}(\nu)}{\pi}}
\end{equation}
where
\begin{equation}
\label{eq:sigma}
    \sigma(\omega) = -\varepsilon_0 \omega{\chi}_i(\omega)
\end{equation}
returns the constitutive relation \ref{eq:constrel} in the region $V$ (see Appendix \ref{sec:susceptibility}). 
The term $\mathbf{P}^{free}$ is given by
\begin{equation}
   \label{eq:Ptimefree}
    \mathbf{P}^{free}\rpt = \int_0^\infty d\nu\, \sqrt{\frac{2\sigma(\nu)}{\pi}} \mathbf{Y}_\nu^{free}\rpt
\end{equation}
where
\begin{equation}
   \mathbf{Y}_\nu^{free}\rpt=\mathbf{Y}_\nu^{(0)}(\mathbf{r})\cos(\nu t) + \frac{1}{\nu}\dot{\mathbf{Y}}_\nu^{(0)}(\mathbf{r})\sin(\nu t).
\end{equation}
$\mathbf{Y}_\nu^{(0)}(\mathbf{r})$ and $\dot{\mathbf{Y}}_\nu^{(0)}(\mathbf{r})$ denote, respectively, the vector field $\mathbf{Y}_\nu$ and its partial derivative with respect to the time evaluated at $t = 0$.

\subsection{Lagrangian in the Coulomb gauge}

It is convenient to represent the electric field $\mathbf{E}$ in $V_\infty$ as
\begin{equation}
   \mathbf{E}= \mathbf{E}_s+\mathbf{E}_c,
\end{equation}
where $\mathbf{E}_s(\rb;t)$ is its solenoidal component (radiation field) and $\mathbf{E}_c(\rb;t)$ is its irrotational component (Coulomb field). The vector field $\mathbf{E}_s$ and the vector field $\mathbf{E}_c$ are orthogonal according to the scalar product $\langle \mathbf{E}_s, \mathbf{E}_c \rangle_{V_\infty}$. We introduce the vector potential $\bf A(\rb;t)$ in the Coulomb gauge,
\begin{align}
    \nabla \cdot \bf A = 0 \quad \text{in}\, V_\infty.
\end{align}
The solenoidal component of the electromagnetic field is given by
\begin{align}
    \Eb_s&=-\dot{\bf A},\\
    \Bb &=\nabla \times \bf A,
\end{align}
where the dot above ${\bf A}$ denotes the partial derivative with respect to time. 
The Coulomb electric field is given by
\begin{equation}
\label{eq:scalp}
    \mathbf{E}_c\rpt =-\frac{1}{4\pi\varepsilon_0} \nabla\oint_{\partial V} \dS'\,\frac{ {P}_{n} \left( \rb ';t\right)}{\left| {\bf r} - {\bf r}' \right|}  \quad \text{in}\, V_\infty.
\end{equation}
The field $\mathbf{E}_c$ is solenoidal in $V$ and $V_\infty \backslash V$ but its normal component to $\partial V$ is discontinuous due to the surface polarization charge $P_n$.

The degrees of freedom of the whole system are the matter fields $\{\mathbf{Y}_\nu(\mathbf{r};t)\}$ and the vector potential $\bf A(\mathbf{r};t)$.
The Lagrangian in the Coulomb gauge is the sum of four terms: the matter term $\Lgr_{mat}=\Lgr_{mat}(\mathbf{Y}_\nu,\dot{\mathbf{Y}}_\nu)$, the Coulomb term $\Lgr_{Coul}=\Lgr_{Coul}(\mathbf{P})$, the radiation term $\Lgr_{rad}=\Lgr_{rad}(\Ab, \dot{\bf A})$, and the interaction term between matter and radiation $\Lgr_{int}=\Lgr_{int}(\dot{\mathbf{P}}, \Ab)$ (e.g., \cite{suttorp_field_2004, philbin_canonical_2010, forestiere_time-domain_2021}). The expression of the Lagrangian is
\begin{equation}
    \Lgr =  \Lgr_{mat} + \Lgr_{Coul} + \Lgr_{rad} + \Lgr_{int},
\end{equation}
where
\begin{subequations}
\begin{eqnarray}
\label{eq:Lmat}
     &&\Lgr_{mat} = \int_V \dV\,\int_0^\infty d\nu \left(\frac{1}{2} \dot{\mathbf{Y}}_\nu ^2-\frac{\nu^2}{2} \mathbf{Y}_\nu^2\right),\\
\label{eq:Lcoul}
   &&\Lgr_{Coul} = -\oint_{\partial V}\dS\oint_{\partial V}\dS'\, \frac{P_{n} (\mathbf{r};t) P_{n} (\mathbf{r'};t)}{8\pi\varepsilon_0|\mathbf{r}-\mathbf{r}'|},\\
\label{eq:Lem}
     &&\Lgr_{rad} = \int_{V_\infty}\dV\left[ \frac{\varepsilon_0}{2} \dot{\bf A}^{2} - \frac{1}{2\mu_0}\left( \nabla\times \mathbf{A}\right)^2 \right],\quad\\
\label{eq:Lint}
    &&\Lgr_{int} =\int_{V} \dV\, \dot{\mathbf{P}} \cdot\Ab.
\end{eqnarray}
\end{subequations}
$P_n $ and $\mathbf{P}$ are functions of ${\mathbf{Y}}_\nu$ through the relation \ref{eq:polfield0}.

\subsection{Canonical variables and Hamiltonian}

We now introduce the conjugate momenta of matter and radiation fields in the Coulomb gauge. The momentum $\mathbf{Q}_\nu \rpt$ conjugated to the matter field $\mathbf{Y}_\nu \rpt$ is 
\begin{equation}
   \label{eq:polfield}
    \mathbf{Q}_\nu = 
        \dot{\mathbf{Y}}_\nu 
    +\alpha_\nu\mathbf{A} \quad \text{in} \; V.
\end{equation}
The momentum $\mathbf{\Pi} \rpt$ conjugated to the vector potential $\mathbf{A} \rpt$ is 
\begin{equation}
    \Pib =\varepsilon_0 \dot{\bf A} \quad \text{in} \; V_\infty.
\end{equation}
The Hamiltonian has three terms: the contribution of the matter ${H}_{mat}={H}_{mat}(\mathbf{Q}_\nu,\mathbf{Y}_\nu,\mathbf{A})$, the contribution of the Coulomb interaction ${H}_{Coul}={H}_{Coul}(\mathbf{Y}_\nu)$ and the contribution of the radiation field ${H}_{rad}={H}_{rad}(\Pib,\mathbf{A})$. The expression of the Hamiltonian is
\begin{equation}
\label{eq:ham1}
    {H} = {H}_{mat}+{H}_{Coul}+{H}_{rad},
\end{equation}
where
\begin{subequations}
\begin{eqnarray}
\label{eq:hmat}
   &&{H}_{mat} = \int_V \dV \int_0^\infty d\nu \left [\frac{1}{2} \left( \mathbf{Q}_\nu - \alpha_\nu {\bf A} \right)^2+\frac{\nu ^2}{2}\mathbf{Y}_\nu^2\right],\quad \\
\label{eq:hcou}
   &&{H}_{Coul} = \oint_{\partial V}\dS \oint_{\partial V}\dS'\, \frac{P_{n} (\mathbf{r};t) P_{n} (\mathbf{r'};t)}{8\pi\varepsilon_0|\mathbf{r}-\mathbf{r}'|},\\
\label{eq:hrad}
   &&{H}_{rad} = \int_{V_\infty} \dV \left[ \frac{1}{2\varepsilon_0} \Pib^{2} + \frac{1}{2\mu_0} \left( \nabla \times {{\bf A}} \right)^2 \right].
\end{eqnarray}
\end{subequations}

The Hamilton’s equations for the matter field and the conjugate momentum are in $V$ and for $0\le\nu<\infty$
\begin{subequations}
\label{eq:YQc}
\begin{eqnarray}
    \label{eq:Yc}
    \dot{\mathbf{Y}}_\nu&=& \mathbf{Q}_\nu - \alpha_\nu{\bf A}, \quad \\
    \label{eq:Qc}
    \dot{\mathbf{Q}}_\nu &=&-\nu^2\mathbf{Y}_\nu+\alpha_\nu \mathbf{E}_c\{P_n\}.
\end{eqnarray}
\end{subequations}
The Hamilton’s equations for the radiation field and the conjugate momentum are in $V_\infty$
\begin{subequations}
\label{eq:perpc}
\begin{eqnarray}
    \label{eq:Aperpc} 
    \dot{\A} &=&\frac{1}{\varepsilon_0} \Pib \;, \\
    \label{eq:Pperpc}
    \dot{\Pib} &=& \frac{1}{\mu_0} \nabla^2 \A + \left(\dot{\mathbf{P}} + \varepsilon_0 \dot{\mathbf{E}}_c\{P_n\}\right).
\end{eqnarray}
\end{subequations}
The Coulomb field $\mathbf{E}_c$ is a function of the normal component of $\mathbf{P}$ on $\partial V$ through relation \ref{eq:scalp}; $\mathbf{P}$, in turn, is a function of $\mathbf{Y}_\nu $ through the relation \ref{eq:polfield0}. 
The vector field $\dot{\mathbf{P}}$ is the polarization current density and the vector field $\varepsilon_0 \dot{\mathbf{E}}_c$ is the displacement current density due to the Coulomb electric field: their sum, $(\dot{\mathbf{P}}+\varepsilon_0 \dot{\mathbf{E}}_c)$, is solenoidal in $V_\infty$.

Equations \ref{eq:YQc} and \ref{eq:perpc} give, for $0\le\nu<\infty$,
\begin{subequations}
\begin{eqnarray}
\label{eq:Ydotdot01}
\ddot{\mathbf{Y}}_\nu + \nu^2 {\mathbf{Y}}_\nu &=\alpha_\nu\left( -\dot{\A}+\mathbf{E}_c\{P_n\}\right) &\text{\;in } V, \\
    \label{eq:Adotdot01}
\ddot{\A} - c_0^2 \nabla^2 \A &= \frac{1}{\varepsilon_0}\left(\dot{\mathbf{P}} + \varepsilon_0 \dot{\mathbf{E}}_c\{P_n\}\right) &\text{\;in } V_\infty.
\end{eqnarray}
\end{subequations}

Equation \ref{eq:Ydotdot01} governs the classical motion of the matter field. Equation \ref{eq:Adotdot01} governs the classical evolution of the vector potential. They have to be solved with the initial conditions for the matter field $\mathbf{Y}_\nu^{(0)}(\mathbf{r})$ and $\dot{\mathbf{Q}}_\nu^{(0)}(\mathbf{r})$, and with the initial conditions for the radiation field $\mathbf{A}^{(0)}(\mathbf{r})$ and $\dot{\mathbf{\Pi}}^{(0)}(\mathbf{r})$. The initial conditions are the sources of the problem. The solution of Eq. \ref{eq:Ydotdot01} gives the constitutive relation \ref{eq:constrel} (see Appendix \ref{sec:susceptibility}).

\section{Quantization}
\label{sec:bases}

First, we quantize the matter and radiation fields in a standard fashion (e.g., \cite{cohen-tannoudji_photons_1997, suttorp_field_2004, philbin_canonical_2010}) by enforcing the canonical commutation relations between the field operators and their conjugate momenta. Then, we introduce the Heisenberg equations.

\subsection{Operators and Commutation Relations }

 The vector field operators $\hQb(\mathbf{r};t)$ and $\hYb_\nu(\mathbf{r};t)$ correspond, respectively, to the canonically conjugate matter vector fields $\mathbf{Q}_\nu$ and $\mathbf{Y}_\nu$; the vector field operators $\hPib(\mathbf{r};t)$ and $\hA(\mathbf{r};t)$ correspond, respectively, to the canonically conjugate radiation vector fields $\Pib$ and $\Ab$. These are the vector field operators that describe the fundamental observables of the problem, as sketched in Fig. 1. They obey the commutation relations for $\nu, \nu' \in [0,\infty)$
\begin{equation}
\label{eq:commut_1}
\left[ \hQb(\mathbf{r};t), \hYb_{\nu'} (\mathbf{r'};t) \right]  =  -i \hbar \I \delta \left( \nu - \nu' \right) \delta \left( \rb - \rb' \right) \quad \rb, \rb' \in V,
\end{equation}
\begin{equation}
\left[ \hPib (\mathbf{r};t), \hA (\mathbf{r'};t) \right] = -i \hbar \boldsymbol\delta^\perp \left( \rb - \rb' \right)  \quad \rb, \rb' \in V_\infty,
\end{equation}
while all remaining commutators vanish; here $\I$ is the three-dimensional unit tensor,  $\boldsymbol\delta^\perp(\rb)=\I\delta(\rb)- \boldsymbol\delta^\parallel(\rb)$ and $\boldsymbol\delta^\parallel(\rb)=\boldsymbol\nabla\boldsymbol\nabla(1/4\pi r)$ (e.g., \cite{cohen-tannoudji_photons_1997}).
According to relation \ref{eq:polfield0} we introduce the vector field operator $\hat{\mathbf{P}}(\mathbf{r};t)$ corresponding to the polarization density field $\mathbf{P}$,
\begin{align}
\label{eq:Polq}
\hat{\mathbf{P}}(\rb;t) &= \int_0^\infty d\nu\, \alpha_\nu
\hat{\mathbf{Y}}_\nu(\rb;t).
\end{align}
The Hamiltonian operator is given by
\begin{equation}
\label{eq:hamiltonian1}
    \hat{H} = \hat{H}_{Coul}+\hat{H}_{rad} +\hat{H}_{mat}
\end{equation}
where $\hat{H}_{mat}$, $\hat{H}_{Coul}$ and $\hat{H}_{rad}$ are obtained from \ref{eq:hmat}-\ref{eq:hrad} by substituting each physical variable with the corresponding operator.

\subsection{Heisenberg Picture}

In this paper, we study the evolution of the matter and radiation field observables in the Heisenberg picture. With an abuse of notation, we indicate with $\hat{O}(t)$ the operator $\hat{O}$ in the Heisenberg picture and with $\hat{O}^{(S)}$ the same operator in the Schr\"odinger picture, thus $\hat{O}(t=0)=\hat{O}^{(S)}$.

The Heisenberg equation for a time-invariant operator $\hat{O}$ is (e.g., \cite{cohen-tannoudji_photons_1997})
\begin{equation}
    \dot{\hat{O}} = \frac{1}{i \hbar} [ \hat{O}, \hat{H} ],
\end{equation}
where the Hamiltonian operator is given by \ref{eq:hamiltonian1}. This equation has to be solved with the initial condition $\hat{O}(t=0)=\hat{O}^{(S)}$. The time evolution of the expectation value of the observable $\hat{O}(t)$ is given by $\langle{\hat{{O}}}\rangle_{\psi_0} =  \langle\psi_0|\hat{O}(t)\psi_0\rangle$ where $|\psi_0\rangle$ is the initial quantum state of the system.

The equations of motion for the matter field operators
$\hQb(\mathbf{r};t), \hYb_\nu(\mathbf{r};t)$ and for the radiation field operators $\hPib(\mathbf{r};t), \hA(\mathbf{r};t)$ follow by evaluating their commutators
with the Hamiltonian. They have the same algebraic structure of the equation governing the corresponding classical quantities (e.g., \cite{cohen-tannoudji_photons_1997}, \cite{suttorp_field_2004}): they coincide with Eqs. \ref{eq:YQc} and \ref{eq:perpc} as long as we substitute the classical vector fields with the corresponding operators in the Heisenberg picture.

Due to the intrinsic spatial inhomogeneity of the problem, a direct solution of the Heisenberg equations for the matter field operators and the radiation field operators is very challenging. We overcome this problem in the following way. First, we expand the matter field operators and the radiation field operators in terms of suitable sets of vector fields depending only on space (modal expansion). Then, we express the Hamiltonian operator in terms of the coordinate operators of the matter fields and the coordinate operators of the radiation fields and their conjugate momenta. Eventually, we derive the Heisenberg equations for the coordinate and conjugate momentum operators that we solve using standard techniques. Once the polarization density field operator has been evaluated, the electric field operator is evaluated using the dyadic Green's function for vacuum.

\section{Modal expansion of the vector field operators}
In this section, we introduce the bases that we use to expand the fundamental vector field operators of the problem.
\label{sec:modal}
\subsection{Matter field operators}

We represent the matter field operators, which are defined in $V$, by applying the Helmholtz decomposition for vector fields defined in a bounded domain. The vector field operator $\hat{\mathbf{Y}}_\nu\rp$, for any $\nu$, is expressed as
\begin{equation}
\hat{\mathbf{Y}}_\nu\rp= \hat{\mathbf{Y}}_\nu^\parallel\rp + \hat{\mathbf{Y}}_\nu^\perp\rp,
\end{equation}
where $\hat{\mathbf{Y}}_\nu^\parallel$ is the longitudinal component of $\hat{\mathbf{Y}}_\nu$ and $\hat{\mathbf{Y}}_\nu^\perp$ is the transverse component. The longitudinal component is irrotational and solenoidal in $V$, and its normal component to $\partial V$ is equal to $\hat{\mathbf{Y}}_\nu^\parallel\cdot \n$. The transverse component is solenoidal in $V$, its normal component on $\partial V$ is equal to zero, and its curl in $V$ is equal to the curl of $\hat{\mathbf{Y}}_\nu$. This decomposition is unique. The vector fields $\hat{\mathbf{Y}}_\nu^\parallel$ and $\hat{\mathbf{Y}}_\nu^\perp$ are orthogonal according to the scalar product $\langle \hat{\mathbf{Y}}_\nu^\parallel, \hat{\mathbf{Y}}_\nu^\perp \rangle$. We represent the vector field operators $\{\hat{\mathbf{Q}}_\nu\}$ in the same way. 

By following \cite{forestiere_time-domain_2021}, we now expand the longitudinal components of $\hat{\mathbf{Y}}_\nu$ and $\hat{\mathbf{Q}}_\nu$ in terms of the static longitudinal modes of the dielectric object, and the transverse components in terms of the static transverse modes. The static longitudinal modes are the eigenfunctions of the electrostatic integral operator defined in Eq. \ref{eq:EQS} of the appendix \ref{sec:StaticModes}. The static transverse modes are the eigenfunctions of the magnetostatic integral operator defined in Eq. \ref{eq:MQS} of the Appendix \ref{sec:StaticModes}. Both integral operators have a discrete spectrum. Both the longitudinal modes and the transverse modes are orthonormal according to the scalar product $\langle {\bf F}, {\bf G} \rangle$. The set of static longitudinal modes $\{\Wl_m\rp, m=1,2,...\}$ is a base for the space of longitudinal vector fields defined on $V$, and the set of static transverse modes $\left\{\mathbf{U}_m^\perp\rp, m=1,2,...\right\}$ is a base for the space of transverse vector fields defined in $V$. They satisfy the closure relation
\begin{multline}
\label{eq:closure}
    \sum_{m} [\mathbf{U}_m^\parallel(\rb)\mathbf{U}_m^\parallel(\rb')+\mathbf{U}_m^\perp(\rb)\mathbf{U}_m^\perp(\rb')]=\\ = \overleftrightarrow{I}\delta(\rb-\rb') \quad \rb, \rb' \in V\,.
\end{multline}
Both sets of modes are dimensionless quantities. 

The vector field operators $\hat{\mathbf{Q}}_\nu^a$ and $\hat{\mathbf{Y}}_\nu^a$, with $a=\parallel, \perp$, are represented as:
\begin{align}
\label{eq:ylong}
\hat{\mathbf{Y}}_\nu^a(\rb;t) &=  \displaystyle\sum_{m} \hat{y}_{\nu,m}^a(t) \mathbf{U}_m^a\rp , \\ 
\label{eq:qlong}
\hat{\mathbf{Q}}_\nu^a(\rb;t) &=  \displaystyle\sum_{m} \hat{q}_{\nu,m}^a(t)\mathbf{U}_m^a\rp,
\end{align}
where $\left\{\hat{y}_{\nu,m}^a\right\}$ is the set of the coordinate operators of $\hat{\mathbf{Y}}_\nu^a$ and $\left\{\hat{q }_{\nu,m}^a\right\}$ is the set of the coordinate operators of $\hat{\mathbf{Q}}_\nu^a$. Since the static longitudinal and transverse modes are real functions, the coordinate operators are Hermitian. They obey the equal time commutation relations 
\begin{align}
\label{eq:etcryl}
\left[ \hat{q}_{\nu, m}^a, \hat{y}_{\nu', m'}^{a'} \right] &= -i \hbar \delta \left( \nu - \nu' \right) \delta_{m m'}\delta_{a a'},
\end{align}
for $m, m' = 1,2,\ldots$ and $a, a'=\parallel, \perp$, while all other commutators vanish. The operator $\hat{q}_m^a$ is canonically conjugate to the operator $\hat{y}_m^a$.

The polarization field operator $\hat{\mathbf{P}}$ is expressed as
\begin{align}
\label{eq:plong}
\hat{\mathbf{P}}(\rb;t) &=  \displaystyle\sum_{m} [\hat{p}_{m}^\parallel(t) \mathbf{U}_m^\parallel\rp +\hat{p}_{m}^\perp(t) \mathbf{U}_m^\perp\rp]
\end{align}
where
\begin{align}
 \label{eq:pola}
    \hat{p}_m^a = \int_0^ \infty d\nu\, \alpha_\nu \hat{y}_{\nu,m}^a,
\end{align}
with $a=\parallel, \perp$; $\{\hat{p}_m^\parallel\}$ and $\{\hat{p}_m^\perp\}$ are the sets of coordinate operators of $\hat{\mathbf{P}}$.

\subsection{Radiation field operators}

We use the transverse-plane wave modes
\begin{equation}
\label{eq:planewave}
    \mathbf{w}_\mu \rp = \frac{1}{\left(2\pi\right)^{3/2}} \eps_{s,\mathbf{k}} e^{i \mathbf{k} \cdot \rb}
\end{equation}
to represent the radiation field operators $\hat{\mathbf{A}}(\rb)$ and $\hPib(\rb)$; ${\bf k} \in \mathbb{R}^3$ is the propagation vector,  $\left\{ \eps_{s,\mathbf{k}}\right\}$ are the polarization unit vectors with $\eps_{s,\mathbf{k}} = \eps_{s,-\mathbf{k}}$ and $s=1,2$.
The two polarization vectors are orthogonal among them,  $\eps_{1,\mathbf{k}} \cdot \eps_{2, \mathbf{k}} = 0$, and are both transverse to the propagation vector, $\eps_{1,\mathbf{k}} \cdot \mathbf{k} = \eps_{2, \mathbf{k}} \cdot \mathbf{k}=0$.
We introduce the multi-index $\mu$ that represents the pair of parameters $\mathbf{k}$ and $s$, $\mu = \left(\mathbf{k},s \right)$, and we denote the set of all possible $\mu$ by $\mathcal{M}$. Furthermore, we denote $\displaystyle\sum_{s} \int_{\mathbb{R}^3} \dk\,\left( \cdot \right)$ by $\sum_\mu \left( \cdot \right)$. The modes $\left\{ \mathbf{w}_\mu \right\}$ are orthonormal in $V_{\infty}$,
\begin{equation}
    \langle \mathbf{w}_{\mu}, \mathbf{w}_\mu'  \rangle_{V_{\infty}} = \delta_{s,s'} \delta \left( \mathbf{k} - \mathbf{k}' \right).
\end{equation}
These modes are also dimensionless quantities.

We represent $\hat{\Ab}(\rb)$ and $\hPib(\rb)$ as:
\begin{align}
\label{eq:aperp0}
     \hat{\Ab}(\rb;t) &= \sum_{\mu} \hat{A}_\mu(t) \mathbf{w}_\mu \rp, \\
\label{eq:piperp} 
     \hPib(\rb;t) &= \sum_{\mu} \hat{\Pi}_\mu(t) \mathbf{w}_\mu \rp,
\end{align}
where $\{\hat{A}_\mu\}$ is the set of coordinate operators of $\hat{\Ab}$ and $\{\hat{\Pi}_\mu\}$ is the set of coordinate operators of $\hPib$. Since $\hat{\Ab}$ and $\hPib$ are Hermitian and the modes $\{\mathbf{w}_\mu\}$ are complex with $\mathbf{w}_\mu^* = \mathbf{w}_{-\mu}$, we have $\hat{A}_\mu^\dagger$ = $\hat{A}_{-\mu}$ and $\hat{\Pi}_\mu^\dagger$ = $\hat{\Pi}_{-\mu}$ where the multi-index $-\mu$ denotes the set $\left(-\mathbf{k},s \right)$. 
The coordinate operators $\{\hat{A}_{\mu}\}$ and $\{\hat{\Pi}_{\mu}\}$ obey the equal time commutation relations
\begin{equation}
\label{eq:etcra}
\left[ \hat{\Pi}_{\mu'}, \hat{A}_{\mu}^\dagger\right] = -i \delta_{s,s'} \delta \left( \mathbf{k} - \mathbf{k}' \right),
\end{equation}
for any couple $\mu, \mu'\in \mathcal{M}$, while all other commutators vanish. The operator $\hat{\Pi}_{\mu}$ is canonically conjugate to the operator $\hat{A}_{\mu}^\dagger$. The coordinate operators of the radiation field commutate with the coordinate operators of the matter fields.

\section{Modal expansion of the Hamiltonian Operator}
\label{sec:modalh}

The terms of the Hamiltonian operator \ref{eq:hamiltonian1} are given by the expressions \ref{eq:hmat}-\ref{eq:hrad} by substituting each classical physical variable with the corresponding operator. Now, we express the individual terms of $\hat H$ as functions of the coordinate operators and their conjugate momenta introduced in the previous Section.

\subsection{Coulomb energy} 
Only the longitudinal component of the matter field contributes to the Coulomb interaction energy $\hat{H}_{Coul}$. We have
\begin{equation}
\label{eq:Hc1}
    \hat{H}_{Coul} = \frac{1}{2\varepsilon_0} \sum_{m}  \frac{1}{ \kappa_m^\parallel} \int_0^\infty d\nu \int_0^\infty d\nu'  \alpha_{\nu}\alpha_{\nu'} \hat{y}_{\nu,m}^{\parallel}\hat{y}_{\nu',m}^{\parallel},
\end{equation}
where $\kappa_m^\parallel$ is the eigenvalue associated with $ \mathbf{U}_m^\parallel$ (see Eq. \ref{eq:EQS} of appendix \ref{sec:StaticModes}). By using \ref{eq:pola} we obtain
\begin{equation}
\label{eq:Hc2}
    \hat{H}_{Coul} = \frac{1}{2\varepsilon_0} \sum_{m}  \frac{1}{\kappa_m^\parallel} {\hat{p}_{m}}^{\parallel \,2}.
\end{equation}
The static longitudinal modes of the dielectric object diagonalize the Coulomb interaction energy \cite{forestiere_quantum_2020}.

\subsection{Radiation energy} 

The expression of $\hat{H}_{rad}$ in terms of the canonically conjugate coordinate operators of the radiation field is
\begin{equation}
    \hat{H}_{rad} = \sum_{\mu} \left( \frac{1}{2 \varepsilon_0} \hat{\Pi}_\mu^\dagger \hat{\Pi}_\mu  + \frac{\varepsilon_0 \omega_\mu^2}{2} \hat{A}_\mu^\dagger \hat{A}_\mu \right),
\end{equation}
where 
\begin{equation}
    \omega_\mu = c_0 k.
\end{equation}
The transverse plane wave modes diagonalize $ \hat{H}_{rad}$.

\subsection{Matter energy} 

The matter term $\hat{H}_{mat}$ has three contributions:
\begin{equation}
\label{eq:Hk}
    \hat{H}_{mat} = \hat{H}_{mat}^{'}+\hat{H}_{mat}^{''}+\hat{H}_{mat}^{'''}
\end{equation}
where
\begin{equation}
    \hat{H}_{mat}^{'} = \sum_{m}\int_0^\infty d\nu  \left[\frac{1}{2}(\hat{q}_{\nu,m}^{\parallel \,2} + \hat{q}_{\nu,m}^{\perp \, 2})+\frac{\nu^2}{2}(\hat{y}_{\nu,m}^{\parallel \, 2}+\hat{y}_{\nu,m}^{\perp \, 2})\right],
\end{equation}
\begin{equation}
    \hat{H}_{mat}^{''} =- \sum_
    {\substack{m,\mu}}\int_0^\infty d\nu\,  \alpha_\nu \left(\hat{q}_{\nu,m}^\parallel R_{m\mu}^\parallel + \hat{q}_{\nu,m}^{\perp} R_{m\mu}^\perp\right)\hat{A}_\mu,
\end{equation}
\begin{equation}
    \hat{H}_{mat}^{'''} = \int_0^\infty d\nu\, \frac{\alpha_\nu^2}{2}  \sum_{\mu', \mu}W_{\mu'\mu}\hat{A}_{\mu'}^\dagger \hat{A}_\mu,
\end{equation}
and
\begin{align}
    R_{m\mu}^{a}&=\langle {\mathbf{U}}_m^{a},{\mathbf{w}_\mu} \rangle, \\
   W_{\mu'\mu}&=\langle {\mathbf{w}_{\mu'}}, {\mathbf{w}_\mu} \rangle,
\end{align}
with $a=\parallel, \perp$. 

The terms $\hat{H}_{Coul}$, $\hat{H}_{rad}$ and $\hat{H}_{mat}^{'}$ are diagonal because of the expansion bases we have used, whereas $\hat{H}_{mat}^{''}$ and $\hat{H}_{mat}^{'''}$ are not diagonal. The term $ \hat{H}''_{mat}$ takes into account the interaction between matter and radiation fields; $ \hat{H}'''_{mat}$ is the diamagnetic term, which is also called the $A^2$ term (e.g. \cite{frisk_kockum_ultrastrong_2019}). As we shall see, our approach allows us to take into account the diamagnetic term $\hat{H}_{mat}^{'''}$ without making any approximation. Therefore, we only have to address the difficulties arising from the nondiagonal term $\hat{H}_{mat}^{''}$. The use of static longitudinal and transverse modes of the dielectric object allows us to overcome these difficulties as in the classical framework \cite{forestiere_time-domain_2021}.

\section{Heisenberg Equations}
\label{sec:heisenberg}

In this section, we first formulate the equations of motion for the coordinate operators of the matter fields and the radiation field in the Heisenberg picture. Then, we eliminate the coordinate operators of the radiation field and derive the equation of motion for the coordinate operators of the matter fields. 

\subsection{Matter and radiation fields}

\subsubsection{Matter}

The equations governing the time evolution of $\hat{y}_{\nu,m}^\parallel$ and $\hat{q}_{\nu,m}^\parallel$, with $m=1,2,3\ldots$ and $0\le\nu<\infty$, are
\begin{subequations}
\label{eq:heispar}
\begin{align}
    \label{eq:heisxpar}
    \dot{\hat{y}}_{\nu,m}^\parallel&=\hat{q}_{\nu,m}^\parallel -\alpha_\nu\sum_{\mu}R_{m\mu}^\parallel \hat{A}_\mu,\\
    \label{eq:heisppar}
    \dot{\hat{q}}_{\nu,m}^\parallel&=-\nu^2\hat{y}_{\nu,m}^\parallel - \frac{\alpha_\nu }{\varepsilon_0\kappa_m^\parallel}\int_0^ \infty d\nu'\, \alpha_{\nu'} \hat{y}_{\nu',m}^\parallel\,.
    \end{align}
    \end{subequations}
The equations governing the time evolution of $\hat{y}_{\nu,m}^\perp$ and $\hat{q}_{\nu,m}^\perp$, with $m=1,2,3\ldots$ and $0\le\nu<\infty$, are 
\begin{subequations}
\label{eq:heisperp}
\begin{align}
    \label{eq:heispperp1}
    \dot{\hat{y}}_{\nu,m}^\perp&=\hat{q}_{\nu,m}^\perp -\alpha_\nu \sum_{\mu}R_{m\mu}^\perp \hat{A}_\mu,\\
    \label{eq:heispperp2}
    \dot{\hat{q}}_{\nu,m}^\perp&=-\nu^{2}\hat{y}_{\nu,m}^\perp.
     \end{align}
    \end{subequations}
These equations are solved with the initial conditions
$\hat{y}_{\nu,m}^a(t=0)=\hat{y}_{\nu,m}^{a \,(S)}$, $\hat{q}_{\nu,m}^a(t=0)=\hat{q}_{\nu,m}^{a \,(S)}$ where $a=\parallel, \perp$. 

Combining equations \ref{eq:heispar} and \ref{eq:heisperp}, we eliminate conjugate momenta $\hat{q}_{\nu,m}^\parallel$ and $\hat{q}_{\nu,m}^\perp$, and obtain for $m=1,2,3\ldots$ and $0\le\nu<\infty$:
\begin{subequations}
\begin{align}
    \label{eq:heisppar3}
    &\ddot{\hat{y}}_{{\nu,m}}^\parallel+\nu^2\hat{y}_{\nu,m}^\parallel + \frac{\alpha_\nu }{\varepsilon_0\kappa_m^\parallel}\int_0^ \infty d\nu'\, \alpha_{\nu'} \hat{y}_{\nu',m}^\parallel\, = -\alpha_\nu\sum_{\mu}R_{m\mu}^\parallel \dot{\hat{A}}_\mu, \\
    &\label{eq:heispperp3}
    \ddot{\hat{y}}_{{\nu,m}}^\perp+\nu^{2} \hat{y}_{{\nu,m}}^\perp=-\alpha_\nu\sum_{\mu}R_{m\mu}^\perp \dot{\hat{A}}_\mu.
\end{align}
\end{subequations}
These equations are solved with the initial conditions
$\hat{y}_{\nu,m}^a(t=0)=\hat{y}_{\nu,m}^{a \, (S)}$ and $\dot{\hat{y}}_{\nu,m}^a(t=0)=\dot{\hat{y}}_{\nu,m}^{a \,(S)}$ where
\begin{equation}
    \label{eq:heisxpar10}
    \dot{\hat{y}}_{\nu,m}^{a \,(S)}=\hat{q}_{\nu,m}^{a \, (S)} -\alpha_\nu\sum_{\mu}R_{m\mu}^a \hat{A}_\mu^{(S)},
\end{equation}
with $a=\parallel, \perp$. 
 Once $\{\hat{y}_{\nu,m}^\parallel\}$, $\left\{\hat{y}_{\nu,m}^\perp\right\}$ and $\{\hat{A}_\mu\}$ have been evaluated, equations \ref{eq:heisxpar} and \ref{eq:heispperp1} allow us to calculate conjugate momenta $\{\hat{q}_{\nu,m}^\parallel\}$ and $\left\{\hat{q}_{\nu,m}^\perp\right\}$.

\subsubsection{Radiation}

The equations that govern the time evolution of $\hat{A}_\mu$ and $\hat{\Pi}_\mu$, with $\mu$ belonging to $\mathcal{M}$, are
\begin{subequations}
    \begin{align}
    \label{eq:heisa}
    \dot{\hat{A}}_\mu =&\frac{1}{\varepsilon_0}\hat{\Pi}_\mu,\\
      \nonumber
    \dot{\hat{\Pi}}_\mu=&-\varepsilon_0\omega_\mu^2\hat{A}_\mu+\sum_{\substack{{m}, {a}}}\int_0^\infty d\nu\,{\alpha_\nu}R_{\mu m}^a\hat{q}_{\nu,m}^{a}+
\\ \label{eq:heispi}
-& \int_0^\infty \alpha_\nu^2d\nu \sum_{\mu'} W_{\mu \mu'} \hat{A}_{\mu'}
\end{align}
\end{subequations}
where
\begin{equation}
    R_{\mu m}^{a}=\langle {{\mathbf{w}_\mu, \mathbf{U}}_m^{a}} \rangle=(R_{m\mu}^{a})^*.
    \end{equation}
These equations are solved with the initial conditions
$\hat{A}_{\mu}(t=0)=\hat{A}_{\mu}^{(S)}$ and $\hat{\Pi}_{\mu}(t=0)=\hat{\Pi}_{\mu}^{(S)}$.

Combining equations \ref{eq:heisxpar}, \ref{eq:heispperp1}, \ref{eq:heisa}, \ref{eq:heispi} and using the closure relation \ref{eq:closure} we eliminate the conjugate momenta $\hat{\Pi}_\mu$, $\hat{q}_{\nu ,m}^\parallel$ and $\hat{q}_{\nu,m}^\perp$. We obtain for any $\mu \in \mathcal{M}$:
\begin{equation}
    \label{eq:rad}
    \ddot{\hat{A}}_\mu+\omega_\mu^2\hat{A}_\mu=\frac{1}{\varepsilon_0}\sum_{\substack{{m}, {a}}} R_{\mu m}^a \int_0^\infty d\nu\, \alpha_{\nu} \dot{\hat{y}}_{\nu,m}^{a}.
\end{equation}
These equations are solved with the initial conditions
$\hat{A}_{\mu}(0)=\hat{A}_{\mu}^{(S)}$ and $\dot{\hat{A}}_{\mu}(0)=\hat{\Pi}_{\mu}^{(S)}/\varepsilon_0$. 
The coupling terms involving $W_{\mu \mu'}$, originating from the diamagnetic term $\hat{H}_{mat}^{'''}$, cancel out. This is a mere consequence of elimination of the conjugate momenta.
Once the operators $\{\hat{A}_\mu\}$ have been evaluated, equation \ref{eq:heisa} allow to calculate the conjugate momenta operators $\{\hat{\Pi}_\mu\}$. Solving the equation \ref{eq:rad} we obtain
\begin{equation}
\label{eq:rad1}
    \dot{\hat{A}}_\mu = \frac{1}{\varepsilon_0}\sum_{\substack{{m}, {a}}} R_{\mu m}^a\int_0^\infty d\nu\,\alpha_\nu \, w_{\mu}(t)* \dot{\hat{y}}_{\nu,m}^a(t) - \hat{\epsilon}_\mu(t)
\end{equation}
where
\begin{equation}
   w_{\mu}\tp = u \tp \cos(\omega_\mu t),
\end{equation}
$u \tp$ is the Heaviside function, and
\begin{equation}
   \hat{\epsilon}_\mu(t) =\omega_\mu \hat{A}_{\mu}^{(S)} \sin(\omega_\mu t) - \frac{1}{\varepsilon_0}{\hat{\Pi}}_\mu^{(S)} \cos(\omega_\mu t).
\end{equation}

The operators $\{\hat{\epsilon}_\mu(t)\}$ take into account the contribution of the initial conditions of the vector potential operator $\hat{\mathbf{A}}\rpt$ and the conjugate momentum $\hat{\mathbf{\Pi}}\rpt$. They are the coordinates of the vector field operator
\begin{equation}
\label{eq:Efree}
\hat{\mathbf{E}}^{free}_s\rpt =\sum_{\mu} [\omega_\mu \hat{A}_{\mu}^{(S)} \sin(\omega_\mu t) - \frac{{\hat{\Pi}}_\mu^{(S)}}{\varepsilon_0} \cos(\omega_\mu t)] \mathbf{w}_\mu\rp.
\end{equation}
The operator $\hat{\mathbf{E}}^{free}_s$ would describe the evolution of the solenoidal component of the electric field operator if the coupling between matter and electric field was absent. For this reason, throughout the paper we call it ``free solenoidal electric field operator".

\subsection{Equations of Motion for the Matter Coordinate Operators}
\label{sec:EqMotion}
 We now derive the equations governing the dynamics of the coordinate operators of the matter field in the time domain, then, we rewrite them in the Laplace domain.
 
\subsubsection{Time domain}
Using expressions \ref{eq:rad1} we eliminate the operators $\{\hat{A}_\mu\}$ in the systems of equations \ref{eq:heisppar3} and \ref{eq:heispperp3}. Therefore, the coordinate operators of the matter fields are governed by the system of integro-differential equations of convolution type (for $m=1,2,3\ldots$ and $0\le\nu<\infty$),
\begin{widetext}
\begin{align}
\label{eq:heisppar4}
    &\ddot{\hat{y}}_{\nu,m}^\parallel+\nu^2 \hat{y}_{\nu,m}^\parallel+\frac{\alpha_\nu }{\varepsilon_0\kappa_m^\parallel}\int_0^ \infty d\nu'\, \alpha_{\nu'} \hat{y}_{\nu',m}^\parallel  + \frac{\alpha_{\nu}}{\varepsilon_0}\sum_{\substack{{m'}, {a' }}}\int_0^\infty d\nu' \,\alpha_{\nu'} s_{mm'}^{\parallel \, a'}(t)* \dot{\hat{y}}_{{\nu'},m'}^{a'}(t) =\frac{\alpha_\nu}{\varepsilon_0} \hat{{d}}_{m}^{\parallel},\\
\label{eq:heispperp4}
    &\ddot{\hat{y}}_{\nu,m}^\perp + \nu^2\hat{y}_{\nu,m}^\perp + \frac{\alpha_\nu}{\varepsilon_0}\sum_{\substack{{m'}, {a'}}} \int_0^\infty d\nu' \,\alpha_{\nu'} s_{mm'}^{\perp \, a'}(t)* \dot{\hat{y}}_{\nu',m'}^{a'}(t) =\frac{\alpha_\nu}{\varepsilon_0}\hat{{d}}_{m}^{\perp},
\end{align}
\end{widetext} 
where
\begin{align}
 \label{eq:kerneltime}
   s_{mm'}^{a\,a'}(t)&=\sum_{\mu}\langle \Ua_m,\mathbf{w}_{\mu}\rangle \langle \mathbf{w}_{\mu}, \Ub_{m'}\rangle w_{\mu}(t),\\
\label{eq:forcing}
      \hat{{d}}_{m}^{\,a}(t)  & = \varepsilon_0\sum_{\mu}R_{m\mu}^a\hat{\epsilon}_{\mu}(t),
\end{align}
and $a,b = \parallel, \perp$. The operators $\{\hat{d}_{m}^{a }\}$ take into account the initial conditions of the radiation field operators.

The kernel $s_{mm'}^{a\,a'}$ in the convolution integrals can be expressed as (Appendix \ref{sec:GreenFunction})
\begin{equation}
\label{eq:Sinttime}
   s_{mm'}^{a\,a'}(t)=\frac{1}{c_0^2} \int_{V}\dV\int_{V}\dV'\, \Ua_m(\rb)\dot{{\overleftrightarrow{g}}}\,^\perp(\rb-\rb';t)\Ub_{m'}(\rb')
\end{equation}
where $\overleftrightarrow{g}^\perp(\rb;t)$ is the transverse dyadic Green's function for the vector potential, in the Coulomb gauge and in free space; the dot indicates the partial derivative with respect to time. The expression of $\dot{{\overleftrightarrow{g}}}\,^\perp(\rb;t)$ is given by Eq. \ref{eq:DerGreenTravTime} of Appendix \ref{sec:GreenFunction}. The convolution integrals describe the energy exchange between the longitudinal and transverse coordinate operators of the matter fields that is mediated by the radiation field. This is a non-conservative process because of the energy radiated toward infinity.

\subsubsection{Laplace domain}
To algebrize the system of equations \ref{eq:heisppar4} and \ref{eq:heispperp4} we use the unilateral Laplace transform. We denote the unilateral Laplace transform of ${x}\left(t\right)$ by $X(s)$ (namely $X(s)=\mathcal{L}\{x(t)\}= \int_{0}^{\infty} {x}\left(t\right)e^{-st}dt$), and the inverse Laplace transform by $\mathcal{L}^{-1}\{X(s)\}$. In our problem, the region of convergence of the Laplace transform includes the imaginary axis because of the loss due to matter and radiation. 

Equations \ref{eq:heisppar4} and \ref{eq:heispperp4} become in the Laplace domain
\begin{widetext}
\begin{align}
 \label{eq:yparallel5}
    \left(s^2+\nu^2\right)\hat{Y}_{\nu,m}^{\parallel} +\frac{\alpha_\nu }{\varepsilon_0\kappa_m^\parallel}\int_0^ \infty d\nu' \alpha_{\nu'} \hat{Y}_{{\nu'},m}^\parallel\,  + \frac{\alpha_\nu}{\varepsilon_0} \sum_{\substack{{m'}, {a'}}} \,s\, S_{mm'}^{\parallel \, a'} \int_0^\infty d\nu' \alpha_{\nu'} \hat{Y}_{{\nu'}, m'}^{a'} & =\frac{\alpha_\nu}{\varepsilon_0}\hat{D}_{\nu,m}^\parallel+\hat{C}_{\nu,m}^\parallel, \\
\label{eq:yperp5}
\left(s^2+\nu^2\right)\hat{Y}_{\nu,m}^{\perp} + \frac{\alpha_\nu}{\varepsilon_0} \sum_{\substack{{m'}, {a'}}}\,s\, S_{mm'}^{\perp \, a'} \int_0^\infty d\nu' \alpha_{\nu'} \hat{Y}_{{\nu'}, m'}^{a'}& = \frac{\alpha_\nu}{\varepsilon_0}\hat{D}_{\nu, m}^\perp+\hat{C}_{\nu, m}^\perp.
\end{align}
\end{widetext}
The unknown operators $\hat{Y}_{\nu, m}^\parallel(s)$ and $\hat{Y}_{\nu,m}^\perp(s)$ are the Laplace transform of $\hat{y}_{\nu,m}^\parallel(t)$ and $\hat{y}_{\nu,m}^\perp(t)$, respectively. The c-function
\begin{equation}
\label{eq:Sint}
   S_{mm'}^{a\,a'}=\frac{s}{c_0^2} \int_{V}\dV\int_{V}\dV'\, \Ua_m(\rb)\overleftrightarrow{G}^\perp(\rb-\rb';s)\Ub_{m'}(\rb')
\end{equation}
is the Laplace transform of $s_{mm'}^{a\,a'}(t)$, where $\overleftrightarrow{G}^\perp(\rb;s)$ is the Laplace transform of $\overleftrightarrow{g}^\perp(\rb;s)$, whose expression is given by \ref{eq:GreenT} in Appendix \ref{sec:GreenFunction}. The operators $\hat{D}_{\nu, m}^a(s)$ are the Laplace transforms of $\hat{d}_{\nu, m}^a(t)$ and the operators
\begin{multline}
 \label{eq:initmat}
    \hat{C}_{\nu,m}^a(s) = [s\hat{{y}}_{\nu,m}^{a\,(S)}
    +  \hat{\dot{y}}_{\nu,m}^{a\,(S)}] \\ + \frac{\alpha_\nu}{\varepsilon_0} \sum_{\substack{{m'}, {b}}} S_{mm'}^{a \, b} (s)\int_0^\infty d\nu'\, \alpha_{\nu'}\hat{{y}}_{\nu',m'}^{b\,(S)}
\end{multline}
 take into account the contribution due to the initial conditions of the coordinate operators of the matter fields.

\section{Evolution of the Polarization Density Field Operator}
\label{sec:polarization}

In this section, we first obtain the equations governing the evolution of the polarization coordinate operators and then we give the expressions for the polarization density field operator in terms of the driving operators. 

\subsection{Coordinate operators in the Laplace domain}
The coordinate operators of the polarization field in the Laplace domain $\{\hat{P}_m^\parallel(s)\}$ and $\{\hat{P}_m^\perp(s)\}$ are related to the coordinates of the matter field by $\hat{P}_m^a = \int_0^ \infty \alpha_\nu \hat{Y}_{\nu,m}^a d\nu$ (see relation \ref{eq:pola}). 

\subsubsection{Governing equations}

Multiplying both sides of Eqs. \ref{eq:yparallel5} and \ref{eq:yperp5} by $\alpha_\nu/(s^2+\nu^2)$, and integrating each term with respect to $\nu$ over $(0,\infty)$ we obtain the system of equations governing the coordinate operators of the polarization for $m=1,2,3\ldots$,
\begin{align}
 \label{eq:xparallel5}
    \left(\frac{1}{\tilde{\chi}}+ \frac{1}{\kappa_m^\parallel}\right)\hat{P}_{m}^{\parallel} + \sum_{m'} \,s\left(S_{mm'}^{\parallel \, \parallel} \hat{P}_{m'}^\parallel + S_{mm'}^{\parallel\perp}\hat{P}_{m'}^\perp\right)& =\hat{F}_{m}^\parallel, \\
\label{eq:xperp5}
\frac{1}{\tilde{\chi}}\hat{P}_{m}^{\perp}+\sum_{m'} s \left(S_{mm'}^{\perp\parallel} \hat{P}_{m'}^\parallel + S_{{m}m'}^{\perp\perp}\hat{P}_{m'}^\perp\right)& =\hat{F}_{m}^\perp,
\end{align}
where
\begin{equation}
\tilde{\chi}(s) = \frac{1}{\varepsilon_0}\int_0^\infty d\nu\, \frac{\alpha_\nu ^2}{ s^2+\nu^2}.
\end{equation}

The function $\tilde{\chi}(s)$ is the susceptibility of the dielectric in the Laplace domain (see Appendix \ref{sec:susceptibility}). In this paper we use the Drude-Lorentz model for the susceptibility of the medium,
\begin{equation}
 \label{eq:chi}
    \tilde{\chi}(s) = \frac{\omega_P^2}{s^2 +s \Gamma+ \omega_0^2},
\end{equation}
where $\omega_P$ is the plasma frequency of the medium, $\omega_0$ is the resonance frequency, and $\Gamma$ is the damping rate of the material. 

The operators $\hat{F}_{m}^\parallel(s)$ and $\hat{F}_{m}^\perp(s)$, with $a=\parallel, \perp$, are known. They are given by 
\begin{equation}
    \hat{F}_{m}^a(s)=\hat{{F}}_{m}^{a\, (e)}+\hat{{F}}_{m}^{a\, (p)},
\end{equation}
where
\begin{subequations}
\begin{align}
      \hat{{F}}_{m}^{a \, (e)}  & = \varepsilon_0\sum_{\mu}R_{m\mu}^a\hat{\mathcal{E}}_{\mu}(s), \\
   \nonumber
   \hat{{F}}_{m}^{a\,(p)}&= \frac{1}{\tilde{\chi}}\int_0^\infty d\nu\, \frac{\alpha_\nu}{s^2+\nu^2} [s\hat{y}_{\nu,m}^{a\,(S)}+ \dot{\hat{y}}_{\nu,m}^{a\,(S)}] + \\ & \sum_{\substack{{m'}, {a'}}} S_{mm'}^{a \, a'} \, \int_0^ \infty d\nu\, \alpha_\nu \hat{y}_{\nu,m'}^{a'\,(S)},
\end{align}
\end{subequations}
and $\hat{\mathcal{E}}_{\mu}(s)$ is the Laplace transform of $\hat{\epsilon}_{\mu}(t)$,
\begin{equation}
\label{eq:freecords}
   \hat{\mathcal{E}}_\mu(s) = \hat{A}_{\mu}^{(S)} \frac{c_0^2k^2}{s^2+c_0^2k^2} - \frac{1}{\varepsilon_0}{\hat{\Pi}}_\mu^{(S)}\frac{s}{s^2+c_0^2k^2}.
\end{equation}

 Equations \ref{eq:xparallel5} and \ref{eq:xperp5} govern the evolution of the coordinate operators of polarization in the Laplace domain. The coefficients $\{sS_{mm'}^{a\,b}\}$ and the susceptibility $\tilde{\chi}(s)$ are c-functions. The known operators, which take into account the initial conditions of the matter field operators and the radiation field operators, are the driving terms of the coordinate operators of polarization. These equations have the same algebraic structure as the corresponding classical problem \cite{forestiere_time-domain_2021}.

\subsubsection{Transfer matrix}

In this context, it is convenient to express the transverse dyadic Green's function $\overleftrightarrow{G}^\perp(\rb;s)$ as $\overleftrightarrow{G}^\perp(\rb;s)=\overleftrightarrow{g}_0^\perp(\rb)+\overleftrightarrow{G}_d^\perp(\rb;s)$
(see Appendix \ref{sec:GreenFunction}) where $\overleftrightarrow{g}_0^\perp$ is the static transverse dyadic Green's function in free space and $\overleftrightarrow{G}_d^\perp$ is the dynamic part: $\overleftrightarrow{g}_0^\perp$ diverges as $1/r$ for $r\rightarrow0$, while $\overleftrightarrow{G}_d^\perp$ is a regular function of $\mathbf{r}$. From the definition of the static transverse modes of the dielectric object (see Appendix \ref{sec:StaticModes}), we obtain
\begin{equation}
\label{eq:extractsing}
   S_{mm'}^{\perp \perp}(s)=\frac{a^2 s}{c_0^2 \kappa_m^\perp}\delta_{mm'}+\delta S_{mm'}^{\perp \perp}(s),
\end{equation}
where $\kappa_m^\perp$ is the eigenvalue associated to the transverse mode $\mathbf{U}^{\perp}_m(\rb)$ and
\begin{equation}
\label{eq:deltaSint}
   \delta S_{mm'}^{\perp \perp}=\frac{s}{c_0^2} \int_{V}\dV\int_{V}\dV'\, \Ua_m(\rb) \overleftrightarrow{G}_d^\perp(\rb-\rb';s)\Ub_{m'}(\rb').
\end{equation}
Equation \ref{eq:extractsing} is a consequence of the orthogonality of the static transverse modes. Extracting the singularity $1/r$ also allows us to adopt a simpler numerical scheme for the computation of the coefficients $\delta S_{mm'}^{a\,b}$.

We now rewrite Eqs.  \ref{eq:xparallel5} and \ref{eq:xperp5} using a matrix notation. We have
\begin{equation}
    \label{eq:polmatr}
    \dss{M}(s)
    \left|
    \begin{array}{c}
    {\hat{\ds{P}}^\parallel} \\  
    {\hat{\ds{P}}^\perp}
    \end{array}
    \right|
    = \left|
    \begin{array}{c}
    \hat{\ds{F}}^\parallel  \\  
    \hat{\ds{F}}^\perp
    \end{array}
    \right|,
\end{equation}
where $\hat{\ds{P}}^\parallel= \left| {\hat{P}}_1^\parallel,\hat{P}_{2}^\parallel, \ldots \right|^\intercal$ is the column vector of the longitudinal coordinate operators of the polarization, $\hat{\ds{P}}^\perp=\left|\hat{P}_{1}^\perp, \hat{P}_{2}^\perp, \ldots\right|^\intercal$ is the column vector of the transverse coordinate operators, and $\dss{M}$ is the block matrix
\begin{equation}
    \dss{M} =
    \left|
    \begin{array}{cc}
    \dss{M}^{\parallel \parallel}     & \dss{M}^{\parallel \perp} \\
    \dss{M}^{\perp \parallel} & \dss{M}^{\perp \perp}
    \end{array}\right|.
    \label{eq:MatrixM}
\end{equation}
The elements of the blocks $\dss{M}^{\parallel \parallel}$,  $\dss{M}^{\perp \perp}$, $\dss{M}^{\parallel \perp}$, $\dss{M}^{\perp \parallel}$ are given by
\begin{equation}
   {M}^{\parallel \parallel}_{mm'}(s) = \left\{
    \begin{array}{cc}
     \frac{1}{\tilde{\chi}(s)} + \frac{1}{\kappa_m^\parallel} + s\, S_{mm}^{\parallel \parallel}(s)    & m=m'  \\
     s \,S_{mm'}^{\parallel\, \parallel}(s)    & m\ne m'
    \end{array}
    \right.
    \label{eq:MatrixMll}
\end{equation}
\begin{equation}
   {M}^{\perp \perp}_{mm'}(s) = \left\{
    \begin{array}{cc}
     \frac{1}{\tilde{\chi}(s)} +\,\frac{a^2 s^2}{c_0^2\kappa_m^\perp}+ s\, \delta S_{mm}^{\perp \perp}(s)    & m=m'  \\
     s\, \delta S_{mm'}^{\perp \perp}(s)    & m\ne m'
    \end{array}
    \right.
    \label{eq:MatrixMtt}
\end{equation}
\begin{equation}
   {M}^{\parallel \perp}_{mn}(s) = 
     s\, S_{mn}^{\parallel \perp}(s), \qquad    {M}^{\perp \parallel}_{mn}(s) = 
     s\, S_{mn}^{ \perp \parallel}(s).
      \label{eq:MatrixMlt}
\end{equation}
The vectors $\hat{\ds{F}}^\parallel = \left| \hat{{F}}^\parallel_1, \hat{{F}}^\parallel_2, \ldots \right|^\intercal
$ and $ \hat{\ds{F}}^\perp = \left| \hat{{F}}_1^\perp, \hat{{F}}_2^\perp, \ldots \right|^\intercal $ are column vectors describing the driving coordinate operators of the polarization. 

The coordinate operators of the polarization field operator are obtained by inverting \ref{eq:polmatr}. We have
\begin{equation}
    \label{eq:coupleds}
    \left|
    \begin{array}{c}
    {\hat{\ds{P}}^\parallel} \\
    {\hat{\ds{P}}^\perp}
    \end{array}
    \right|
    =  \dss{H}(s) \left|
    \begin{array}{c}
    \hat{\ds{F}}^\parallel  \\  
    \hat{\ds{F}}^\perp
    \end{array}
    \right|
\end{equation}
where $ \dss{H} = \dss{M}^{-1}$ is the transfer matrix of the dielectric object. It is equal to the Laplace transform of the impulse responses of the dielectric object in the corresponding classical problem, which has been extensively studied in \cite{forestiere_time-domain_2021}. The product between the element $H_{mm'}^{a\,a'}(s)$ of the matrix $\dss{H}(s)$ and the driving coordinate operator $\hat{F}_{m'}^{a'}(s)$ gives the contribution of $\hat{F}_{m'}^{a'}(s)$ to the coordinate operator $\hat{P}_m^a(s)$ with $a,a'=\parallel,\perp$, and $m, m'=1,2,...$\,.

\subsubsection{General properties}

The susceptibility of the material $\tilde{\chi}(s)$ accounts for the strength of the coupling between the matter and the electric field. It appears in the diagonal elements of the matrix $\dss{M}$ and in the expressions of the driving terms. 

The susceptibility $\tilde{\chi}(s)$ tends to zero for $|s|\rightarrow\infty$, while the amplitudes of $sS_{mm'}^{a\,a'}(s)$ and $s\delta S_{mm'}^{a\,a'}(s)$ tend to finite limits different from zero. For $|s|\rightarrow\infty$ the diagonal elements of the matrix $\dss{M}(s)$ diverge as $1/\tilde{\chi}(s)$ and the off-diagonal elements remain bounded, hence the elements of the matrix $\dss{H}(s)$ tend to zero as $\tilde{\chi}(s)$ for $|s|\rightarrow\infty$. Furthermore, the driving operators diverge as the square root of susceptibility. As a consequence, the coordinate operators of the polarization tend to the null operator for $|s|\rightarrow\infty$. Therefore, we must consider only a limited frequency interval $\left(0, \omega_{max} \right)$ to evaluate the impulse response $\dss{h}(t)$.

The second term on the left-hand side of Eq. \ref{eq:xparallel5} is responsible for the electroquasistatic (plasmon) oscillations of the medium. The first term in the expression \ref{eq:extractsing} is responsible for the magnetoquasistatic oscillations of the medium. The eigenvalues $\kappa_m^\parallel$ and $\kappa_m^\perp$ are positive dimensionless quantities that depend only on the shape of the object, they do not depend on its size; furthermore, $\kappa_m^\parallel \ge 2$ \cite{fredkin_resonant_2003,mayergoyz_electrostatic_2005,forestiere_magnetoquasistatic_2020}.

The coefficients $\{sS_{mm'}^{a\,a'}\}$, with $a,a'=\{\parallel, \perp\}$, and $\{s\,\delta S_{mm'}^{\perp\,\perp}\}$ describe the coupling between the longitudinal and the transverse coordinate operators due to the interaction of the polarization with the radiation. They account for the exchange of electromagnetic energy between the modes $\Ua_m$ and $\Ub_{m'}$, which is a non-conservative process due to the radiated energy toward infinity. 

We introduce the dimensionless parameter $\gamma = |s|a/c_0$.  The amplitude of $sS_{mm'}^{a\,b}$ tends to zero as $\gamma^2$ for $\gamma\rightarrow0$, and the amplitude of $s\,\delta S_{mm'}^{\perp\,\perp}$ tends to zero as $\gamma^4$ (Appendix \ref{sec:small size}). The dimensionless parameter $\gamma$ allows to discriminate the regime in which the effects of the coupling between the coordinates operators of the polarization are negligible from the one in which the coupling role is important. For $|\tilde{\chi}|\gamma^2\ll1$ we can disregard the coupling terms in Eqs. \ref{eq:xparallel5} and \ref{eq:xperp5}, and Eq. \ref{eq:coupleds} reduces to
\begin{subequations}
\begin{align}
 \label{eq:xparallel5sm}
  \hat{P}_{m}^{\parallel} &\cong\frac{\kappa_m^\parallel}{\kappa_m^\parallel+\tilde{\chi}(s)} \tilde{\chi}(s)\hat{F}_{m}^\parallel, \\
\label{eq:xperp5sm}
\hat{P}_{m}^{\perp} &\cong\frac{c_0^2\kappa_m^\perp}{c_0^2\kappa_m^\perp+a^2 s^2\tilde{\chi}(s)} \tilde{\chi}(s)\hat{F}_{m}^\perp.
\end{align}
\end{subequations}
The constraint $|\chi|\gamma^2\ll1$ is certainly satisfied in the small size limit $a \ll \lambda_c$ where
$\lambda_c=\min\limits_{\omega}(c_0/[\omega\sqrt{|\chi(\omega)|}])$ and $\chi(\omega)=\tilde\chi(s=i\omega)$.

The static longitudinal modes diagonalize the contribution to the Hamiltonian of the electroquasistatic (Coulomb) interaction energy between the longitudinal modes, while the static transverse modes diagonalize the magnetostatic (Ampere) interaction energy between the transverse modes. The other interaction energy terms between the modes are not diagonalized. However, in the small-size limit $a \ll \lambda_c$, the contribution of these terms becomes negligible and the matrix $\dss{M}$ is quasi diagonal. Therefore, we expect that only a few static longitudinal and transverse modes are needed to calculate each element of the transfer matrix $\dss{H}$ of a dielectric object even when its size $2a$ is of the order of the characteristic length $\lambda_c$.

\subsection{Polarization density field operator in the Laplace domain}

In the Laplace domain the polarization density field operator is given by
\begin{equation}
    \boldsymbol{\hat{\mathcal{P}}}\left(\mathbf{r};s \right) =  \displaystyle\sum_{m} [\hat{P}_{m}^\parallel(s) \mathbf{U}_m^\parallel\rp +\hat{P}_{m}^\perp(s) \mathbf{U}_m^\perp\rp].
\end{equation}
To express this operator in terms of the driving operators, it is useful to introduce the driving vector field operator
\begin{equation}
  \boldsymbol{\hat{{F}}}(\mathbf{r};s) =    \boldsymbol{\hat{{F}}}^{\,(e)}(\mathbf{r};s)  +  \boldsymbol{\hat{{F}}}^{\,(m)}(\mathbf{r};s)
 \end{equation}
 where
\begin{subequations}
\begin{align}
 \boldsymbol{\hat{{F}}}^{\,(e)} &=\varepsilon_0
 \boldsymbol{\hat{\mathcal{E}}}^{free}_s, \\
  \label{eq:Pfree}
 \boldsymbol{\hat{{F}}}^{\,(m)} &= \frac{1}{\tilde{\chi}}\boldsymbol{\hat{\mathcal{P}}}^{\,free}+\frac{1}{c_0^2} \int_{V} \dV'\, s\,\overleftrightarrow{G}^\perp(\rb-\rb';s) \hat{\mathbf{P}}^{(S)}(\mathbf{r}),
\end{align}
\end{subequations}
with
\begin{equation}
\boldsymbol{\hat{\mathcal{P}}}^{\,free}(\mathbf{r};s) = \int_0^\infty d\nu \frac{\alpha_\nu}{s^2+\nu^2} [s\hat{\mathbf{Y}}_\nu^{(S)}(\mathbf{r})+\dot{\hat{\mathbf{Y}}}_\nu^{(S)}(\mathbf{r})],
\end{equation}
\begin{equation}
    \dot{\hat{\mathbf{Y}}}_\nu^{(S)}\rp = 
        \mathbf{Q}_\nu^{(S)}\rp
    -\alpha_\nu\mathbf{A}^{(S)}\rp,
\end{equation}
and
\begin{equation}
\hat{\mathbf{P}}^{(S)}(\mathbf{r})=\int_0^\infty d\nu \alpha_\nu \hat{\mathbf{Y}}_\nu^{(S)}(\mathbf{r}).
\end{equation}
The vector field operator $\boldsymbol{\hat{\mathcal{E}}}^{free}_s(\mathbf{r};s)$ is the Laplace transform of the free solenoidal electric field operator $\hat{\mathbf{E}}^{free}_s(\mathbf{r};t)$,
\begin{equation}
\boldsymbol{\hat{\mathcal{E}}}^{free}_s(\mathbf{r};s) =\sum_{\mu}\hat{\mathcal{E}}_\mu(s) \mathbf{w}_\mu\rp.
\end{equation}
where $\hat{\mathcal{E}}_\mu(s)$ is given by \ref{eq:freecords}.

Using \ref{eq:coupleds} we obtain for the polarization density field operator
\begin{equation}
\label{eq:Pols}
\boldsymbol{\hat{\mathcal{P}}}\left(\mathbf{r};s \right)=\ds{\mathbf{U}}^\intercal (\mathbf{r}) \dss{H}(s)\, \langle \ds{\mathbf{U}} \left(\mathbf{r}'\right), \boldsymbol{\hat{{F}}}\left(\mathbf{r}';s \right) \rangle
\end{equation}
where $\ds{\mathbf{U}}= \left| \mathbf{U}^\parallel_1, \mathbf{U}^\parallel_2, \ldots, \mathbf{U}^\perp_1, \mathbf{U}^\perp_2, \ldots\right|^\intercal$. We rewrite this relation as follows
\begin{equation}
\label{eq:Polf}
\boldsymbol{\hat{\mathcal{P}}}\left(\mathbf{r};s \right) = \int_V \dV'\, \overleftrightarrow{\Theta} (\mathbf{r},\mathbf{r'};s)\boldsymbol{\hat{{F}}}(\mathbf{r'};s)
\end{equation}
where the c-dyad $\overleftrightarrow{\Theta} (\mathbf{r},\mathbf{r}';s)$ is defined as
\begin{multline}
    \overleftrightarrow{\Theta} (\mathbf{r},\mathbf{r'};s) = \sum_{\substack{{m,m'}, {a,a' }}} H_{m m'}^{a\,a'}(s) \mathbf{U}^a_{m}  (\mathbf{r}) \mathbf{U}^{a'}_{m'}  (\mathbf{r}').
\end{multline}

The relation \ref{eq:Polf} is one of the most important results we have obtained with our approach. It allows to evaluate directly in the Laplace domain statistical functions like the expectation values of the polarization observable, the uncertainty and the correlation functions. For example, the expectation value of the polarization density field observable is given by
\begin{multline}
    \langle\boldsymbol{\hat{\mathcal{P}}}\left(\mathbf{r};s \right)\rangle_{\psi_0} = \int_V \dV'\, \overleftrightarrow{\Theta} (\mathbf{r},\mathbf{r'};s)\ \langle \boldsymbol{\hat{{F}}}(\mathbf{r'};s)\rangle_{\psi_0}
\end{multline}
where $\langle \boldsymbol{\hat{{F}}}(\mathbf{r'};s)\rangle_{\psi_0} =  \langle\psi_0|\boldsymbol{\hat{{F}}}(\mathbf{r'};s)|\psi_0\rangle$
is the expectation value of the driving field operator. The c-dyadic field $\overleftrightarrow{\Theta} (\mathbf{r},\mathbf{r'};s)$ and, hence, the transfer matrix play a crucial role.
\begin{figure*}
    \centering
    \includegraphics[width=\textwidth]{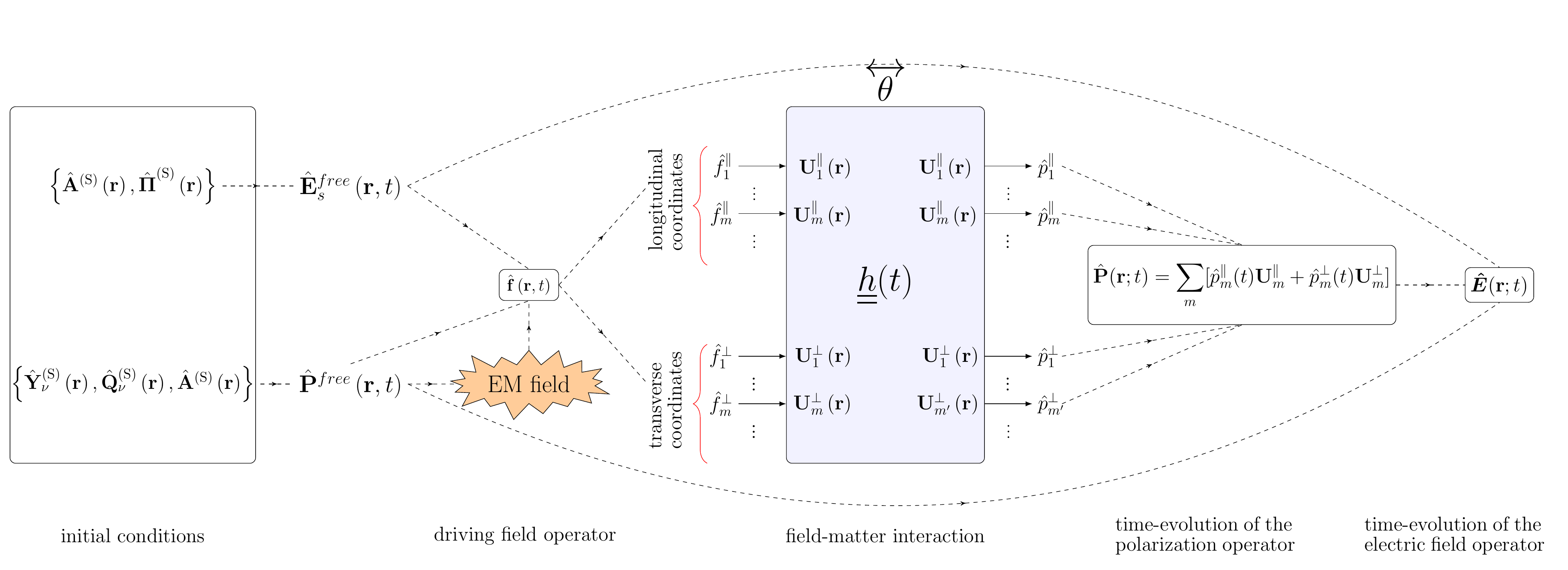}
     \caption{The above scheme summarizes the proposed approach. The conjugate operators in the Schr\"odinger picture $\left\{ \hat{\mathbf{A}}^{\left(S\right)} \left( \rb\right), \hPib^{\left(S\right)}\left( \rb \right)\right\}$ and $\left\{ \hat{\mathbf{Y}}^{\left(S\right)}_\nu\left( \rb \right), \hat{\mathbf{Q}}^{\left(S\right)}_\nu\left( \rb \right) \right\}$ play the role of initial conditions for the radiation field and the matter field operators, respectively. They determine the ``free solenoidal electric field operator" $\hat{\mathbf{E}}^{free}_s\rpt$ and the ``free polarization field operator" $\hat{\mathbf{P}}^{free}\rpt$, which in turn determine the ``driving field operator" $\hat{\mathbf{f}}(\rb;t) $. In particular, $\hat{\mathbf{P}}^{free}\rpt$ contributes to $\hat{\mathbf{f}}$ both directly and through its contribution to the radiation field operators. 
     The operator $\hat{\mathbf{f}}$ is then decomposed in terms of its ``coordinate operators" along the longitudinal $\mathbf{U}_m^\parallel$ and transverse $\mathbf{U}_m^\perp$ modes of the objects. The coordinates of polarization density field operator $\hat{\mathbf{P}}(\rb;t)$ are given by the convolution between the impulse response matrix $\dss{h}(t)$ defined in Eq. \ref{eq:ht} and the coordinate operators of $\hat{\mathbf{f}}\rpt$. $\dss{h}(t)$ is the impulse response matrix that we obtain in the framework of classical electrodynamics. $\overset\leftrightarrow{\theta}$ is the kernel of the integral operator that gives the polarization density field operator as functions of $\hat{\mathbf{f}}$ (Eq. \ref{eq:Polft}), and it is a linear combinations of the elements of $\dss{h}$. Eventually, $\hat{\mathbf{P}}(\rb;t)$ and $\hat{\mathbf{E}}^{free}\rpt$ determine the electric field operator $\hat{\mathbf{E}}\rpt$ through the dyadic Green's function for the free space, see Eqs. \ref{eq:ElectricField2} and \ref{eq:ELapl4}.}
    \label{fig:Scheme}
\end{figure*}

\subsection{Polarization density field operator in the time domain}

The polarization density field operator in the time domain has the expression
\begin{equation}
    \hat{\mathbf{P}}(\rb;t) =  \displaystyle\sum_{m} [\hat{p}_{m}^\parallel(t) \mathbf{U}_m^\parallel\rp +\hat{p}_{m}^\perp(t) \mathbf{U}_m^\perp\rp]
\end{equation}
where $\hat{\ds{p}}^\parallel(t)$ and $\hat{\ds{p}}^\perp(t)$ are the inverse Laplace transforms of $\hat{\ds{P}}^\parallel(s)$ and $\hat{\ds{P}}^\perp(t)$, respectively. The operators  $\hat{\ds{p}}^\parallel(t)$ and $\hat{\ds{p}}^\perp(t)$ are given by
\begin{equation}
    \label{eq:coupledt}
    \left|
    \begin{array}{c}
    {\hat{\ds{p}}^\parallel}(t) \\
    {\hat{\ds{p}}^\perp}(t)
    \end{array}
    \right|
    =  \dss{h}(t) * \left|
    \begin{array}{c}
    \hat{\ds{f}}^\parallel(t)  \\  
    \hat{\ds{f}}^\perp (t)
    \end{array}
    \right|
\end{equation}
where 
\begin{equation}
    \dss{h}(t)=\mathcal{L}^{-1}\{\dss{H}(s)\}
    \label{eq:ht}
\end{equation} 
is the impulse response matrix of the dielectric object; $\hat{\ds{f}}^\parallel(t)$ and $\hat{\ds{f}}^\perp(t)$ are the inverse Laplace transforms of $\hat{\ds{F}}^\parallel(s)$ and $\hat{\ds{F}}^\perp(s)$, respectively.
The convolution product between the element $h_{mm'}^{a\,a'}(t)$ of the matrix $\dss{h}(t)$ and the driving coordinate operator $\hat{f}_{m'}^{a'}(t)$ gives the contribution of $\hat{f}_{m'}^{a'}(t)$ to the coordinate operator of the polarization $\hat{p}_{m}^a(t)$ with $a,a'=\parallel,\perp$ and $m,m'=1,2,...$\,.

The matrix $\dss{h}(t)$, whose elements are c-functions, is the impulse response matrix of the dielectric object in the classical framework. Therefore, the representation of the polarization field operator in terms of the static longitudinal and transverse modes of the dielectric object leads to the same advantages \cite{forestiere_time-domain_2021}.

By applying the inverse Laplace transform to \ref{eq:Polf} we immediately obtain
\begin{equation}
\label{eq:Polft}
\boldsymbol{\hat{{P}}}\left(\mathbf{r};t \right) = \int_V \dV'\, \overleftrightarrow{\theta} (\mathbf{r},\mathbf{r'};t) \oast {\hat{\bold{f}}}(\mathbf{r'};t)
\end{equation}
where the c-dyad $\overleftrightarrow{\theta} (\mathbf{r},\mathbf{r}';t)$ is given by
\begin{multline}
    \overleftrightarrow{\theta} (\mathbf{r},\mathbf{r'};t) = \sum_{\substack{{m,m'}, {a,a' }}} h_{m m'}^{a\,a'}(t) \mathbf{U}^a_{m}  (\mathbf{r}) \mathbf{U}^{a'}_{m'}  (\mathbf{r}');
\end{multline}
$\oast$ denotes the time convolution product between a dyad and a vector field. The driving operator $\hat{\bold{f}} (\mathbf{r};t)$ has two contributions,
\begin{equation}
  \hat{\bold{f}} (\mathbf{r};t) =   \hat{\bold{f}}^{\,(e)}(\mathbf{r};t)  +    \hat{\bold{f}}^{\,(m)}(\mathbf{r};t).
 \end{equation}
The first contribution is given by
\begin{equation}
   \hat{\bold{f}}^{\,(e)} =
 \varepsilon_0 \hat{\mathbf{E}}^{free}_s.
\end{equation}
The expression of the second contribution is
\begin{equation}
\begin{aligned}
 \bold{\hat{{f}}}^{\,(m)}(\mathbf{r};t) &= \eta(t)*\boldsymbol{\hat{{P}}}^{free}(\mathbf{r};t) \\ &+ \frac{1}{c_0^2} \displaystyle\int_{V}\dV'\,\dot{\overleftrightarrow{g}}^\perp(\rb-\rb';t) \boldsymbol{\hat{{P}}}^{free}(\mathbf{r};0)
\end{aligned}
\end{equation}
where
\begin{equation}
\label{eq:pfreet}
\boldsymbol{\hat{{P}}}^{free}(\mathbf{r};t) = \int_0^\infty d\nu\, \alpha_\nu \hat{\mathbf{Y}}_\nu^{free}\rpt,
\end{equation}
\begin{equation}
   \hat{\mathbf{Y}}_\nu^{free}\rpt=\hat{\mathbf{Y}}_\nu^{(S)}(\mathbf{r})\cos(\nu t) + \frac{1}{\nu}\dot{\hat{\mathbf{Y}}}_\nu^{(S)}(\mathbf{r})\sin(\nu t),
\end{equation}
and $\eta(t)=\mathcal{L}^{-1}\{1/\tilde{\chi}\}$.

The vector field operator $\hat{\mathbf{P}}^{free}$ takes into account the contribution of the initial conditions of the matter field operator $\hat{\mathbf{Y}}_\nu \rpt$ and the conjugate momentum $\hat{\mathbf{Q}}_\nu\rpt$ for $0\le\nu<\infty$. $\hat{\mathbf{P}}^{free}$ would describe the evolution of the polarization density field operator if the interaction between the matter and the electromagnetic field was absent. For this reason, throughout the paper, we call it the ``free polarization field operator". The free polarization field operator $\hat{\mathbf{P}}^{free}$ 
coincides with the \lq\lq fluctuating dipole density distribution\rq\rq, and the free solenoidal electric field operator $\hat{\mathbf{E}}^{free}$ coincides with the \lq\lq purely fluctuating \rq\rq electric field introduced in \cite{drezet_quantizing_2017}. They account for the initial quantum state of the system in the Heisenberg picture and guarantee the unitarity of the full evolution of the system, as pointed out in \cite{drezet_quantizing_2017}. In particular, the contribution of the free solenoidal electric field operator is very important in the inhomogeneous medium problems \cite{drezet_equivalence_2017,dorier_canonical_2019,dorier_critical_2020}. 

The scheme shown in Figure \ref{fig:Scheme} summarizes the approach that this paper proposes. Once the evolution of the polarization density field operator has been evaluated, the evolution of the electric field operator is determined as described in the next section. We highlight that in the limit of zero interaction between the matter field and the electric field we obtain the expressions corresponding
to the uncoupled matter and the vacuum electromagnetic field. This is a very important check of consistency as already pointed out in the recent literature \cite{drezet_quantizing_2017, drezet_equivalence_2017,dorier_canonical_2019,dorier_critical_2020}. 

\subsection{Statistical functions of the polarization density field operator}
\label{sec:Stat}
The knowledge of the impulse responses allows to evaluate the statistical functions such as the expectation values of the polarization observable, the uncertainty, and the correlation functions. In the following, as examples, we consider the expectation value and the correlation.

When the system is in the initial quantum state $|\psi_0\rangle$ the time evolution of the expectation value of the polarization density field operator $\langle\bold{\hat{{P}}}\left(\mathbf{r};t \right)\rangle_{\psi_0}$ is given by
\begin{equation}
    \langle\bold{\hat{{P}}}\left(\mathbf{r};t \right)\rangle_{\psi_0} = \int_V \dV'\, \overleftrightarrow{\theta} (\mathbf{r},\mathbf{r'};t)\oast \langle \bold{\hat{{f}}}(\mathbf{r'};t)\rangle_{\psi_0}
\end{equation}
where $\langle \bold{\hat{{f}}}(\mathbf{r};t)\rangle_{\psi_0} =  \langle\psi_0|\bold{\hat{{f}}}(\mathbf{r};t)|\psi_0\rangle$
is the expectation value of the driving field operator in the time domain. 

We now introduce the c-functions representing the two-time correlation function between the coordinates of the driving field operators
\begin{widetext}
\begin{equation}
    F_{{m_1},{m_2}}^{a_1,a_2}(t_1,t_2)=\int_V\int_V \dV_1\dV_2\mathbf{U}_{m_1}^{a_1}(\bold{r}_1) \langle \bold{\hat{{f}}}(\mathbf{r}_1;t_1)\bold{\hat{{f}}}(\mathbf{r}_2;t_2)\rangle_{\psi_0}\mathbf{U}_{m_2}^{a_2}(\bold{r}_2)
\end{equation}
\end{widetext}
and the function 
\begin{equation}
    Z_{m_1,m_2}^{a_1,a_2}(\bold{r}_1,\bold{r}_2)=\mathbf{U}_{m_1}^{a_1}(\bold{r}_1) \cdot \mathbf{U}_{m_2}^{a_2}(\bold{r}_2),
\end{equation}
where $ \langle \bold{\hat{{f}}}(\mathbf{r}_1;t_1)\bold{\hat{{f}}}(\mathbf{r_2};t_2)\rangle_{\psi_0}=\langle\psi_0| \bold{\hat{{f}}}(\mathbf{r}_1;t_1)\bold{\hat{{f}}}(\mathbf{r_2};t_2)|\psi_0\rangle$ is a dyad. The correlation of the polarization density field operator $\langle\bold{\hat{{P}}}\left(\mathbf{r}_1;t_1 \right)\cdot\bold{\hat{{P}}}\left(\mathbf{r}_2;t_2 \right)\rangle_{\psi_0}=\langle\psi_0| \bold{\hat{{P}}}^\dagger(\mathbf{r}_1;t_1)\cdot \bold{\hat{{P}}}(\mathbf{r_2};t_2)|\psi_0\rangle$ is given by
\begin{widetext}
\begin{equation}
   \langle\bold{\hat{{P}}}\left(\mathbf{r}_1;t_1 \right)\cdot\bold{\hat{{P}}}\left(\mathbf{r}_2;t_2 \right)\rangle_{\psi_0} = \sum_{\substack{{m_1,{m'}_1, m_2,{m'}_2}\\{a_1,a'_1,a_2,a'_2}}}Z_{m_1,m_2}^{a_1,a_2}(\bold{r}_1,\bold{r}_2)\int_{0}^{\infty}\int_{0}^{\infty}d\tau_1d\tau_2 h_{m_1,m'_1}^{a_1,a'_1}(t_1-\tau_1) h_{m_2,m'_2}^{a_2,a'_2}(t_2-\tau_2) F_{{m'_1},{m'_2}}^{a'_1,a'_2}(\tau_1,\tau_2).
\end{equation}
\end{widetext}
The time evolution of these statistical functions depends on the convolutions between the impulse responses and the statistical functions of the driving field operator.

\section{Electric Field Operator}
\label{sec:Electric}

The electric field operator $\hat{\mathbf{E}}\left(\mathbf{r};t \right)$ is given by
\begin{equation}
    \hat{\mathbf{E}}\left(\mathbf{r};t \right)=-\frac{1}{\varepsilon_0}\hat \Pib\left(\mathbf{r};t \right) + \hat{\mathbf{E}}_c\left(\mathbf{r};t \right)
\end{equation}
where $\hat{\mathbf{E}}_c\left(\mathbf{r};t \right)$ is given by \ref{eq:scalp} with the operator $\hat{P}_n$ instead of the classical variable $P_n$. 
We now give the expression of the electric field operator $\hat{\mathbf{E}}$ as function of the polarization field density operator $\hat{\mathbf{P}}$. In Appendix G we give the expression of $\hat{\mathbf{E}}\rpt$ at any point of the space. However, it is convenient to distinguish between the region $V$ occupied by the dielectric object and the external region $V_\infty \backslash V$ in order to avoid dealing with principal value integrals. 

\subsection{Inside the dielectric object}

 The polarization density field operator is related to the electric field operator by equation
\begin{equation}
   \label{eq:constrelop}
    \hat{\mathbf{P}}\rpt = \varepsilon_0\displaystyle \zeta(t)*\hat{\mathbf{E}} \left(\mathbf{r};t \right)+ \hat{\mathbf{P}}^{free}(\mathbf{r};t).
\end{equation}
From this relation we obtain
\begin{equation}
\hat{\mathbf{E}}\rpt =\frac{1}{\varepsilon_0} \eta(t)*[\hat{\mathbf{P}} \left(\mathbf{r};t \right)-\hat{\mathbf{P}}^{free} \left(\mathbf{r};t \right)] \quad\text{in} \,V.
\label{eq:Einside}
\end{equation}

\subsection{Outside the dielectric object}

In the region outside the dielectric object $V_\infty \backslash V$ the electric field operator is given by (Appendix \ref{sec:electricfield})
\begin{equation}
    \boldsymbol{\hat{{E}}}(\mathbf{r};t) =-\mu_0\int_{V} \dV'\, \dot{\overleftrightarrow{g}} \left(\mathbf{r} - \mathbf{r}';t\right) \oast \dot{\hat{\mathbf{P}}}(\mathbf{r}';\tau)+ \hat{\mathbf{E}}^{free}(\mathbf{r};t),
    \label{eq:ElectricField2}
\end{equation}
where $\overleftrightarrow{g}$ is the time domain dyadic Green's function for the vector potential in the temporal gauge and in the free space,
\begin{equation}
    \hat{\mathbf{E}}^{free} = \hat{\mathbf{E}}^{free}_s +\hat{\mathbf{E}}^{free}_c,
\end{equation}
\begin{equation}
    \hat{\mathbf{E}}^{free}_c(\mathbf{r};t) = -\frac{u(t)}{\varepsilon_0}\int_{V} \dV'\, {\overleftrightarrow{g}}_0 (\mathbf{r} - \mathbf{r}') {\hat{\mathbf{P}}}^{free}(\mathbf{r}';0)
\end{equation}
and ${\overleftrightarrow{g}}_0 (\mathbf{r})$ is the static dyadic Green's function. The expression of $\dot{\overleftrightarrow{g}}\left(\mathbf{r} ;t\right)$ is given by \ref{eq:gtotdin} and the expression of ${\overleftrightarrow{g}}_0 (\mathbf{r})$ is given by \ref{eq:gtotstat}.

The statistical functions of any operator, including the electric field, can be derived starting from the expression of the polarization density field operator.

\section{Numerical procedure for the calculation of the impulse responses}
\label{sec:Num}

 Any desired statistical function of the observables can be expressed as an integral operator of the driving field operators, whose kernel is a multilinear expressions of the elements of the impulse response matrix (see Sec. \ref{sec:Stat}). In this section, we summarize the main steps of the numerical procedure for the calculation of the impulse response matrix of the dielectric object.

The procedure for the calculation of the impulse response matrix consists of five steps.

Step a) \textit{Numerical calculation of the static modes of the object} following appendix \ref{sec:Computation}. The static modes and the corresponding eigenvalues are independent of the size of the object thus, for any given shape, they must be computed only once. The integral operators are real, symmetric and positive. 
The linear dimensions of the mesh elements must be much smaller than the minimum spatial variation of the highest order mode we consider. 
LAPACK routines \cite{anderson_lapack_1999}, specialized for symmetric eigenvalue problems, require a computational time $T_{mode}$ that scales as $N^3$ where $N$ is the dimension of the matrix \cite{demmel_applied_1997}. 

Step b) \textit{Numerical calculation of the coupling coefficients} $\{S_{mm'}^{a\,b}(s)\}$ at $s=i\omega + \epsilon$ with $\epsilon \downarrow 0$ following appendix \ref{sec:CoefficientS}. We indicate with $t_a$ the time required for the calculation of a coupling coefficient at a single frequency. Even though $t_a$ is negligible compared to the time required for the calculation of the static modes, many frequency samples are required to accurately compute the impulse responses. 

Step c) \textit{Assembly of the matrix} $\dss{M}$. If $Q$ is the number of modes required to describe the matter field, the matrix $\dss{M}$ has dimension $Q \times Q$.  If $N_F$ is the number of required frequency samples, the overall assembly time is $T_a = N_F \times Q^2 \times t_a$.

Step d) \textit{Calculation of the element} $H_{m m'}^{a a'}$ of the matrix $\dss{H}$ solving the system of equations \ref{eq:xparallel5} and \ref{eq:xperp5} with $\hat{F}_{m'}^{a'} = 1$ and $\hat{F}_{m''}^c =0$ for $m''\ne m'$ and $c\ne a'$, using the standard LU decomposition. The inversion time $t_i$ scales as $Q^3$ for each frequency step, thus the overall inversion time $T_i = N_F t_i$ scales as $N_F Q^3$.


Step e) \textit{numerical calculation of the inverse Fourier transform } of $\dss{H}(\omega)$. The computational time $t_{fft}$ for the inverse Fourier transform of the single element of $\dss{H}$ scales as $N_F \times\log N_F$, thus the total computational time to calculate the impulse response matrix $T_{fft} = Q^2  \times t_{fft}$ scales as $ Q^2 \times N_F \times\log N_F$.

Summarizing, the total computational time is 
$T = T_{mode} + Q^2 \times \left( N_F \times  t_a  + t_{fft} \right) + N_F \times t_i$. The static mode must be calculated only one time, at the beginning, thus $T_{mode}$ is negligible with respect to the remaining terms. Furthermore, the computational time for the evaluation of coupling coefficient $N_F \times t_a$ is dominant compared to $t_{fft}$. Therefore we have $T \approx N_F \times (Q^2 \times t_a  +  t_i)$.

We point out that the set of static modes we need to represent adequately the matter field operators depends on the initial conditions of the state of the entire system. Let us indicate with $\lambda_s$ the smallest spatial length on which the statistical functions of the driving field operator varies. When the sizes of the object is of the order up to $\min\limits_{\omega}\{c_0/[\omega \sqrt{|{\chi}(\omega)|}]\}$ and $\lambda_s$ the number of required static modes $Q$ is of the order of unity. In these cases, $t_i$ is negligible with respect to $t_a$ because we need to invert very small matrices, and the total computational time reduces to $ T \approx  N_F \times Q^2 \times t_a$.

\section{Infinite homogeneous dielectric, dielectric slab and sphere} 
\label{sec:Infinite}

Even if the proposed approach has been developed to deal with finite size dielectric objects of arbitrary shapes, it is instructive to apply it to elementary cases such as an infinite homogeneous dielectric, a dielectric slab, and a sphere. We use the sphere to validate the numerical procedure presented in the previous section  by comparing it against the semi-analytical expression of the impulse response \cite{forestiere_time-domain_2021}.
 
\subsection{Infinite homogeneous dielectric}

We first consider an infinite homogeneous dielectric. In this case, the polarization density field and the electric field have the same support, $V_\infty$. The irrotational components of both fields are equal to zero because the electric field is solenoidal everywhere in $V_\infty$ due to homogeneity. In finite size objects, the normal component of the electric field is discontinuous on the boundary of the object due to the discontinuity of the permittivity. In infinite homogeneous dielectric, we exclude the presence of charges at infinity; otherwise the energy stored in the electromagnetic field would be infinite. As a consequence, we only need the transverse modes to represent the polarization density field. A complete set of transverse modes in $V_\infty$ is composed by the transverse plane waves, the same set we have used to represent the vector potential. Therefore, we represent the polarization density field operator in the Laplace domain as
\begin{equation}
\label{eq:aperp}
     \hat{\Pb}(\rb;s) = \sum_{\mu} \hat{P}_{\mu}^{\perp}(s) \mathbf{w}_\mu \rp,
\end{equation}
where $\hat{P}_{\mu}^\perp(s)$ are the coordinate operators of the polarization. The coordinate operators of the matter field and the conjugate momentum operators obey the same commutation relation between the coordinate operators of the vector potential and the conjugate momentum operators.

The coefficient $S_{\mu\mu'}^{\perp \perp}(s)$ is given by
\begin{equation}
    S_{\mu\mu'}^{\perp \perp}(s)=\frac{s}{s^2+\omega_\mu^2} \delta_{s,s'} \delta \left( \mathbf{k} - \mathbf{k}' \right).
\end{equation}
where $\omega_\mu = c_0 k$. Equation \ref{eq:xperp5}, which governs the transverse coordinate operator of the polarization, reduces to 
\begin{equation}
\left(\frac{1}{\tilde{\chi}}+\frac{s^2}{s^2+\omega_\mu^2}\right) \hat{P}_{\mu}^{\perp} =\hat{{F}}_{\mu}^\perp(s)
\end{equation}
where 
\begin{equation}
\hat{{F}}_{\mu}^\perp(s) =\varepsilon_0\hat{\mathcal{E}}_{\mu}(s) + \frac{1}{\tilde{\chi}} \hat{\mathcal{P}}^{free}_{\mu}(s),
\end{equation}
$\hat{\mathcal{E}}_{\mu}(s)$ is given by Eq. \ref{eq:freecords} and the coordinate operator of the free polarization field is given by 
\begin{equation}
\hat{\mathcal{P}}^{free}_{\mu}(s) = \langle{\bf w_\mu}, \boldsymbol{\hat{\mathcal{P}}}^{\,free} \rangle_{V_\infty}.
\end{equation}
The transfer function ${{H}}_{\mu}^{\perp\,\perp}(s)$ is given by
\begin{equation}
\label{eq:HtrI}
{{H}}_{\mu}^{\perp\,\perp}(s) =\tilde{\chi}(s) \frac{s^2+\omega_\mu^2}{[1+\tilde{\chi}(s)]s^2+\omega_\mu^2}.
\end{equation}
The impulse response ${{h}}_{\mu}^{\perp\,\perp}(t)$ is the inverse Laplace transform of this expression.

Using Eq. \ref{eq:Einside} in the Laplace domain, we obtain the following expression for the coordinate operator of the electric field $\hat{E}_{\mu}(s)$
\begin{equation}
\hat{E}_{\mu}(s) = \frac{1}{(1+\tilde{\chi})s^2 + \omega_\mu^2}\frac{1}{\varepsilon_0}\left(\hat{j}^{(p)}_{\mu} + \hat{j}^{(e)}_{\mu}\right),
\end{equation}
where
\begin{equation}
\hat{j}^{(p)}_{\mu}  = s\hat{\mathcal{P}}^{free}_{\mu}(s)-\hat{\mathcal{P}}^{(0)}_{\mu},
\end{equation}
and
\begin{equation}
\hat{j}^{(e)}_{\mu} = \frac{k^2}{\mu_0} \hat{A}_{\mu}^{(0)} +s {\hat{\Pi}}_{\mu}^{(0)}.
\end{equation}
This result coincides with the solution we would obtain by solving directly in the Laplace-wavenumber domain the equations of motion for the polarization density field operator and the vector potential operator, 
\begin{subequations}
\begin{align}
\hat{\mathbf{P}}\rpt =&-\varepsilon_0\displaystyle \zeta(t)*\dot{\hat{\mathbf{A}}} \left(\mathbf{r};t \right)+ \hat{\mathbf{P}}^{free}(\mathbf{r};t), \\
\ddot{\hat{\A}} - c_0^2 \nabla^2 \hat{\A} =& \frac{1}{\varepsilon_0}\dot{\hat{\mathbf{P}}}.
\end{align}
\end{subequations}
These equations must be solved with the initial conditions for the radiation field operators. 
Differently from Huttner and Barnett's paper, we have the additional contribution of $\hat{j}_\mu^{\left(e\right)}$, which takes into account the fluctuation of the electromagnetic field. This is consistent with what Drezet had already observed \cite{drezet_quantizing_2017}. The origin of this discrepancy is in the fact that in
the Huttner and Barnett model only the scattered modes are taken into account, while the free modes are disregarded. Drezet has also pointed out that this discrepancy ``does not impact too much the homogeneous medium case considered by Huttner and Barnett" \cite{drezet_quantizing_2017}, while it has a strong impact in the inhomogeneous medium problems \cite{drezet_equivalence_2017,dorier_canonical_2019,dorier_critical_2020}.

\subsection{Dielectric slab}

We now consider a homogeneous dielectric slab of thickness $2a$ and a linearly polarized electromagnetic waves that propagate normally to the slab. The problem is one-dimensional. We introduce a Cartesian coordinate system $(x,y,z)$ with the $x$ axis orthogonal to the slab and the $z$ axis parallel to the electric field. In this case, also, the electric field and the polarization density fields are everywhere solenoidal, therefore we only need the transverse modes to represent the polarization density field. A complete set of transverse modes is composed by
\begin{equation}
{\bf U}^\perp_m(x) =  \frac{1}{2a} e^{ik_mx} \hat{{\bf z}}
\end{equation}
for $-\infty<m<+\infty$ and $-a\le x\le+a$, where $\hat{{\bf z}}$ denotes the unit vector parallel to the $z$ axis and $k_m=m(\pi/a)$. 
The polarization density field operator in the Laplace domain is expressed as
\begin{equation}
     \hat{\Pb}(\rb;s)  = \sum_{m} \hat{P}_{m}^{\perp}(s) {\bf U}_m(x).
\end{equation}
In this case, the coordinates of the matter field operator and the conjugate momentum operator obey commutation relations similar to the commutation relation between the coordinates of the vector potential operator and the conjugate momentum operator.

The expression of the Green's function $\overleftrightarrow{G}^\perp(\rb;s)$ is given by
\begin{equation}
\overleftrightarrow{G}^\perp(x;s) = \frac{c_0}{2s} e^{-s|x|/c_0}.
\end{equation}
We now evaluate the coefficients $S_{m m'}^{\perp\,\perp}$. We obtain
\begin{multline}
    S_{m m'}^{\perp\,\perp} = \frac{a}{4c_0}[B_{m m'}^{+}(s)+B_{m m'}^{-}(s)]\times \\
    [(-1)^{m+m'} e^{-as/c_0}- (-1)^{m'-m}]
\end{multline}
where
\begin{equation}
B_{m m'}^{\pm}= \frac{c_0}{as\pm im\pi c_0}\frac{c_0}{as \pm im'\pi c_0}.
\end{equation}
The elements of the matrix $\dss{M}^{\perp \perp}$ are given by
\begin{equation}
   {M}^{\perp \perp}_{mm'}(s) = \frac{1}{\tilde{\chi}(s)} \delta_{m m'}+ s\, S_{mm'}^{\perp \perp}(s).
\end{equation}
The transfer matrix $\dss{H}^{\perp \perp}$ is the inverse of $\dss{M}^{\perp \perp}$. In this case, unlike the infinite homogeneous dielectric case, we have an infinite discrete set of coupled equations governing the coordinate operators. In the small size limit, $a \ll \lambda_c$, the modes are decoupled, and
\begin{equation}
   {H}^{\perp \perp}_{mm'}(s) \cong \frac{\tilde{\chi}(s)}{1 + s\, S_{mm}^{\perp \perp}(s) \tilde{\chi}(s)} \delta_{mm'}.
\end{equation}
When $a$ is of the order of $\lambda_c$ the modes are coupled, nevertheless each mode is coupled to a few modes. To calculate the elements of the transfer matrix $\dss{H}^{\perp \perp}$ we have to resort to the step d) of the numerical procedure summarized in Sec. \ref{sec:Num}. The expressions of the polarization density operator and the electric field operator are consistent with those described in the literature (e.g., \cite{matloob_electromagnetic_1995}). 

Alternatively, the dielectric slab problem can also be solved analytically by applying standard techniques (e.g., \cite{matloob_electromagnetic_1995}, \cite{van_bladel_electromagnetic_2007}) to the one-dimensional system of equations (the subscript $z$ denotes the z-component)
\begin{subequations}
\begin{eqnarray}
\label{eq:Ydotdot1d}
\hat{{P}}_z =-\varepsilon_0\displaystyle \zeta*\dot{\hat{{A}}}_z+ \hat{{P}}^{free}_z \quad -a \le x \le a, \\
    \label{eq:Adotdot1d}
\ddot{\hat{A_z}} - c_0^2 \frac{\partial^2 \hat{A_z} }{\partial x^2} = \frac{1}{\varepsilon_0}\dot{\hat{{P_z}}} \quad -\infty \le x \le \infty .
\end{eqnarray}
\end{subequations}
These equations must be solved with the initial conditions for the radiation field operators. 

\subsection{Sphere and Validation}

We now compare the impulse responses of a sphere obtained with the numerical procedure with those obtained semi-analytically following \cite{forestiere_time-domain_2021}. In particular, we consider the elements of the impulse response matrix associated with the modes with lowest multipolar order. Unlike the infinite dielectric
and the dielectric slab, for a sphere we need both longitudinal and
transverse modes to represent the polarization because
the Coulomb electric field is different from zero. These
modes can be expressed analytically in terms of the vector
spherical functions, and the coefficient $S_{mm'}^{ab}$ can be calculated semi-analytically \cite{forestiere_time-domain_2021}. We use the Drude-Lorentz model given by \ref{eq:chi} for the susceptibility of the sphere. We introduce the size parameter $\beta=k_Pa$ where $k_P = \omega_P/c_0=1/\lambda_c$.

\begin{figure}
    \centering
    \includegraphics[width=\columnwidth]{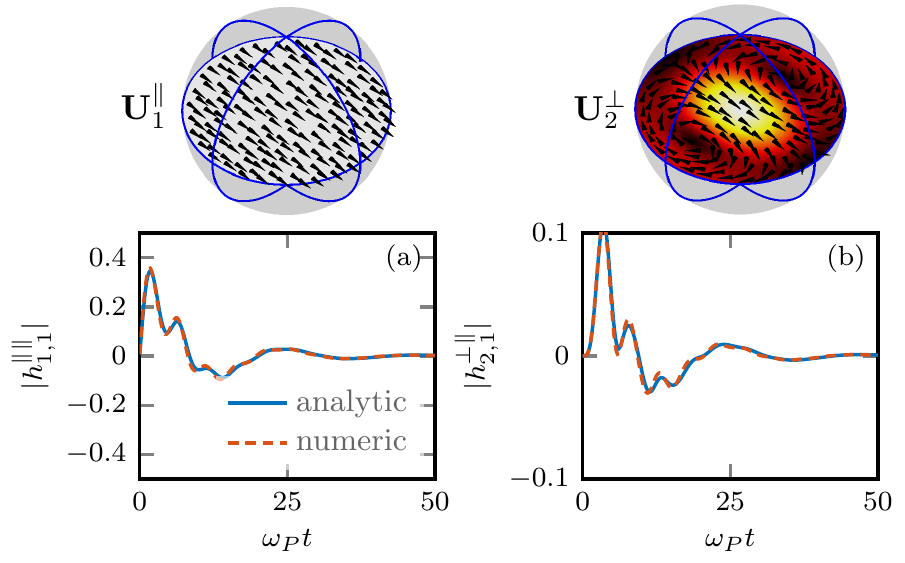}
    \caption{Impulse responses of a lossless metal sphere ($\omega_0=0$, $\Gamma=0$) with $\beta=\pi$, where $\beta=k_P a$, $a$ is the radius and $k_P=\omega_P/c_0$ obtained by the semi-analytical and numerical procedures. 
    The impulse responses $h^{\parallel \parallel}_{1,1}\left( t \right)$ and $h^{\perp \parallel}_{2,1}$ (b) are associated with the modes $\mathbf{U}_1^\parallel$ and $\mathbf{U}_2^\perp$, which are shown in the inset above.}    
    \label{fig:ValidationSphere}
\end{figure}

In Fig. \ref{fig:ValidationSphere} we show the impulse response of a lossless metal sphere ($\omega_0=0$, $\Gamma=0$)  with $\beta=\pi$  obtained by the semi-analytical and the numerical procedures. In particular, we investigate the coupling of the electric dipole mode to other modes, beyond the small size limit. From numerical analysis we have found that the electric
dipole mode $\mathbf{U}_1^\parallel$ significantly couples only to the mode $\mathbf{U}_2^\perp$ due to the symmetries of the sphere. The distribution of the electric dipole mode $\mathbf{U}_1^\parallel$ and of the mode $\mathbf{U}_2^\perp$ are shown on the equatorial plane of the sphere on the top of Fig. \ref{fig:ValidationSphere}. Impulse responses $h^{\parallel \parallel}_{1,1}$ and $h^{\perp \parallel}_{2,1}$ are shown in Fig. \ref{fig:ValidationSphere} (a) and (b), respectively. Very good agreement is found.

\begin{figure}
    \centering
    \includegraphics[width=\columnwidth]{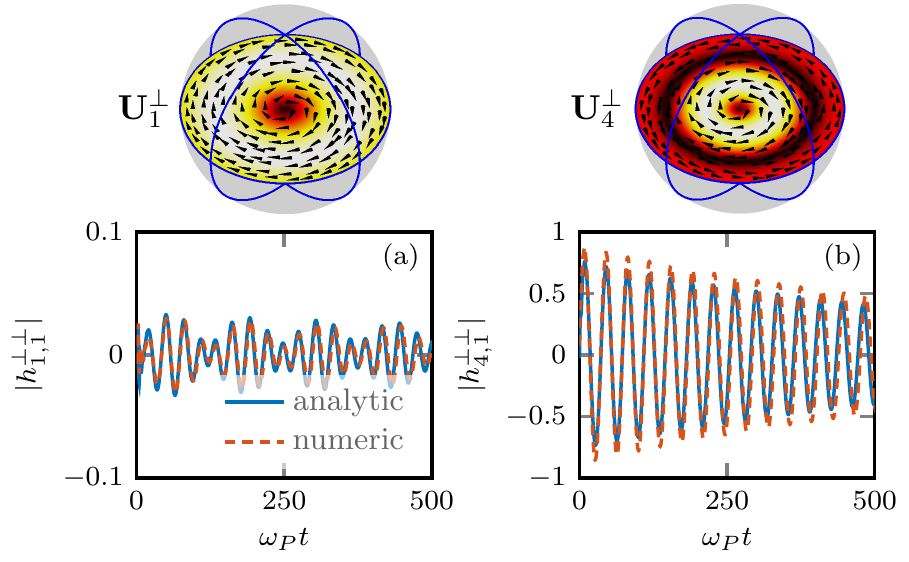}
    \caption{Impulse responses of a lossless dielectric sphere ($\omega_0 = \omega_p/4$, $\Gamma=0$) with $\beta=\pi$, where $\beta=k_P a$, $a$ is the radius and $k_P=\omega_P/c_0$ obtained by the semi-analytical and numerical procedures. 
    The impulse responses $h^{\perp \perp}_{1,1}\left( t \right)$ and $h^{\perp \perp}_{4,1}$ (b) are associated with the modes $\mathbf{U}_1^\perp$ and $\mathbf{U}_{4}^\perp$, which are shown in the inset above. }
    \label{fig:ValidationTE}
\end{figure}

Next, we investigate the impulse response of a dielectric sphere with $\omega_0 = \omega_P/4$ and $\beta=\pi$.   First, we investigate the coupling of the magnetic dipole mode to other modes.  From numerical analysis we have found that for $\beta=\pi$ the magnetic dipole $\mathbf{U}_{1}^\perp$ significantly couples only with the higher order magnetic dipole $\mathbf{U}_{4}^\perp$ due to the symmetry of the problem.  The transverse modes $\mathbf{U}_{1}^\perp$ and $\mathbf{U}_{4}^\perp$ are shown on the top of Fig. \ref{fig:ValidationTE}. Both modes have a non vanishing magnetic dipole moment: the first is a current loop, the second mode is made by two counter-rotating current loops. In Figures \ref{fig:ValidationTE} (a) and (b) we show the impulse responses $h^{\perp \perp}_{1,1} \left( t \right)$ and $h^{\perp \perp}_{4,1} \left( t \right)$, which are obtained by using the numerical and the semi-analytic calculations. We found very good agreement between them.

\begin{figure}
    \centering
    \includegraphics[width=\columnwidth]{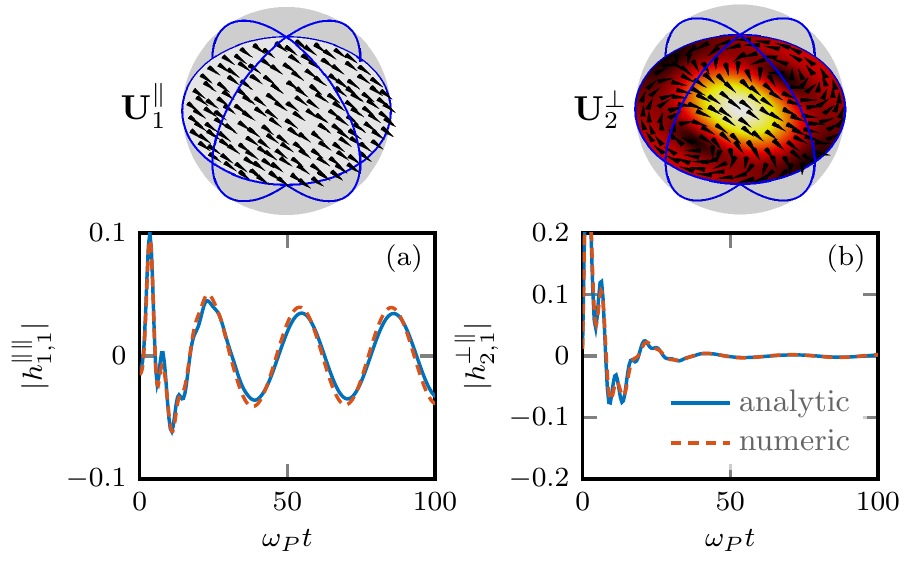}
    \caption{Impulse responses of a lossless dielectric sphere ($\omega_0 = \omega_p/4$, $\Gamma=0$)  with $\beta=\pi$, where $\beta=k_P a$, $a$ is the radius and $k_P=\omega_P/c_0$ obtained by the semi-analytical and numerical procedures. 
    The impulse responses $h^{\parallel \parallel}_{1,1}\left( t \right)$ and $h^{\perp \parallel}_{2,1}$ (b) are associated with the modes $\mathbf{U}_1^\parallel$ and $\mathbf{U}_2^\perp$. }
    \label{fig:ValidationTM}
\end{figure}

Eventually, we consider the same dielectric sphere, but we now focus on the coupling of the electric dipole mode with other modes. As in the case of the metal sphere, from the numerical analysis we have found that the electric dipole mode $\mathbf{U}_{1}^\parallel$ significantly couples to the mode $\mathbf{U}_{2}^\perp$ due to the symmetry of the problem. The two modes are shown at the top of Fig. \ref{fig:ValidationTM}.  Impulse responses $h^{\parallel \parallel}_{1,1}$ and $h^{\perp \parallel}_{2,1}$ are shown in Fig. \ref{fig:ValidationTM} (a) and (b), respectively. Also in this case, we found very good agreement between the numerical and the semi-analytic solutions.

\section{Disk}
\label{sec:disk}

In this section, we evaluate the elements of the transfer matrix $\dss{H}(s)$ and the impulse response matrix $\dss{h}(t)$ for a disk with rounded edges, radius $a$, height $a/2$ and radius of curvature $a/4$. The behaviour of the disk material is described by the Drude-Lorentz model \ref{eq:chi} with plasma frequency $\omega_P$, resonant frequency $\omega_0$ and damping rate of the material $\Gamma$. We study how the elements of $\dss{H}(s)$ and $\dss{h}(t)$ associated with the modes with lowest multipolar order change in terms of the size parameter $\beta=k_P a$, for different values of $\Gamma$. We consider two types of materials, a metal disk, $\omega_0 =0$, and a dielectric disk, $\omega_0 = \omega_P/4$. 

\begin{figure*}
    \centering
    \includegraphics[width=0.8\textwidth]{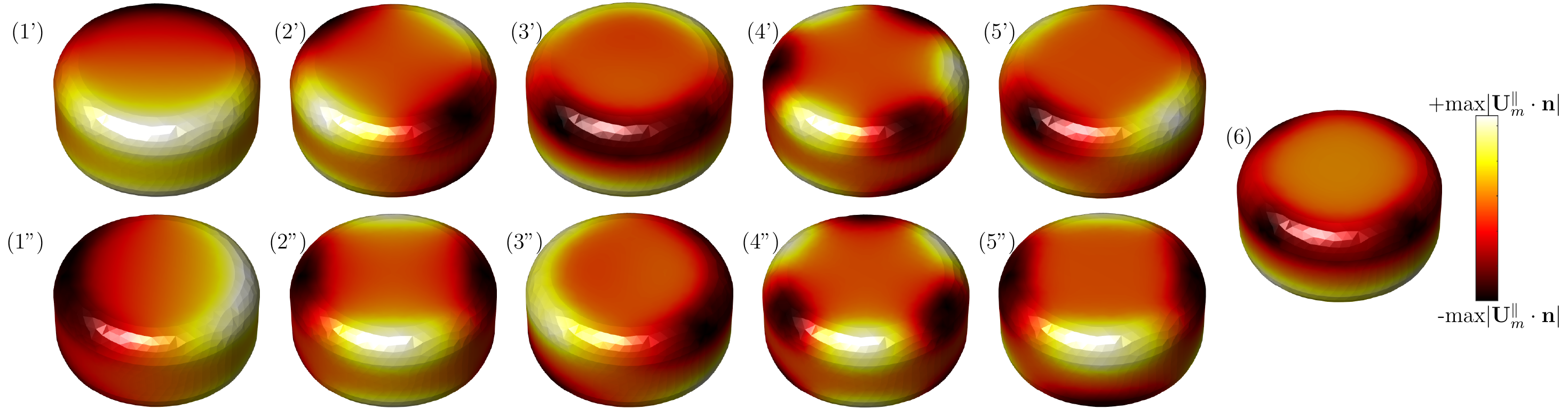} 
    \caption{Longitudinal static modes $\mathbf{U}^\parallel_m$  of a dielectric disk with rounded edges (radius a, height a/2, radius of curvature a/4). The modes are sorted according to the associated eigenvalue $\kappa^\parallel$ in descending order. The surface charge density is shown on the boundary of disk. The vertically aligned modes are degenerate. Only the first 11 eigenmodes are shown. In the colorbar, $\text{max} |  \mathbf{U}^\parallel_m \cdot \mathbf{n} |$ is such that $\| \mathbf{U}^\parallel_m \|=1$. }
    \label{fig:LongModes}
\end{figure*}

\begin{figure*}
    \centering
    \includegraphics[width=\textwidth]{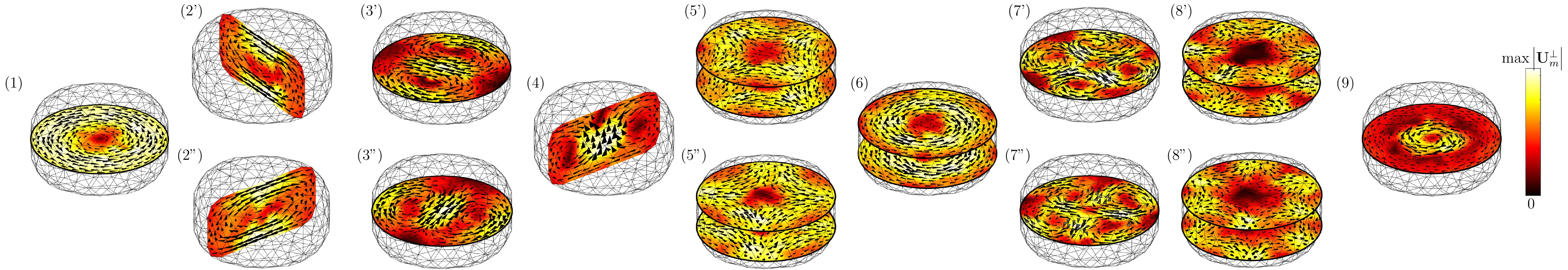}
    \caption{Transverse static modes $\mathbf{U}_m^\perp$ of a dielectric disk with rounded edges (radius a, height a/2, radius of curvature a/4). The modes are sorted according to the associated eigenvalue $\kappa^\perp$ in ascending order. The magnitude and the direction of the vector field $\mathbf{U}^\perp_m$ are shown in representative section planes of the disk. The vertically aligned modes are degenerate. Only the first 14 eigenmodes are shown. In the colorbar, $\text{max}| \mathbf{U}_m^\perp |$ is such that $\| \mathbf{U}^\perp_m \|=1$, where $|\mathbf{U}_m^\perp | = \sqrt{\mathbf{U}_m^\perp \cdot \mathbf{U}_m^\perp }$.}
    \label{fig:TransModes}
\end{figure*}

\begin{figure}
    \centering
    \includegraphics[width=0.7\columnwidth]{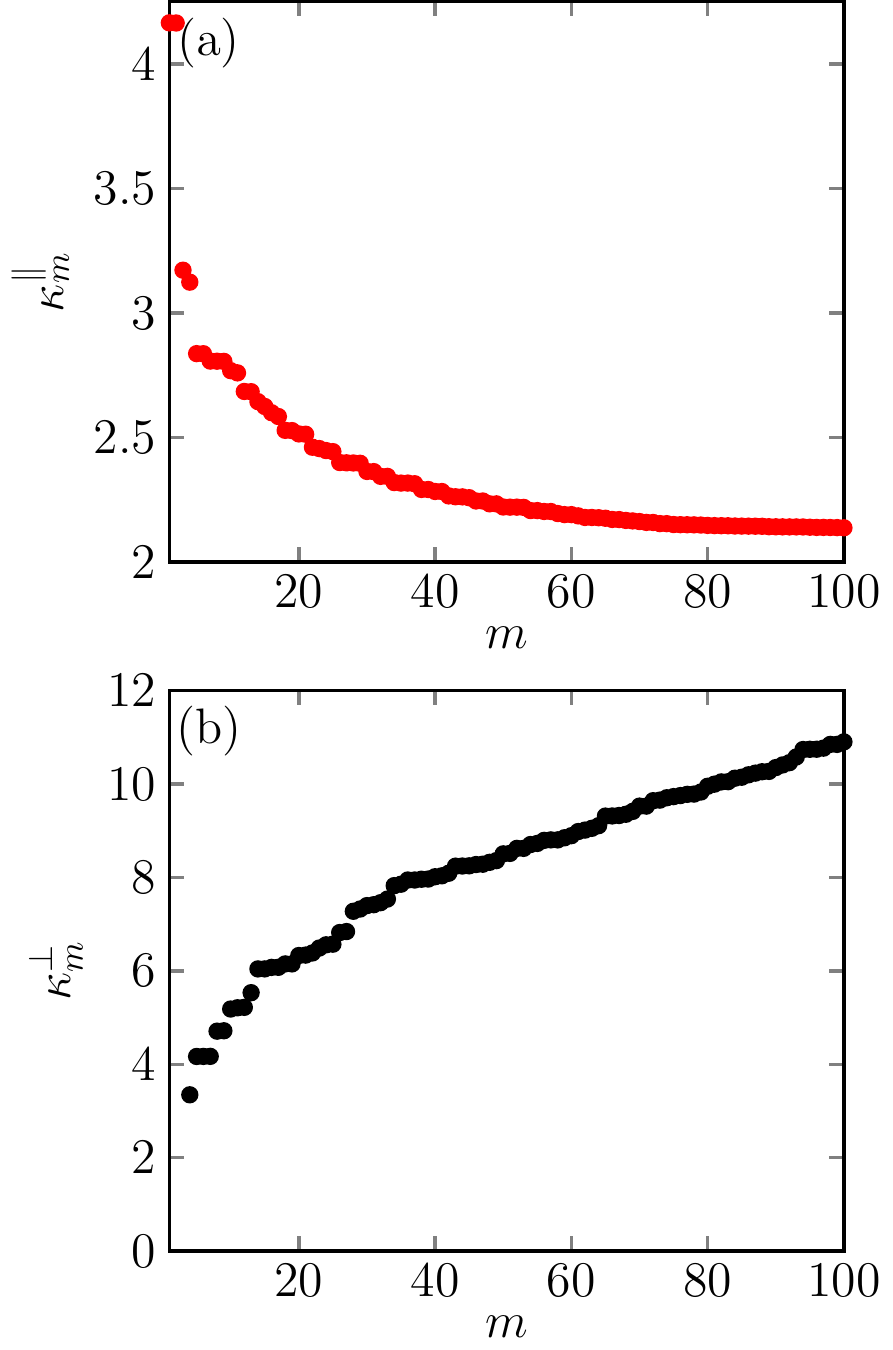}
    \caption{Eigenvalues $\kappa_m^\parallel$ (a) and $\kappa_m^\perp$ (b) associated to the longitudinal and transverse static modes of a dielectric disk with rounded edges (radius a, height a/2, radius of curvature a/4). Only the first $100$ eigenvalues are shown in each case.}
    \label{fig:Eigenvalues}
\end{figure}

\subsection{Longitudinal and transverse static modes}

We compute numerically the static modes following appendix \ref{sec:Computation}. We discretize the domain of the disk $V$ together with its boundary $\partial V$. The volume and the surface meshes have been generated by the mesh generator Gmsh \cite{geuzaine_gmsh_2009}.

 The longitudinal modes are shown in Fig. \ref{fig:LongModes}, while the corresponding eigenvalues $\kappa_m^\parallel$ are shown in Fig. \ref{fig:Eigenvalues} (a). In particular, the eigenvalues and the corresponding modes are ordered in descending order: in this way the first modes are associated with lower multipolar order (electric dipole, electric quadrupole, ...). 
 
 The transverse modes are shown in Fig. \ref{fig:TransModes}, while the corresponding eigenvalues $\kappa_m^\perp$ are shown in Fig.  \ref{fig:Eigenvalues} (b).
In particular, the eigenvalues and the corresponding modes are ordered in ascending order: in this way the first modes are associated with lower multipolar order (magnetic dipole, $\mathbf{P}_{e2}$ dipole moment \cite{bladel_hierarchy_1988,bladel_electromagnetic_2007}, magnetic quadrupole, ...).

\subsection{Small size limit}

In the small size limit $\beta = k_P a \ll 1$ the static longitudinal and the transverse modes of the disk are the natural modes of the polarization. The matrix $\dss{H}$ is quasi diagonal. Combining \ref{eq:xparallel5sm}, \ref{eq:xperp5sm} and \ref{eq:chi} we obtain
\begin{subequations}
\begin{align}
  H^{\parallel \parallel}_{m m}(s) &\cong\frac{\omega_P^2}{s^2 + s \Gamma+ (\omega_0^2 + \omega_P^2/\kappa_m^\parallel)}, \\
H^{\perp \perp}_{m m}(s) &\cong\frac{\omega_P^2}{ (1+\beta^2/\kappa_m^\perp)s^2 + \Gamma s + \omega_0^2}.
\end{align}
\end{subequations}
The corresponding impulse responses in the time domain are given by
\begin{subequations}
\begin{align}
  h^{\parallel \parallel}_{m m}(t) &\cong\frac{\omega_P^2}{\Omega_m^\parallel}u(t)e^{-t/\tau_m^\parallel} \sin(\Omega_m^\parallel t), \\
h^{\perp \perp}_{m m}(t) &\cong\frac{\omega_P^2}{ \delta_m\Omega_m^\perp}u(t)e^{-t/\tau_m^\perp} \sin(\Omega_m^\perp t),
\end{align}
\label{eq:ImpulseSmall}
\end{subequations}
where $\tau^\parallel=2/\Gamma$, $\tau_m^\perp=2\delta_m/\Gamma$, $\delta_m=(1+\beta^2/\kappa_m^\perp)$, $\Omega_m^\parallel=\sqrt{(\omega_0^2 + \omega_P^2/\kappa_m^\parallel)-\Gamma^2/4}$ and $\Omega_m^\perp=\sqrt{\omega_0^2/\delta_m -\Gamma^2/4\delta_m^2}$ ($u(t)$ is the Heaviside function). Since the eigenvalue $\kappa_m^\parallel$ decreases as $m$ increases and the eigenvalue $\kappa_m^\perp$ increases as $m$ increases, both the natural frequencies $\Omega_m^\parallel$ and $\Omega_m^\perp$ increase as the mode index $m$ increases. The electric dipole mode and the magnetic dipole mode have the smallest natural frequencies: they are the fundamental natural modes of polarization in the small size limit. 

In the small size limit, the decay rate of the impulse response \ref{eq:ImpulseSmall} depends only on material losses, because the radiation damping rate goes to zero at least as $\beta^3$ as $\beta\rightarrow0$ \cite{forestiere_resonance_2020}.

Beyond the small size limit, the coupling among longitudinal and transverse modes may become significant due to the radiation. As in the case of the sphere \cite{forestiere_time-domain_2021}, also in the case of a rotationally invariant disk, symmetry prevents some modes from coupling. In the following, we investigate the coupling of the electric dipole mode and the magnetic dipole mode with higher order modes. 

Beyond the small size limit, the decay rates increase due to the radiative losses, the natural frequencies shift due to the coupling among modes, the impulse responses may show beatings due to the interplay among coupled modes, additional peaks arise due to the transverse electromagnetic standing wave modes of the object.

\begin{figure}
    \centering
    \includegraphics[width=\columnwidth]{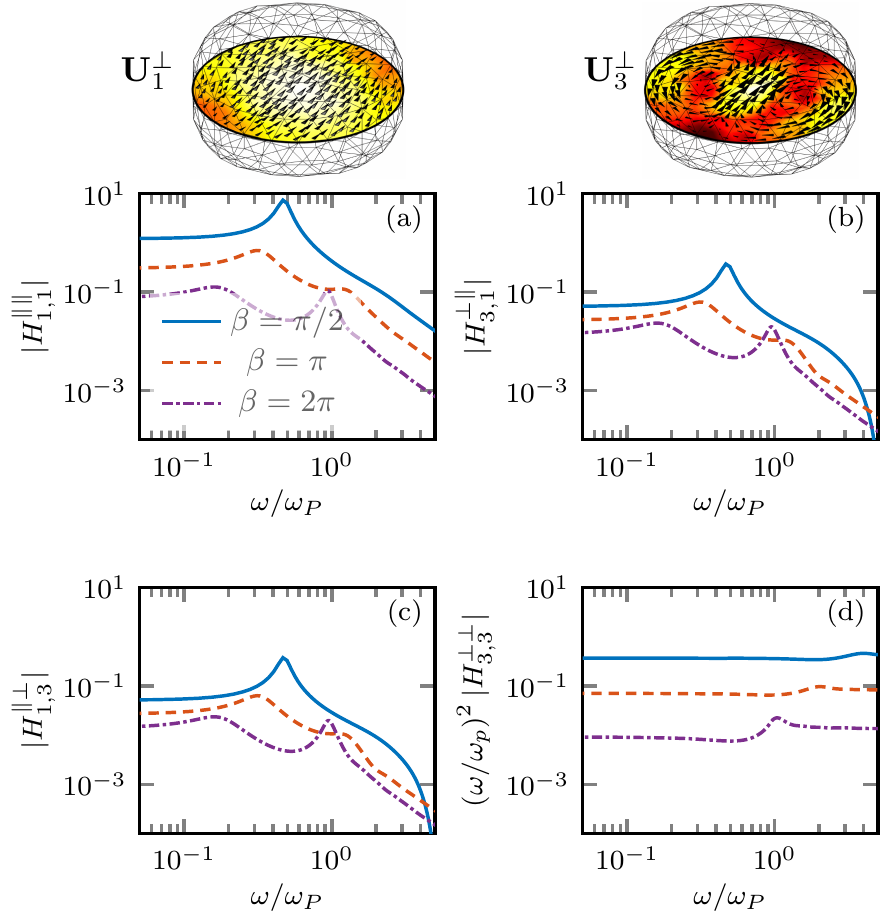}
    \caption{Amplitude responses of a lossless metal disk ($\omega_0 =0$, $\Gamma=0$) with $\beta=\pi/2$, $\pi$, $2\pi$, where $\beta=k_P a$, $a$ is the radius, and $k_P=\omega_P/c_0$. The elements $H^{\parallel \parallel}_{1,1}$ (a), $H^{\perp \parallel}_{3,1}$ (b), $H^{\parallel \perp}_{1,3}$ (c) and $H^{\perp \perp}_{3,3}$ (d) are associated to the modes $\mathbf{U}_{1}^\parallel$ and $\mathbf{U}_{3}^\perp$ shown in the inset above; $H_{1,3}^{\parallel \perp}=H_{3,1}^{\perp \parallel}$ for the reciprocity.}
    \label{fig:FreqMetal}
\end{figure}

\subsection{Metal}

We now investigate the coupling of the electric dipole mode to other modes for a metal disk, $\omega_0=0$, beyond the small size limit, as the size parameter $\beta$ varies in the interval $\left[0,2\pi\right]$. To highlight the role of radiation losses, we initially disregard material losses, $\Gamma=0$. In this limit case, we have $\Omega_m^\parallel=\omega_P\sqrt{1/\kappa_m^\parallel}$ and $\Omega_m^\perp=0$. The impulse response $h^{\perp\,\perp}_{m\,m}$ of the transverse modes degenerates in a ramp function in the small size limit.  
From numerical analysis we have found that the electric dipole mode $\mathbf{U}_{1}^\parallel$ significantly couples to the mode $\mathbf{U}_{3}^\perp$, which carries a $\mathbf{P}_{e2}$ dipole moment \cite{bladel_hierarchy_1988}, due to the symmetries of the disk. Their interaction with other modes is negligible as the size parameter $\beta$ varies in the interval $\left[0,2\pi\right]$. The magnitudes of $\mathbf{U}_{1}^\parallel$, $\mathbf{U}_{3}^\perp$ and their field lines are shown in the equatorial plane of the disk at the top of Fig. \ref{fig:FreqMetal}.

In Figures \ref{fig:FreqMetal} (a) and (b) we show the amplitude of $H_{1,1}^{\parallel \parallel}$ and $H_{3,1}^{\perp \parallel}$ as function of the normalized frequency $\omega/\omega_P$. They account for the contribution of the driving coordinate operator $\hat{F}_1^\parallel$ to the coordinate operators $\hat{P}_1^\parallel$ and $\hat{P}_3^\perp$ of the polarization as the frequency $\omega$ varies.  In Fig. \ref{fig:FreqMetal} (c) and (d) we show the amplitude responses of $H_{1,3}^{\parallel \perp}$ and $H_{3,3}^{\perp \perp}$. They account for the contribution of the driving coordinate operator $\hat{F}_3^\perp$ to the coordinate operators $\hat{P}_1^\parallel$ and $\hat{P}_3^\perp$ of the polarization as the frequency varies.  For $0\le\beta\le2\pi$ the self and mutual coupling between the modes $\mathbf{U}_{1}^\parallel$ and $\mathbf{U}_{3}^\perp$ dominate the frequency response, being the coupling with the remaining modes negligible. As expected, the curves in panels (b) and (c) are identical because of reciprocity.

First, we describe the amplitude response of $H^{\parallel \parallel}_{1,1}$ in Fig. \ref{fig:FreqMetal} (a). For $\beta=\pi/2$, $H^{\parallel \parallel}_{1,1}$ exhibits a low-frequency peak, which is located close to the quasistatic natural frequency of the electric dipole mode $\mathbf{U}_{1}^\parallel$. A ``bump" is present at higher frequencies, revealing the presence of a second pole in the frequency response. For $\beta = \pi$, the frequency response broadens around the first peak, and the bump becomes a secondary peak: the contribution of the radiation starts to be significant. Eventually, for $\beta= 2 \pi$, the second peak becomes the highest one. Similar considerations also hold for the ``mutual" frequency responses $H_{1,3}^{\parallel \perp}=H_{3,1}^{\perp \parallel}$, shown in Figs. \ref{fig:FreqMetal} (b-c). We show in Fig. \ref{fig:FreqMetal} (d) the frequency response $H_{3,3}^{\perp \perp}$, which has been scaled by $\left(\omega/\omega_p\right)^2$ to obtain a finite value for $\omega\rightarrow0$. Indeed, in the small size limit $H_{3,3}^{\perp\perp}\approx \omega_P^2/\left(\delta_3 s^2 \right)$ has a double pole at the origin. The high frequency bump is due to the transverse electromagnetic standing waves of the particle, as found for the case of a sphere in \cite{forestiere_time-domain_2021}. By increasing the size parameter $\beta$, the bump undergoes a red-shift, and for $\beta=2 \pi$ it eventually becomes a peak.

\begin{figure}
    \centering
    \includegraphics[width=\columnwidth]{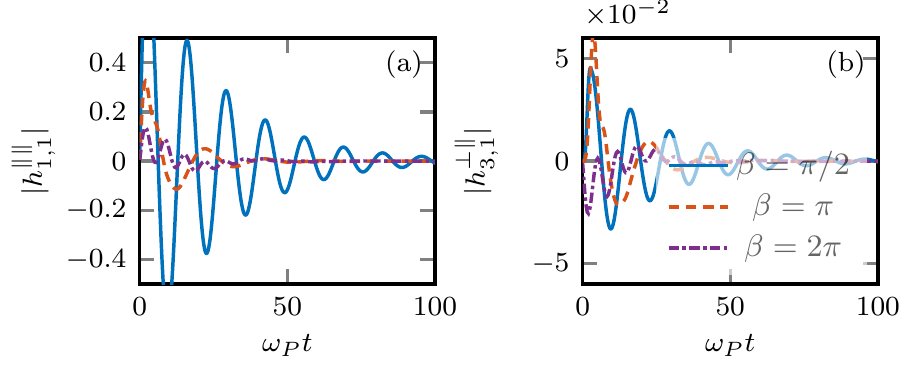}
    \caption{Impulse responses of a lossless metal disk ($\omega_0 =0$, $\Gamma=0$) with $\beta=\pi/2$, $\pi$, $2\pi$, where $\beta=k_P a$, $a$ is the radius and $k_P=\omega_P/c_0$.  The impulse response $h^{\parallel \parallel}_{1,1}\left( t \right)$ (a) corresponds to the amplitude response shown in Fig. 9 (a) and the impulse response  $h^{\perp \parallel}_{3,1}$ (b) corresponds to the amplitude response  shown in Fig. 9 (b).}
    \label{fig:TimeMetal}
\end{figure}

Figure \ref{fig:TimeMetal} shows the impulse responses $h_{1,1}^{\parallel \parallel}(t)$ and $h_{3,1}^{\perp \parallel}(t)$.  For $\beta=\pi/2$, they are dominated by a single harmonic with a frequency corresponding to the low-frequency peak of the amplitude response shown in Fig. \ref{fig:FreqMetal} (a). As $\beta$ increases to $\pi$ the radiative damping determines faster decay. For $\beta=2\pi$ the interaction between the poles associated with the first two peaks of $H^{\parallel \parallel}_{1,1}$ shown in Fig. \ref{fig:FreqMetal} (a) gives rise to a beating. 

\begin{figure}
    \centering
    \includegraphics[width=\columnwidth]{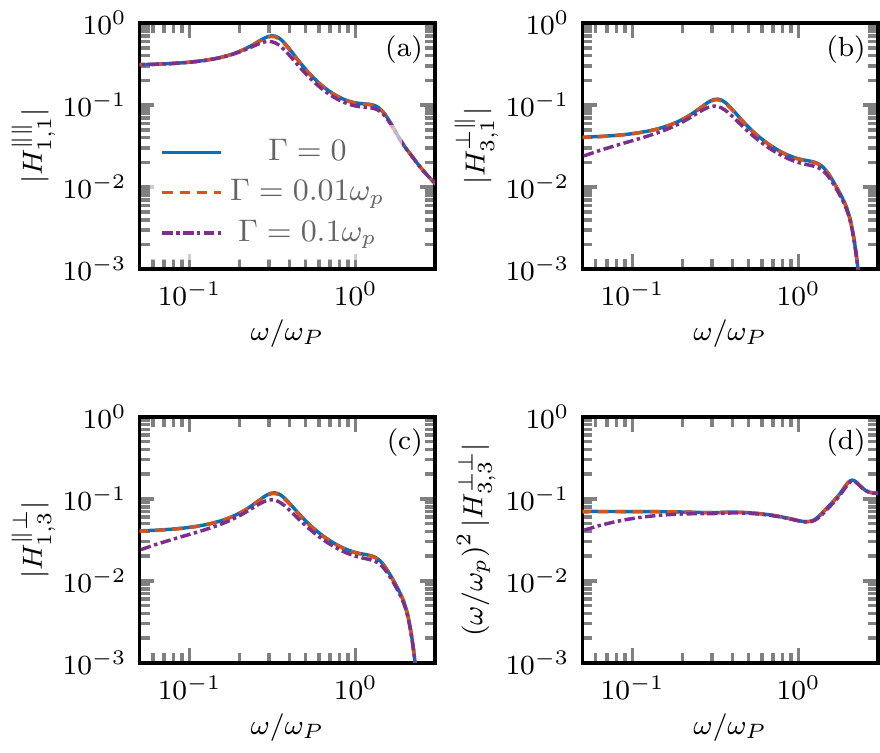}
    \caption{Amplitude responses of a metal disk ($\omega_0 = 0$, $\beta=\pi$) for $\Gamma=0$, $0.01\omega_P$, $0.1\omega_P$, where $\beta=k_P a$, $a$ is the radius and $k_P=\omega_P/c_0$.  The elements $H^{\parallel \parallel}_{1,1}$ (a), $H^{\perp \parallel}_{3,1}$ (b), $H^{\parallel \perp}_{1,3}$ (c) and $H^{\perp \perp}_{3,3}$ (d) are associated to the modes $\mathbf{U}_{1}^\parallel$ and $\mathbf{U}_{3}^\perp$ shown in the inset above Fig. 9; $H_{1,3}^{\parallel \perp}=H_{3,1}^{\perp \parallel}$ for the reciprocity. } 
    \label{fig:FreqMetalGamma}
\end{figure}

\begin{figure}
    \centering
    \includegraphics[width=\columnwidth]{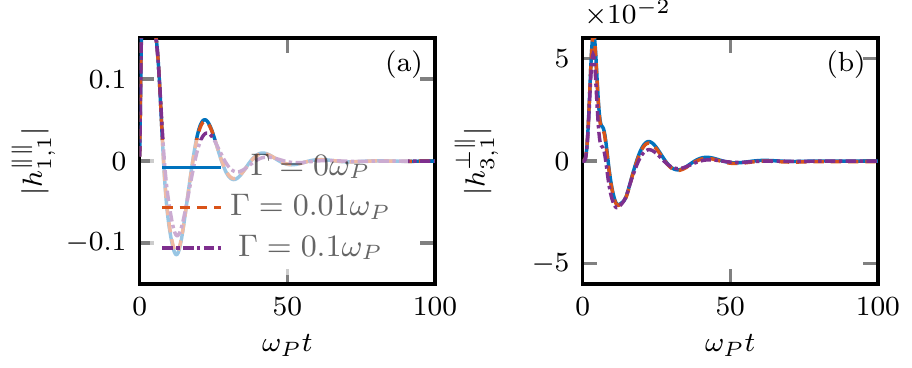}
    \caption{Impulse responses of a metal disk ($\omega_0 =0$, $\beta=\pi$) for $\Gamma=0$, $0.01\omega_P$, $0.1\omega_P$, where $\beta=k_P a$, $a$ is the radius and $k_P=\omega_P/c_0$. The impulse response $h^{\parallel \parallel}_{1,1}\left( t \right)$ (a) corresponds to the amplitude response shown in Fig. 11 (a) and the impulse response  $h^{\perp \parallel}_{3,1}$ (b) corresponds to the amplitude response  shown in Fig. 11 (b). }
    \label{fig:TimeMetalGamma}
\end{figure}

Eventually, we investigate the role that material losses play. Specifically, in Figs. \ref{fig:FreqMetalGamma} and \ref{fig:TimeMetalGamma} we show the frequency and impulse responses of the disk for $\beta = \pi$ by varying the matter damping rate $\Gamma$. It is apparent that an increase of $\Gamma$ up to $0.1\omega_p$ only determines a slightly modification of the frequency and impulse responses. This is because, for the considered values of $\Gamma$ and $\beta$, the radiative losses are dominant.

\begin{figure}
    \centering
    \includegraphics[width=\columnwidth]{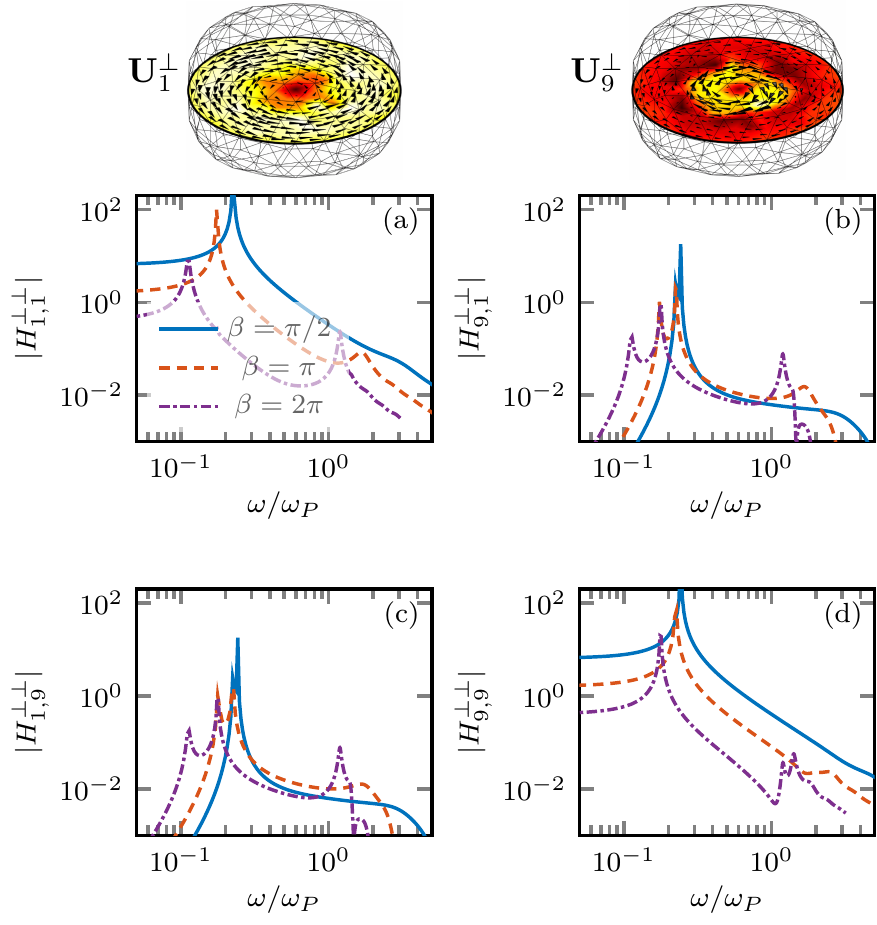}
    \caption{Amplitude responses of a lossless dielectric disk ($\omega_0 = \omega_p/4$, $\Gamma=0$) with $\beta=\pi/2$, $\pi$, $2\pi$, where $\beta=k_P a$, $a$ is the radius and $k_P=\omega_P/c_0$. The elements $H^{\perp \perp}_{1,1}$ (a), $H^{\perp \perp}_{9,1}$ (b), $H^{\perp \perp}_{1,9}$ (c) and $H^{\perp \perp}_{9,9}$ (d) are associated to the modes $\mathbf{U}_{1}^\perp$ and $\mathbf{U}_{9}^\perp$ shown in the inset above; $H_{1,9}^{\perp \perp}=H_{9,1}^{\perp \perp}$ for the reciprocity.}
    \label{fig:FreqTE}
\end{figure}

\subsection{Dielectric}

We now investigate the coupling of the electric dipole mode and the magnetic dipole mode to other modes for a dielectric disk with $\omega_0 = \omega_P/4$ as $\beta$ varies in the interval $\left[0,2\pi\right]$. In the lossless limit, the susceptibility is positive when $\omega < \omega_0$ and negative when $\omega > \omega_0$; when $\omega \rightarrow 0$ the susceptibility is equal to $16$.

\subsubsection{Magnetic dipole coupling}

Once again, to highlight the role of radiation losses, we initially disregard material losses, $\Gamma = 0$. From numerical analysis we have found that the magnetic dipole $\mathbf{U}_{1}^\perp$ significantly couples only with the higher order magnetic dipole $\mathbf{U}_{9}^\perp$ due to the symmetry of the problem. As the size parameter $\beta$ varies in the interval $\left[0,2\pi\right]$ the interaction with other modes is negligible. The transverse modes $\mathbf{U}_{1}^\perp$ and $\mathbf{U}_{9}^\perp$ are shown on the top of Fig. \ref{fig:FreqTE}. Both modes have a non vanishing magnetic dipole moment: the first is a current loop, and the second mode is made by two counter-rotating current loops.

In Fig. \ref{fig:FreqTE} (a), we show the amplitude response of $H^{\perp \perp}_{1,1}$ for $\beta = \pi/2$, $\pi$, $2\pi$, and $\Gamma=0$. For $\beta = \pi/2$ the peak of the amplitude of $H^{\perp \perp}_{1,1}$ is located in the neighborhood of the natural frequency of the mode $\mathbf{U}^\perp_1$. As for the amplitude response associated with the longitudinal dipolar mode in the metal disk, a bump arises at higher frequencies, which is associated with a second pole in the response. Increasing $\beta = \pi$, the first peak undergoes a broadening, while the high-frequency bump becomes a secondary peak. Both peaks experience a red shift. For $\beta = 2 \pi$, the first peak is still dominant, but the second peak increases in intensity. 

As expected, the amplitude responses of $H^{\perp \perp}_{9,1}$ and $H^{\perp \perp}_{1,9}$ shown in Fig. \ref{fig:FreqTE} (b) and (c) are identical due to reciprocity. For $\beta = \pi/2$, these curves exhibit only one peak, which arises from the resonant contribution of the two modes $\mathbf{U}^\perp_1$ and $\mathbf{U}^\perp_{9}$, whose natural frequencies are approximately $\omega_0$ in the small size limit. For $\beta = \pi$, they show two peaks of comparable magnitude, which are located in the neighborhood of the natural frequency of the modes $\mathbf{U}^\perp_1$ and $\mathbf{U}^\perp_{9}$, respectively. A bump appears at higher frequencies due to the standing electromagnetic waves of the object. Increasing $\beta $ to $2 \pi$, the first two peaks undergo a shift and broadening, and the second peak becomes dominant. In addition, the high-frequency bump becomes a third peak. 
 
 In Figure \ref{fig:FreqTE} (d), we show the amplitude response $H^{\perp \perp}_{9,9} \left( \omega \right)$. In this case, the first peak on the left is associated with the natural frequency of the mode $\mathbf{U}^\perp_{9}$. Increasing $\beta$ a second bump arises for $\beta = \pi$; for $\beta = 2 \pi$ a multitude of minor peaks appear corresponding to the natural frequencies of standing transverse electromagnetic waves.
  
\begin{figure}
    \centering
    \includegraphics[width=\columnwidth]{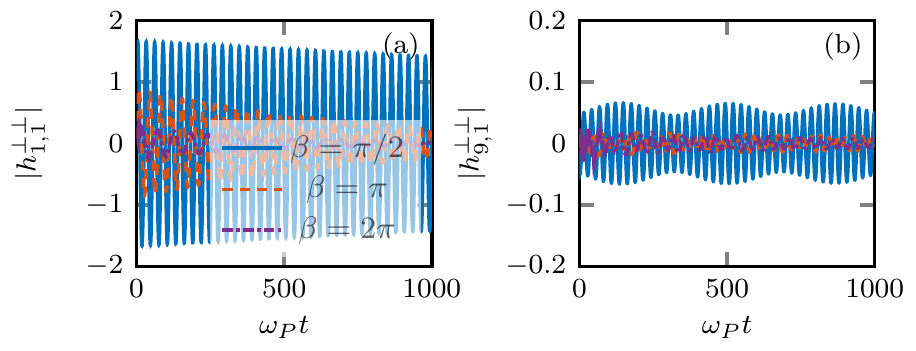}
    \caption{Impulse responses of a lossless dielectric disk ($\omega_0 = \omega_p/4$, $\Gamma=0$) with $\beta=\pi/2, \pi, 2\pi$, where $\beta=k_P a$, $a$ is the radius and $k_P=\omega_P/c_0$.  The impulse response $h^{\perp \perp}_{1,1}\left( t \right)$ (a) corresponds to the amplitude response shown in Fig. 13 (a) and the impulse response $h^{\perp \perp}_{9,1}$ (b) corresponds to the amplitude response shown in Fig. 13 (b).}
    \label{fig:TimeTE}
\end{figure}

In Figures \ref{fig:TimeTE} (a) and (b) we show the impulse responses $h^{\perp \perp}_{1,1} \left( t \right)$ and $h^{\perp \perp}_{9,1} \left( t \right)$ that correspond, respectively, to the amplitude responses shown in Figs. 10 (a) and 10 (b). The impulse response $h^{\perp \perp}_{1,1}$ is dominated by the natural frequency of the mode $\mathbf{U}^\perp_1$. When the value of $\beta$ increases, the impulse response shows a faster decay rate, which is consistent with the broadening observed in the amplitude response. The impulse response $h^{\perp \perp}_{9,1}$ for $\beta=\pi$ and $\beta=2\pi$ shows a beating between the natural frequencies of the first two modes, as expected from the analysis of the amplitude responses.

\begin{figure}
    \centering
    \includegraphics[width=\columnwidth]{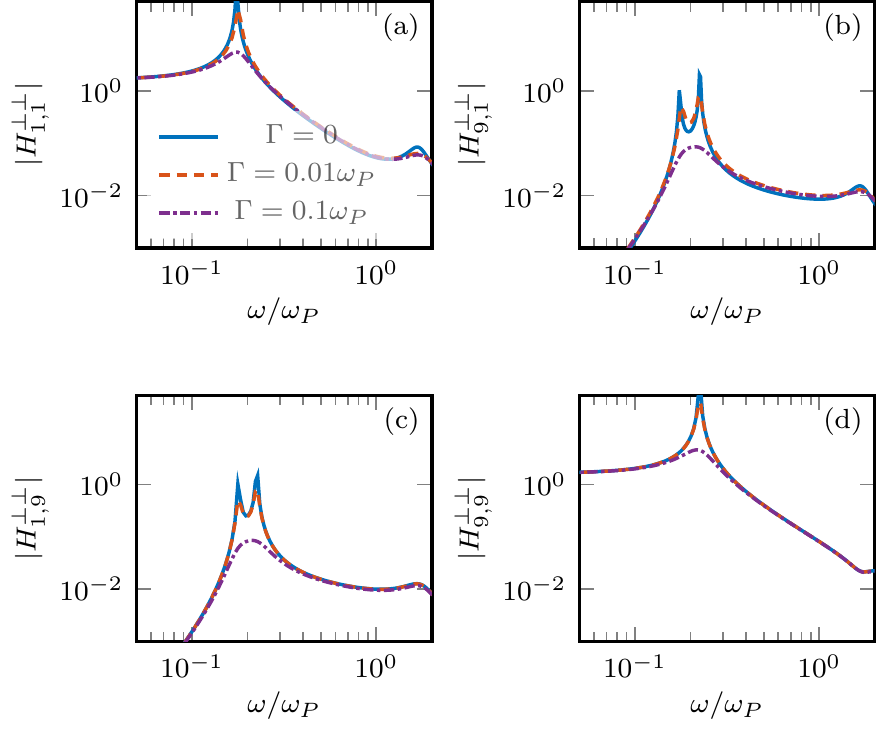}
    \caption{Amplitude responses of a dielectric disk ($\omega_0 = \omega_p/4$, $\beta=\pi$) for $\Gamma=0$, $0.01\omega_P$, $0.1\omega_P$, where $\beta=k_P a$, $a$ is the radius and $k_P=\omega_P/c_0$. The elements $H^{\perp \perp}_{1,1}$ (a), $H^{\perp \perp}_{9,1}$ (b), $H^{\perp \perp}_{1,9}$ (c) and $H^{\perp \perp}_{9,9}$ (d) are associated to the modes $\mathbf{U}_{1}^\perp$ and $\mathbf{U}_{9}^\perp$ shown in the inset above Fig. 13; $H_{1,9}^{\perp \perp}=H_{9,1}^{\perp \perp}$ for the reciprocity.} 
    \label{fig:FreqTE_G}
\end{figure}

\begin{figure}
    \centering
    \includegraphics[width=\columnwidth]{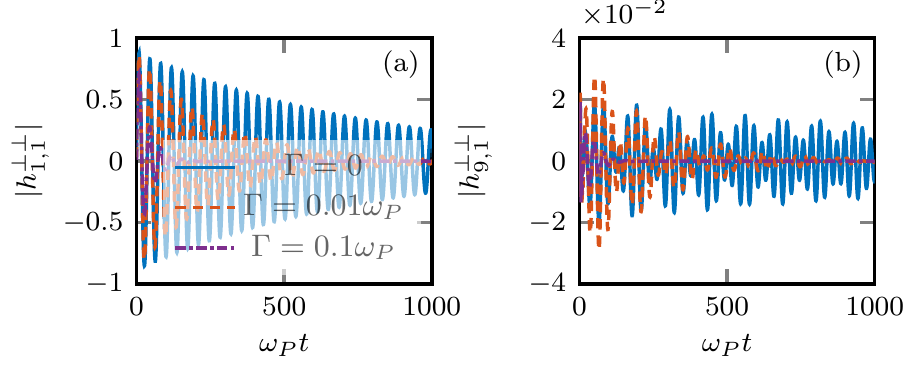}
    \caption{Impulse responses of a dielectric disk ($\omega_0 = \omega_p/4$, $\beta=\pi$) for $\Gamma=0$, $0.01\omega_P$, $0.1\omega_P$, where $\beta=k_P a$, $a$ is the radius and $k_P=\omega_P/c_0$. The impulse response $h^{\perp \perp}_{1,1}\left( t \right)$ (a) corresponds to the amplitude response shown in Fig. 15 (a) and the impulse response $h^{\perp \perp}_{9,1}$ (b) corresponds to the amplitude response shown in Fig. 15 (b). }
    \label{fig:TimeTE_G}
\end{figure}

We now investigate the role of material losses. Specifically, in Figs. \ref{fig:FreqTE_G} and \ref{fig:TimeTE_G} we show the same frequency and impulse responses for $\beta = \pi$ by varying $\Gamma$. In Figures \ref{fig:FreqTE_G}, we observe that an increase of $\Gamma$ determines a smoothing and broadening of the peaks of the amplitude responses. In particular, in Fig. \ref{fig:FreqTE_G} (b) and (c), it is apparent that for $\Gamma=0.1 \omega_P$ the two peaks associated to the two transverse modes merge into one. In Figure \ref{fig:TimeTE_G}, we show that the impulse responses exhibit a faster decay, as expected from the broadening of the corresponding frequency responses. In Figure \ref{fig:TimeTE_G} (b) we show that, while for low losses, ($\Gamma=0$ and $\Gamma=0.01 \omega_P$), the impulse response $h_{9,1}^{\perp,\perp}$ shows a beating between the resonance frequencies of the longitudinal and transverse modes, this beating is no longer visible for $\Gamma=0.1\omega_P$. This is consistent with the fact that in the frequency response the two peaks broaden and merge.

\subsubsection{Electric dipole coupling}

\begin{figure}
    \centering
    \includegraphics[width=\columnwidth]{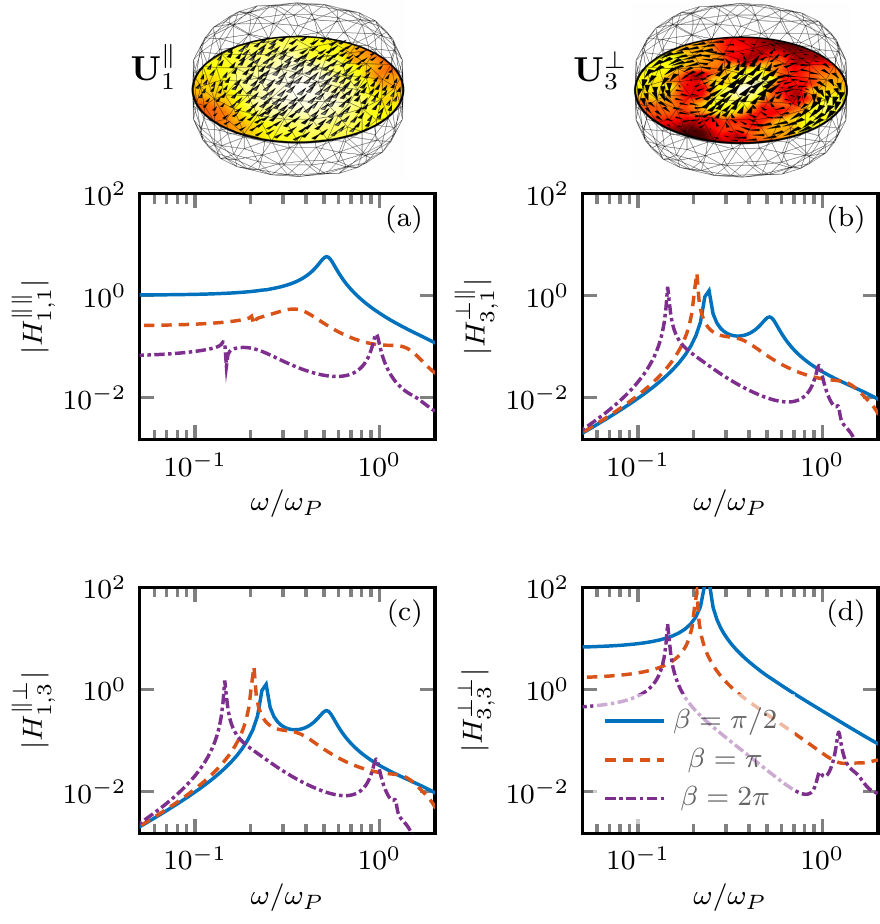}
        \caption{Amplitude responses of a dielectric disk ($\omega_0 = \omega_p/4$, $\Gamma=0$) with $\beta=\pi/2, \pi, 2\pi$, where $\beta=k_P a$, $a$ is the radius and $k_P=\omega_P/c_0$. The elements $H^{\parallel \parallel}_{1,1}$ (a), $H^{\perp \parallel}_{3,1}$ (b), $H^{\parallel \perp}_{1,3}$ (c) and $H^{\perp \perp}_{3,3}$ (d) are associated to the modes $\mathbf{U}_{1}^\perp$ and $\mathbf{U}_{3}^\perp$ shown in the inset above; $H_{1,3}^{\parallel \perp}=H_{3,1}^{\perp \parallel}$ for the reciprocity.}
    \label{fig:FreqTM}
\end{figure}

\begin{figure}
    \centering
    \includegraphics[width=\columnwidth]{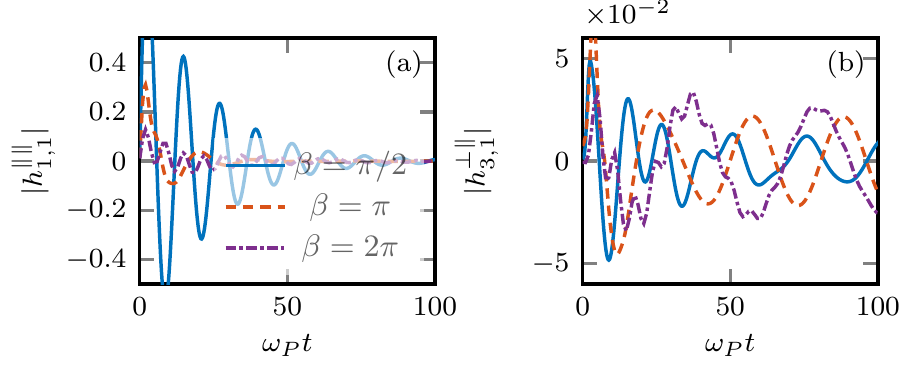}
    \caption{Impulse responses of a lossless  dielectric disk ($\omega_0 = \omega_p/4$, $\Gamma=0$) with $\beta=\pi/2$, $\pi$, $2\pi$, where $\beta=k_P a$, $a$ is the radius and $k_P=\omega_P/c_0$. The impulse response $h^{\parallel \parallel}_{1,1}\left( t \right)$ (a) corresponds to the amplitude response shown in Fig. 17 (a) and the impulse response $h^{\perp \parallel}_{3,1}$ (b) corresponds to the amplitude response shown in Fig. 17 (b).}
    \label{fig:TimeTM}
\end{figure}

We consider the same dielectric disk, but now focus on the coupling of the electric dipole mode to other modes. As in the case of a metal disk, from the numerical analysis we have found that the electric dipole mode $\mathbf{U}_{1}^\parallel$ significantly couples to the mode $\mathbf{U}_{3}^\perp$, which carries a $\mathbf{P}_{e2}$ dipole moment \cite{bladel_hierarchy_1988}, due to the symmetry of the problem. As the size parameter $\beta$ varies in the interval $\left[0,2\pi\right]$ the interaction with other modes is negligible.
The two  modes are shown in the equatorial plane of the disk on top of Fig. \ref{fig:FreqTM}. 

In Figures \ref{fig:FreqTM} (a) and (b), we show the amplitude response of $H_{1,1}^{\parallel \parallel}$ and $H_{3,1}^{ \perp \parallel}$. In Figures \ref{fig:FreqTM} (c) and (d), we show the amplitude responses of $H_{1,3}^{\parallel \perp}$ and $H_{3,3}^{\perp \perp}$.  As expected, the curves in panels (b) and (c) are identical for reciprocity.

First, we describe the behavior of the amplitude response of $H^{\parallel \parallel}_{1,1}$. For $\beta = \pi$, it shows only one peak located in proximity of the natural frequency of the longitudinal mode $\mathbf{U}_{1}^\parallel$. It is apparent that a zero cancels the pole associated with the natural frequency of the transverse mode. At higher frequencies there is a ``bump" due to the transverse electromagnetic standing waves of the particle, as the case of the sphere \cite{forestiere_time-domain_2021}. For $\beta = 2\pi$ this bump becomes a secondary peak.

The amplitudes of $H^{\parallel \perp}_{1,3}$ and $H^{\perp \parallel}_{3,1}$  are identical for reciprocity. For $\beta=\pi/2$, the amplitude of $H^{\perp \parallel}_{3,1}$ has two peaks and one bump. The first peak is located in proximity of the natural frequency of transverse mode $\mathbf{U}_{3}^\perp$, the second one is located in proximity of the natural frequency of the longitudinal mode $\mathbf{U}_{1}^\parallel$. By increasing $\beta$ to $\pi$ the second peak becomes a bump, and a second high-frequency bump arises due to the transverse electromagnetic standing waves of the particle. For $\beta= 2 \pi$, the low-frequency bump disappears, while the second high-frequency bump becomes a peak.

The amplitude response of $H^{\perp \perp}_{3,3}$ for $\beta = \pi/2$ is dominated by the peak associated with the natural frequency of the transverse mode $\mathbf{U}_{3}^\perp$. For $\beta = \pi$ a high frequency bump appears, associated with transverse electromagnetic standing waves. The first peak also dominates the response for $\beta=2\pi$, while the second bump becomes a secondary peak.

The impulse response $h^{\parallel \parallel}_{1,1}$ is shown in Fig. \ref{fig:TimeTM} (a). For $\beta=\pi/2$, it oscillates with the natural frequency of the longitudinal mode. It shows a fast decay rate due to radiation losses. For $\beta=\pi$, the decay rate is even faster. For $\beta=2\pi$, the impulse response oscillates with the frequency associated with the transverse electromagnetic standing waves of the disk as for the sphere \cite{forestiere_time-domain_2021}. It exhibits a slower decay. The impulse response $h^{\perp \parallel}_{3,1}$ is shown in Fig. \ref{fig:TimeTM} (b). For $\beta=\pi/2$, it shows a beating between the natural frequencies of the transverse and longitudinal modes. For $\beta=\pi$, the impulse response oscillates with the natural frequency of the transverse mode. For $\beta= 2 \pi$, the impulse response $h^{\parallel \parallel}_{3,1}$ shows a very slow decay with oscillation given by the natural frequency of the transverse mode. Furthermore, these oscillations are modulated by ripples that oscillate with the natural frequency of the transverse mode.

\begin{figure}
    \centering
    \includegraphics[width=\columnwidth]{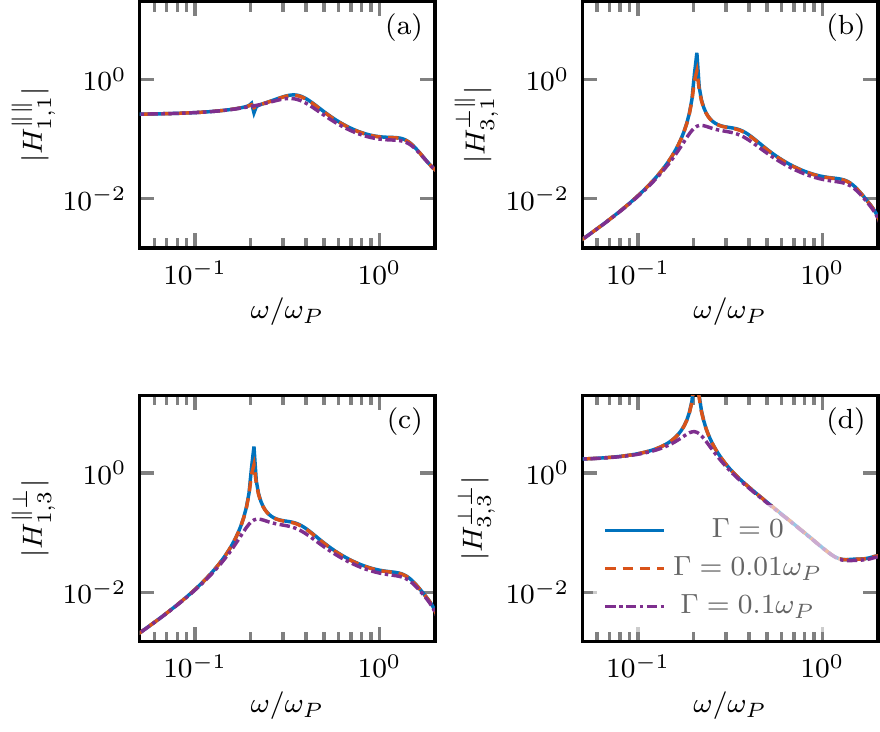}
    \caption{Amplitude responses of a dielectric disk ($\omega_0 = \omega_p/4$, $\beta=\pi$) for $\Gamma=0$, $0.01\omega_P$, $0.1\omega_P$, where $\beta=k_P a$, $a$ is the radius and $k_P=\omega_P/c_0$. The elements $H^{\parallel \parallel}_{1,1}$ (a), $H^{\perp \parallel}_{3,1}$ (b), $H^{\parallel \perp}_{1,3}$ (c) and $H^{\perp \perp}_{3,3}$ (d) are associated to the modes $\mathbf{U}_{1}^\parallel$ and $\mathbf{U}_{3}^\perp$ shown in the inset above Fig. 17; $H_{1,3}^{\parallel \perp}=H_{3,1}^{\perp \parallel}$ for the reciprocity. } 
    \label{fig:FreqTM_G}
\end{figure}

\begin{figure}
    \centering
    \includegraphics[width=\columnwidth]{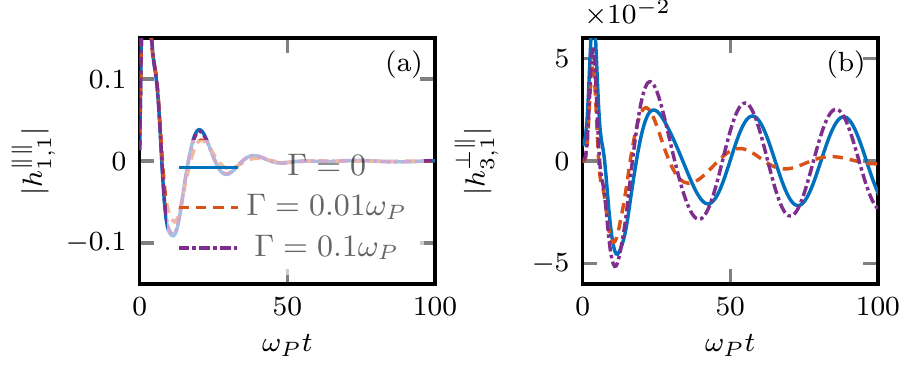}
    \caption{Impulse responses of a dielectric disk ($\omega_0 = \omega_p/4$, $\beta=\pi$) with $\Gamma=0$, $0.01\omega_P$, $0.1\omega_P$, where $\beta=k_P a$, $a$ is the radius and $k_P=\omega_P/c_0$. The impulse response $h^{\parallel \parallel}_{1,1}\left( t \right)$ (a) corresponds to the amplitude response shown in Fig. 19 (a) and the impulse response  $h^{\perp \parallel}_{3,1}$ (b) corresponds to the amplitude response  shown in Fig. 19 (b). }
    \label{fig:TimeTM_G}
\end{figure}

Now, we investigate the role of material losses. Specifically, in Figs. \ref{fig:FreqTM_G} and \ref{fig:TimeTM_G} we study the frequency and impulse responses for $\beta = \pi$ by varying $\Gamma$. In Figures. \ref{fig:FreqTM_G}, we observe that an increase of $\Gamma$ determines a smoothing and broadening of the peaks of the amplitude responses associated with both longitudinal and transverse modes of the disk. In Figure \ref{fig:TimeTM_G} (a), we show that the impulse responses exhibit a faster decay as $\Gamma$ increases, as expected by the broadening of the corresponding frequency response.  

\section{Summary and Conclusions}
\label{sec:Conclusion}

We introduced an operative full-wave approach for modeling the quantum electrodynamics of dispersive dielectric objects of finite size in unbounded space in the Heisenberg picture. It is based on a Hopfield type scheme. Its principal characteristics are: 

i) the matter and the electromagnetic field are kept distinct,  enabling the treatment of the polarization and electromagnetic fluctuations on equal footing; 

ii) the polarization density field observable is expanded in terms of the static longitudinal and transverse modes of the object. In this way, the Coulomb and Ampere interaction energy terms of the Hamiltonian are diagonalized. 

iii) The radiation field observables are expanded in terms of transverse plane waves. 

iv) The equation of motion for the longitudinal and transverse coordinate operators of the polarization field observable are coupled. The coupling is due to the interaction of the polarization with the solenoidal component of the electromagnetic field, which is described through the full-wave transverse dyadic Green's function for the vector potential in free space. The driving terms of the equations are operators that take into account the initial conditions of the matter field observables and the radiation field observables. 

v) When the size of the dielectric object is much smaller than $\lambda_c=\min\limits_{\omega}\{c_0/[\omega \sqrt{|{\chi}(\omega)|}]\}$ the static longitudinal and transverse modes are the natural modes of the polarization, and their mutual coupling due to radiation is weak. As the size increases, the modes become increasingly coupled, but as long as the size of the object is up to $\lambda_c$, each mode couples to a few modes. 

The principal outcomes of this approach, which advances the literature, are:

a) The polarization density field observable is expressed in terms of the driving term operators, through the impulse response of the dielectric object that we obtain in the framework of the classical electrodynamics. 

b) The electric field observable is expressed in terms of the polarization density field observable by means of the dyadic Green's function for the free space. 

c) The statistical functions of the polarization density field observable and the radiation field observables are integral operators of the statistics of the driving term operators. The kernels of the integral operators are linear/multilinear expressions of the impulse responses of the dielectric object. 

d) Few static longitudinal and transverse modes are implicated in the numerical calculation of each element of the impulse response matrix of dielectric objects with sizes of the order up to $\lambda_c$.

We evaluated the impulse response matrix for different object shapes by using the Drude-Lorentz model for the susceptibility. First, we consider the case of an infinite homogeneous dielectric. Differently from the Huttner and Barnett's paper, we have the additional contribution of $\hat{j}_\mu^{\left(e\right)}$, which takes into account the fluctuation of the electromagnetic field. This is consistent with what Drezet had already observed \cite{drezet_quantizing_2017}. Second, we treated the case of a dielectric slab with a linearly polarized electromagnetic wave propagating normally to the slab. Both in this case and in the previous one the longitudinal modes are absent due to homogeneity.
Then, we validated the impulse response matrix against a semi-analytical solution in the case of a sphere. Eventually, we analysed a metal disk and a dielectric disk, which are very relevant for nano-photonics. We investigated the frequency response and the impulse response of modes with low multipolar order. We consider a disk with size parameters $\beta \in [0,2\pi]$ without and with losses. For $\beta=\pi/2$ we verified that the modes are weakly coupled, and the elements of the impulse response matrix are close to the impulse responses of damped harmonic oscillators, while for $\beta = 2 \pi$ the mode becomes coupled, nevertheless the coupling is limited to few modes. 

\newpage
\appendix

\section{Macroscopic classical dielectric susceptibility}
\label{sec:susceptibility}

We show, in the classical framework, that the macroscopic polarization density field is given by Eq. \ref{eq:constrel} when the polarization density field is expressed in terms of matter fields $\{\mathbf{Y}_\nu\}$ through Eq. \ref{eq:polfield0}, and the coupling coefficient $\alpha_{\nu}$ is given by the expression \ref{eq:coupling}. 

The classical time evolution of the matter field is governed by the equation (see Eq. \ref{eq:Ydotdot01})
\begin{equation}
\label{eq:YdotdotA}
\ddot{\mathbf{Y}}_\nu + \nu^2 {\mathbf{Y}}_\nu=\alpha_\nu \mathbf{E} \quad \text{\;in } V
\end{equation}
for $0\le\nu<\infty$. 
In the Laplace domain it becomes
\begin{equation}
\label{eq:YdotdotL}
(s^2+\nu^2) \boldsymbol{{\mathcal{Y}}}_\nu=\alpha_\nu \boldsymbol{{\mathcal{E}}} + [s{\mathbf{Y}}_\nu^{(0)}(\mathbf{r})+\dot{\mathbf{Y}}_\nu^{(0)}(\mathbf{r})]
\end{equation}
where $\boldsymbol{{\mathcal{Y}}}_\nu(\mathbf{r};s)$ is the Laplace transform of ${\mathbf{Y}}_\nu(\mathbf{r};t)$ and $\boldsymbol{{\mathcal{E}}}(\mathbf{r};s)$ is the Laplace transform of ${\mathbf{E}}(\mathbf{r};t)$.  $\mathbf{Y}_\nu^{(0)}(\mathbf{r})$ and $\dot{\mathbf{Y}}_\nu^{(0)}(\mathbf{r})$ denote, respectively, the vector field $\mathbf{Y}_\nu$ and its partial derivative with respect to the time evaluated at $t = 0$. Therefore, in the Laplace domain, the polarization density field is given by
\begin{equation}
\label{eq:PolA}
 \boldsymbol{{\mathcal{P}}}=\epsilon_0\tilde{\chi}(s) \boldsymbol{{\mathcal{E}}}+ \boldsymbol{{\mathcal{P}}}^{free}
\end{equation}
where
\begin{equation}
\label{eq:chiTilde}
\tilde{\chi}(s)=\frac{1}{\epsilon_0}\int_0^\infty d\nu\, \frac{\alpha_\nu^2}{s^2+\nu^2},
\end{equation}
and
\begin{equation}
\label{eq:P0}
 \boldsymbol{{\mathcal{P}}}^{free}(\bold{r};s)=\int_0^\infty d\nu\, \frac{\alpha_\nu}{s^2+\nu^2} [s{\mathbf{Y}}_\nu^{(0)}(\mathbf{r})+\dot{\mathbf{Y}}_\nu^{(0)}(\mathbf{r})].
\end{equation}

The region of convergence of the Laplace transform contains the imaginary axis; therefore, we evaluate $\tilde{\chi}(s)$ for $s=i\omega + \epsilon$ where $\epsilon \downarrow 0$.
By using the relation (e.g., \cite{heitler_quantum_1984})
\begin{equation}
  \frac{1}{x-i\epsilon}= i\pi\delta(x) + \mathcal{P}\frac{1}{x},
\end{equation}
where $\mathcal{P}$ denotes the Cauchy principal value, we obtain for $\chi(\omega)= \tilde{\chi}(s=i\omega+\epsilon)$ the following expression
\begin{equation}
  \varepsilon_0\chi(\omega)= \mathcal{P}\int_0^\infty d\nu\, \frac{\alpha_\nu^2}{\nu^2-\omega^2} -i\frac{\pi}{2}\frac{\alpha_\omega^2}{\omega}.
\end{equation}
Expressing the susceptibility in the frequency domain $\chi(\omega)$ as $\chi=\chi_r+i\chi_i$ we obtain
\begin{equation}
  \alpha_\nu=\sqrt{\frac{2\sigma(\nu)}{\pi}}
\end{equation}
where $ \sigma(\nu) = -\varepsilon_0 \nu{\chi}_i(\nu)$.

In the time domain, we have (in the region $V$)
\begin{equation}
   \label{eq:Ptime}
    \mathbf{P}\rpt = 
        \epsilon_0 \zeta (t) *\mathbf{E} \left(\mathbf{r};t \right) + \mathbf{P}^{free}(\mathbf{r;t})
\end{equation}
where $\zeta(t)$ is the inverse Fourier transform of the susceptibility of the dielectric $\chi(\omega)$,
\begin{equation}
    \mathbf{P}^{free}\rpt = \int_0^\infty d\nu\, \sqrt{\frac{2\sigma(\nu)}{\pi}} \mathbf{Y}_\nu^{free}\rpt,
\end{equation}
and
\begin{equation}
   \mathbf{Y}_\nu^{free}\rpt=\mathbf{Y}_\nu^{(0)}(\mathbf{r}) \cos(\nu t) + \frac{1}{\nu}\dot{\mathbf{Y}}_\nu^{(0)}(\mathbf{r}) \sin(\nu t).
\end{equation}
$\mathbf{P}^{free}(\mathbf{r;t})$ takes into account the contribution of the initial state of the matter field to the polarization dynamics: it would describe the evolution of the polarization density field if the interaction of the dielectric with the electric field was absent. It depends only on the initial state of the matter fields.

\section{Longitudinal and transverse static modes of the dielectric body}
\label{sec:StaticModes}

Following \cite{forestiere_time-domain_2021}, we exploit the static longitudinal (electrostatic) modes and the static transverse (magnetostatic) modes of the dielectric body to represent, respectively, the longitudinal and transverse components of the matter vector field operators. 

The static longitudinal modes of the body are solutions of the eigenvalue problem \cite{fredkin_resonant_2003,mayergoyz_electrostatic_2005, forestiere_resonance_2020}
\begin{equation}
    \nabla_{\bf{r}} \oint_{\partial V} \dS'\, \frac{ \Wl_m \rpp \cdot \n \rpp}{ 4 \pi \left| \rb - \rb' \right| }  =\frac{1}{\kappa_m^\parallel}\Wl_m \rp 
    \quad \text{in} \, V
    \label{eq:EQS}
\end{equation}
where $\kappa_m^\parallel$ is the eigenvalue associated with the eigenmode $\Wl_m\rp$. The eigenvalues $\kappa_m^\parallel$, which are dimensionless quantities, are discrete, real, positive, and equal to or greater than two ($\kappa_m^\parallel \ge 2$). The longitudinal eigenmodes and the corresponding eigenvalues depend only on the shape of the body, they do not depend on its size. The solution of problem \ref{eq:EQS} can be evaluated numerically using the method outlined in \cite{mayergoyz_electrostatic_2005,mayergoyz_plasmon_2012} and summarized in Appendix \ref{sec:Computation}.

The static transverse modes of the body are solutions of the eigenvalue problem \cite{forestiere_electromagnetic_2019,forestiere_magnetoquasistatic_2020}
\begin{equation}
\frac{1}{a^2} \int_{V} \dV'\, \frac{ \mathbf{U}_m^\perp\rpp }{4 \pi \left| \mathbf{r}- \mathbf{r}' \right|} = \frac{1}{\kappa_m^\perp} \mathbf{U}_m^\perp\rp \quad \text{in}\, V,
\label{eq:MQS}
\end{equation}
with
\begin{equation}
 \mathbf{U}_m^\perp\rp \cdot \n \rp=0 \qquad \text{on} \, \partial V,
 \label{eq:boundaryMQS}
\end{equation}
where $a$ is the radius of the smallest sphere that surrounds the dielectric and $\kappa_m^\perp$ is the eigenvalue associated to the eigenmode $\mathbf{U}_m^\perp$. Equation \ref{eq:MQS} with constraint \ref{eq:boundaryMQS} holds in a weak form in the functional space of the solenoidal vector field in $V$ with a normal component to $\partial V$ equal to zero, equipped with the inner product $\langle\mathbf{F}, \mathbf{G} \rangle$. The eigenvalues $\kappa_m^\perp$, which are dimensionless quantities, are discrete, real and positive. As for the longitudinal eigenmodes, the transverse eigenmodes and the corresponding eigenvalues depend only on the shape of the body; they do not depend on its size. The problem \ref{eq:MQS} can be solved by using standard tools of computational electromagnetism as outlined in \cite{forestiere_magnetoquasistatic_2020} and summarized in Appendix \ref{sec:Computation}.

\section{Computational of Longitudinal and Transverse Static Modes}

\label{sec:Computation}

\subsection{Longitudinal Modes}

To compute the longitudinal modes of the object we preliminary solve the eigenvalue problem \cite{mayergoyz_electrostatic_2005,mayergoyz_plasmon_2012}
\begin{equation}
 \sigma_m \left( \mathbf{r} \right) =- {\kappa^\parallel_m}\left[\frac{\sigma_m}{2} \left( \mathbf{r} \right) + \oint_{S} \dS'\, \sigma_m \left( \mathbf{r}' \right) \frac{\left(\mathbf{r}-\mathbf{r}' \right) \cdot \mathbf{n}\left(\mathbf{r}\right)}{4 \pi\left|\mathbf{r}-\mathbf{r}' \right|^{3}} \right]
 \label{eq:EQSmayergoyz}
\end{equation}
where $\mathbf{r} \in \partial V$ and $\sigma_m \left( \mathbf{r} \right)=\Wl_m \rp \cdot \n \rp$ is the eigenfunction. This problem is equivalent to the eigenvalue problem \ref{eq:EQS}. Then, we determine the longitudinal mode $\mathbf{U}_m^\parallel$ of the object that is given by
\begin{equation}
    \mathbf{U}_m^\parallel \left( \mathbf{r} \right) = \frac{1}{4\pi\varepsilon_0} \oint_{\partial V} \dS'\, \sigma_m \left( \mathbf{r}' \right) \frac{\left(\mathbf{r}-\mathbf{r}' \right) \cdot \mathbf{n}\left(\mathbf{r}\right)}{\left|\mathbf{r}-\mathbf{r}' \right|^{3}}.
\end{equation}
For solving numerically Eq. \ref{eq:EQSmayergoyz}, we discretize the boundary $\partial V$ of the object with a triangular surface mesh having triangles $T_j$ with $j=1,2,\ldots,N_t$. The charge density distribution is represented in terms of a basis of piecewise constant functions $ u_j$ on such triangles
\begin{equation}
    \sigma_m\rp = \sum_{j=1}^{N_t} I_j^S u_j \left( \mathbf{r} \right).
\end{equation}
Each basis function is defined as
\begin{eqnarray}
  u_j \left( \mathbf{r} \right) = \left\{ \begin{array}{cc} \frac{1}{A_j} & \text{in} \, T_j, \\ 0   & \text{otherwise}, \end{array} \right.
\end{eqnarray}
where $A_j$ is the area of the triangle $T_j$. Since the surface is overall neutral, the true unknowns are $N_t-1$. The discrete form of the integral equation \ref{eq:EQSmayergoyz} is 
\begin{equation}
    \dss{R} \, \ds{I} = -\frac{1}{2}\left[\dss{R} + \dss{C} \right] \ds{I} 
    \label{eq:EQSnumPro}
\end{equation}
where the elements of the matrices $\dss{R}$ and $\dss{L}$ are 
\begin{align}
    {R}_{ij} &= \oint_S \dS \, u_j \left( \mathbf{r} \right) u_i \left( \mathbf{r} \right)  = \left\{ \begin{array}{cc} \frac{1}{A_j} & i = j, \\ 0   & i \ne j, \end{array} \right. \\
    C_{ij} &= \frac{1}{A_i} \frac{1}{A_j}  \oint_{T_i} \oint_{T_j} \dS' \dS\, \frac{\left(\mathbf{r}-\mathbf{r}' \right) \cdot \mathbf{n}\left(\mathbf{r}\right)}{2 \pi\left|\mathbf{r}-\mathbf{r}' \right|^{3}}.
    \label{eq:Lss}
\end{align}
The singular surface integral in \ref{eq:Lss} involving the static Green function have been computed using the analytical formulas \cite{graglia_numerical_1993}. 

The problem \ref{eq:EQSnumPro} is reduced to a standard symmetric eigenvalue problem by exploiting the LAPACK \cite{anderson_lapack_1999} routine DSYGST. All eigenvalues and eigenvectors of the resulting real symmetric matrices are computed through the routine DSYEV.

\subsection{Transverse Modes}
Equation \ref{eq:MQS} can be numerically solved by drawing on the standard repertoire of computational electromagnetics for volume integral equations \cite{schaubert_tetrahedral_1984,van_bladel_electromagnetic_2007}. We introduce a tetrahedral mesh of the volume $V$ with $N_{node}$ nodes,  $N_{tetra}$ tetrahedral elements, and $N_{edge}$ edges. To discretize the current density field we use loop functions $\left\{ {\bf u}_k\right\}$ as a discrete basis \cite{li_applying_2006}.

The loop basis function associated to the $k$-th edge is given by (e.g. \cite{li_applying_2006}):
\begin{equation}
    {\bf u}_k (\boldsymbol{r}) = \sum_{h=1}^{N_k} \frac{\mathbf{e}_h}{V_h} f_h(\boldsymbol{r})
\end{equation}
where $N_k$ is the total number of tetrahedrons attached to the $k$-th edge. The unit vector $\mathbf{e}_h$ is parallel with the edge of the tetrahedron $T_h$ that is not adjacent to the $k$-th edge. $V_h$ is the volume of the tetrahedron $T_h$ and
\begin{equation}
    {f}_h\rp = \left\{
    \begin{array}{cl}
        1& \text{in} \; T_h,  \\
          0 & \text{otherwise} \;.
    \end{array}
    \right.
\end{equation}

A generic edge may be either a ``boundary" edge or an ``interior" edge. The loop functions associated to the interior edges are inherently solenoidal. We can also associate a ``half-loop"  basis function with each boundary edges (e.g. \cite{li_applying_2006}), but they give a non-vanishing surface charge.  However, because of boundary condition \ref{eq:boundaryMQS}, we disregard the ``half-loop" basis functions.

The basis function associates to the interior edges are not linearly independent.  There are several ways available to set up a maximum and independent loop basis set. We take advantage of the fact that nodes and edges in a geometrical mesh can also be used to set up an undirected graph. We recall some concept from graph theory.

A tree $\mathcal{T}$ of a connected graph $\mathcal{G}$ is a connected subgraph that contains all the nodes of $\mathcal{G}$, but without any loop. For any given graph $\mathcal{G}$, many possible choices of trees are possible. Given a connected graph and a chosen tree, the branches of $\mathcal{G}$ are partitioned in two disjoint sets: the ones belonging to $\mathcal{T}$, called {\it twigs}, and the ones that do not belong to $\mathcal{T}$ that are called {\it links}. 

We introduce the graph $\mathcal{G}$ associated with all the nodes and edges of the mesh of $V$ and the sub-graph $\mathcal{G}_{S}$ associated with the boundary nodes and edges. We consider a tree $\mathcal{T}'$ of $\mathcal{G}$, that has the additional constraint of including a tree of $\mathcal{G}_{S}$. We use as a basis for the current density field the loop functions $\{\bold{u}_h^i\rp\}$ associated with the internal links of the tree $\mathcal{T}'$.

We represent the unknown $\mathbf{J}\rp$ in terms of the $N_L$ loop functions
\begin{equation}
\mathbf{J}\rp = \sum_{h=1}^{N_{L}}I_{h} \mathbf{u}_{h}^i 
\label{eq:representation}
\end{equation}

The discrete generalized eigenvalue problem is obtained by substituting the expansion (\ref{eq:representation}) into Eq. (\ref{eq:MQS}) and applying the Galerkin method
\begin{equation}
\frac{\kappa_m^\perp}{a^2} \dss{ L} \, \ds{I} = \dss{R} \, \ds{I}.
\label{eq:NumEigPro}%
\end{equation}
The elements of these matrices are given by:
\begin{subequations}
\begin{align}
  R_{pq}  & =\int_{V} \dV\, \mathbf{u}%
_{p}^i\left(  \mathbf{r}\right) \cdot \mathbf{u}_{q}^i \left(  \mathbf{r}\right) \\
 {L}_{pq}  & = \int_{V}%
\int_{V}  \dV \, \dV'\, \frac{\mathbf{u}_{p}^i \left(  \mathbf{r}\right)  \cdot
\mathbf{u}_{q}^i \left( \mathbf{r}^{\prime}\right)}{4\pi \left| \mathbf{r} - \mathbf{r}' \right|} .
\end{align}
\end{subequations}
The problem \ref{eq:NumEigPro} is reduced to a standard symmetric eigenvalue problem by exploiting the LAPACK \cite{anderson_lapack_1999} routine DSYGST, then all the eigenvalues and eigenvectors of the resulting real symmetric matrix are computed through the routine DSYEV.


\section{Expression of the kernel $s_{mm'}^{a\,a'}(t)$ in terms of the transverse dyadic Green function in free space}
\label{sec:GreenFunction}

In this Appendix, we express the kernel $s_{mm'}^{a\,a'}(t)$ defined by \ref{eq:kerneltime} in terms of the transverse dyadic Green function for the vector potential in free space.

The Laplace transform of the kernel $s_{mm'}^{a\,a'}(t)$ is
\begin{equation}
\label{eq:ExpS}
    S_{mm'}^{a\,a'}(s)=\sum_{\mu}\frac{s}{s^2+c_0^2 k^2} \langle \Ua_m, \mathbf{w}_\mu \rangle \langle \mathbf{w}_\mu,\Ub_{m'}\rangle .
\end{equation}
We rewrite it as follows:
\begin{equation}
\label{eq:SintLaplace}
   S_{mm'}^{a\,a'}(s)=\frac{s}{c_0^2} \int_{V}\dV\int_{V}\dV'\, \Ua_m(\rb)\overleftrightarrow{G}^\perp(\rb-\rb';s)\Ub_{m'}(\rb'),
\end{equation}
where $\overleftrightarrow{G}^\perp(\rb-\rb';s)$ is the dyad
\begin{equation}
\label{eq:ExpS1}
    \overleftrightarrow{G}^\perp(\rb-\rb';s)=\sum_{\mu}\frac{1}{k^2+s^2/c_0^2} \mathbf w_{\mu}(\rb)\,\mathbf w_{\mu}^*(\rb').
\end{equation}
Using the expression of $\mathbf w_{q}(\rb)$ (see \ref{eq:planewave}) we obtain:
\begin{equation}
\label{eq:ExpS2}
    \overleftrightarrow{G}^\perp(\rb;s)=\frac{1}{(2\pi)^3}\int{} d^3\mathbf{k} \,\overleftrightarrow{\mathcal{G}}^\perp(\mathbf{k};s)\,e^{i\mathbf{k}\cdot\rb} 
\end{equation}
where
\begin{equation}
\label{eq:FourierI}
    \overleftrightarrow{\mathcal{G}}^\perp(\mathbf{k};s)= \frac{1}{k^2+s^2/c_0^2} (\overleftrightarrow{I}-\hat{\mathbf{k}}\,\hat{\mathbf{k}})
\end{equation}
is the transverse dyadic Green function for the electric field in the wavenumber domain and in free space. By evaluating the Fourier integral \ref{eq:ExpS1} we obtain (e.g., \cite{arnoldus_transverse_2003})
\begin{equation}
\label{eq:GreenT}
    \overleftrightarrow{G}^\perp \left(\rb;s \right)=\overleftrightarrow{G} \left(\rb;s \right) - \overleftrightarrow{G}^\parallel \left(\rb;s \right),
\end{equation}
where 
\begin{multline}
    \overleftrightarrow{G} \left(\rb;s \right)=\frac{e^{-sr/c_0}}{4\pi r}\times \\
  \times \left[(\overleftrightarrow{I}-{\mathbf{e}}_r{\mathbf{e}}_r)
  + c_0\frac{(\overleftrightarrow{I}-3\mathbf{e}_r{\mathbf{e}}_r)}{sr}(1+\frac{c_0}{sr})\right],
\end{multline}
$\mathbf{e}_r=\mathbf{r}/r$ and
\begin{equation}
\label{eq:Green0}
  \overleftrightarrow{G}^\parallel \left(\rb;s \right)= \frac{c_0^2}{s^2}\frac{(\overleftrightarrow{I}-3\mathbf{e}_r{\mathbf{e}}_r)}{4\pi r^3}.
\end{equation}
$\overleftrightarrow{G}\left(\rb;s \right)$ is the dyadic Green for the vector potential in the temporal gauge, $\overleftrightarrow{G}^\perp \left(\rb;s \right)$ is the transverse component, and $\overleftrightarrow{G}^\parallel \left(\rb;s \right)$ is the longitudinal component. 

In the time domain, the transverse dyadic Green function in free space is given by
\begin{multline}
\label{eq:GreenTravTime} 
    \overleftrightarrow{g}^\perp \left(\rb;t \right)=\frac{(\overleftrightarrow{I}-\mathbf{e}_r{\mathbf{e}}_r)}{4\pi r}\delta(t-r/c_0)+ \\ 
    c_0\frac{(\overleftrightarrow{I}-3\mathbf{e}_r{\mathbf{e}}_r)}{4\pi r^2}\,u(t-r/c_0)\left[1+\frac{c_0}{r}(t-r/c_0))\right]+\\
    -c_0^2\frac{(\overleftrightarrow{I}-3\mathbf{e}_r{\mathbf{e}}_r)}{4\pi r^3} u(t)t\,. \quad
\end{multline}
The expression of $s_{mm'}^{a\,a'}(t)$ is
\begin{equation}
\label{eq:Sinttime2}
   s_{mm'}^{a\,a'}(t)=\frac{1}{c_0^2} \int_{V}\dV\int_{V}\dV'\, \Ua_m(\rb)\dot{{\overleftrightarrow{g}}}\,^\perp(\rb-\rb';t)\Ub_{m'}(\rb')
\end{equation}
where
\begin{multline}
    \dot{\overleftrightarrow{g}}^\perp \left(\rb;t \right)=\frac{(\overleftrightarrow{I}-\mathbf{e}_r{\mathbf{e}}_r)}{4\pi r}\dot{\delta}(t-r/c_0)+\\
    \label{eq:DerGreenTravTime}
    +c_0\frac{(\overleftrightarrow{I}-3\mathbf{e}_r{\mathbf{e}}_r)}{4\pi r^2}\left[\delta(t-r/c_0)+\frac{c_0}{r}\,u(t-r/c_0))\right]+\\
    -c_0^2\frac{(\overleftrightarrow{I}-3\mathbf{e}_r{\mathbf{e}}_r)}{4\pi r^3} \,u(t). \qquad
\end{multline}

It is convenient to express $\overleftrightarrow{G}^\perp(\rb;s)$ as
\begin{equation}
   \overleftrightarrow{G}^\perp(\rb;s)=\overleftrightarrow{g}_0^\perp(\rb)+\overleftrightarrow{G}_d^\perp(\rb;s),
   \label{eq:GreenDecomposition}
\end{equation}
where 
\begin{equation}
\label{eq:GreenStat}
    \overleftrightarrow{g}_0^\perp \left(\rb \right)=\overleftrightarrow{G}^\perp(\rb;s=0)=(\overleftrightarrow{I}+\hat{\mathbf{r}}\,\hat{\mathbf{r}})\frac{1}{8 \pi r}
\end{equation}
and 
\begin{equation}
\label{eq:GreenDin}
    \overleftrightarrow{G}_d^\perp \left(\rb;s \right) = \overleftrightarrow{G}^\perp \left(\rb; s\right) - \overleftrightarrow{g}_0^\perp \left(\rb \right).
\end{equation}
The term $\overleftrightarrow{g}_0^\perp$ is the static transverse dyadic Green function for the free space, which diverges as $1/r$ for $r\rightarrow0$. The dynamic part $\overleftrightarrow{G}_d^\perp$, which tends to zero as $s$ for $s\rightarrow0$, is a regular function of $r$. Indeed, we have
\begin{multline}
    \overleftrightarrow{G}_d^\perp \left( s \right) =  -\frac{s}{4 \pi c_0} e^{-{sr}/{2c_0}} f_1 \left( {sr}/{2c_0} \right) \left( \overleftrightarrow{I} - \mathbf{e}_r{\mathbf{e}}_r \right)
    + \\ (\overleftrightarrow{I}-3\mathbf{e}_r{\mathbf{e}}_r)\frac{s}{12\pi c_0}  f_2 \left( s r / c_0 \right) 
\end{multline}
where
\begin{align}
f_1 \left( \xi \right) &= \frac{\sinh{\xi}}{\xi}, \\
f_2 \left( \xi \right) &=
\frac{3}{2 x^3} \left( - 2  + \xi^2 + 2 \xi e^{-\xi} + 2 e^{-\xi} \right).
\end{align}
The functions $f_1 \left( \xi \right)$ and $f_2 \left( \xi \right)$ are regular as $\xi \rightarrow 0$, in particular $f_1, f_2 \rightarrow 1$.

\section{Evaluation of the coefficients $ S_{mm'}^{aa'}$}

\label{sec:CoefficientS}
We first consider the coefficients $S_{mm'}^{\parallel \parallel}(s)$ and $S_{mm'}^{\perp \perp}(s)$. Starting from expression \ref{eq:Sint} and applying the Green decomposition \ref{eq:GreenDecomposition} and the integral identity \ref{eq:IntegralIdentityI} we obtain
\begin{multline}
\label{eq:Sint_ll}
  \frac{c_0^2}{s} \, S_{mm'}^{\parallel \parallel}(s)= \\ \int_{V}\dV\int_{V}\dV'\, \mathbf{U}^\parallel_m(\rb)\ \overleftrightarrow{G}_d^\perp(\rb-\rb';s)\mathbf{U}^\parallel_{m'}(\rb') + \\ + \int_{V}\dV\int_{V}\dV'\, \mathbf{U}^\parallel_m(\rb) \cdot   \frac{1}{4 \pi \left| \rb-\rb'\right|} \mathbf{U}^\parallel_{m'}(\rb') + \\  \oint_{\partial V}\dS\oint_{\partial V}\dS'\, \mathbf{U}^\parallel_m(\rb) \cdot \n   \frac{|\rb-\rb'|}{8\pi} \mathbf{U}^\parallel_{m'}(\rb') \cdot \n
\end{multline}
and
\begin{multline}
\label{eq:Sint_tt}
   \delta S_{mm'}^{\perp \perp}(s)= \\ \frac{1}{c_0^2} \int_{V}\dV\int_{V}\dV'\, \mathbf{U}^{\perp}_{m}(\rb)\,s\,\overleftrightarrow{G}_d^\perp(\rb-\rb';s)\mathbf{U}^{\perp}_{m'}(\rb').
\end{multline}

We now consider the coefficient $S_{mm'}^{\parallel \perp}$. Proceeding as above we obtain
\begin{multline}
\label{eq:Sint_lt}
  \frac{c_0^2}{s} \, S_{mm'}^{\parallel \perp}(s)=  \\ \int_{V}\dV\int_{V}\dV'\, \mathbf{U}_m^\parallel(\rb)\ \overleftrightarrow{G}_d^\perp(\rb-\rb';s)\mathbf{U}^{     \perp}_{m'}(\rb') + \\ + \int_{V}\dV\int_{V}\dV'\,  \frac{\mathbf{U}^\parallel_m(\rb) \cdot \mathbf{U}^\perp_{m'}(\rb')  }{4 \pi \left| \rb-\rb'\right|} 
\end{multline}
where we have exploited the fact that the normal component of $\mathbf{U}^\perp_{m'}$ to the boundary $\partial V$ is zero.

\section{Small size limit}
\label{sec:small size}

We here analyze the behavior of the coefficients $s S_{mm'}^{a\,b}(s)$ in the small size limit, i.e., $\gamma=(a|s|/c_0) \ll 1$. We have:
\begin{equation}
   \label{eq:small1}
s S_{mm'}^{\parallel\,\parallel}(s)=\gamma^2\Sigma_{mm'}^{\parallel\,\parallel}+\mathcal{O}\left(\gamma^4\right),
\end{equation}
\begin{equation}
 \label{eq:small2}
s S_{mm'}^{\parallel\,\perp}(s)=\gamma^2\Sigma_{mm'}^{\parallel\,\perp}+\mathcal{O}\left(\gamma^5\right),
\end{equation}
and
\begin{equation}
 \label{eq:small3}
s S_{mm'}^{\perp\,\perp}(s)=\frac{a^2\gamma^2}{\kappa_m^\perp}\delta_{mm'}+\gamma^4\Sigma_{mm'}^{\perp\,\perp}+\mathcal{O}\left(\gamma^4\right),
\end{equation}
where
\begin{equation}
    \Sigma_{mm'}^{\perp\,\perp} =\frac{1}{8 \pi a^4} \int_V \dV \int_V \dV'  \, \mathbf{U}_m^\perp \rp \cdot \mathbf{U}_{m'}^\perp \rpp \left| \rb -\rb'\right|,
    \end{equation}
    \begin{equation}
   \Sigma_{mm'}^{\parallel\,\perp} = \frac{1}{4 \pi a^2} \int_V \dV \int_V \dV'  \,   \frac{ \mathbf{U}_m^\parallel \rp \cdot \mathbf{U}_{m'}^\perp \rpp}{\left| \rb -\rb' \right|},
   \end{equation}
   and
 \begin{multline}
     \Sigma_{mm'}^{\parallel\,\parallel} = \frac{1}{4 \pi a^2} \int_V \dV \int_V \dV' \, \frac{\mathbf{U}_m^\parallel \rp \cdot   \mathbf{U}_{m'}^\parallel \rpp}{\left| \rb -\rb' \right|} +\\
    \frac{1}{8 \pi a^2} \oint_{S} \dS \oint_{S} \dS'
    \mathbf{U}^\parallel_{m} \rp \cdot \n {\left| \mathbf{r} - \mathbf{r}' \right|} \mathbf{U}^\parallel_{m'} \rpp \cdot \n' .
 \end{multline}
The quantities $\Sigma_{mm'}^{\perp\,\perp}$, $\Sigma_{mm'}^{\parallel\,\perp}$, $\Sigma_{mm'}^{\parallel\,\parallel}$ and $1/\kappa_p^\perp$ do not depend on the size of the dielectric object $a$ and on the complex variable $s$, they only depend on the object shape. 

Equations \ref{eq:small1}-\ref{eq:small3} have been obtained by using the identities
\begin{multline}
    \int_V \dV \int_V \dV' \, \Ua_{p} \rp  \frac{ \left( \mathbf{r} - \mathbf{r}' \right) \, \left( \mathbf{r} - \mathbf{r}' \right) } {\left| \mathbf{r} - \mathbf{r}' \right|^3} \Ub_{p'} \rpp = \\
    \oint_{S} \dS \oint_{S} \dS' \left( \Ua_{p} \rp \cdot \n \right)  {\left| \mathbf{r} - \mathbf{r}' \right|} \left(  \Ub_{p'} \rpp \cdot \n'  \right) + \\
    \int_V \dV \int_V \dV' \,  \frac{\Ua_p \rp \cdot \Ub_{p'} \rpp} {\left| \mathbf{r} - \mathbf{r}' \right|}, 
    \label{eq:IntegralIdentityI}
\end{multline}
\begin{multline}
    \int_V \dV \int_V \dV' \, \Ua_{p} \rp  \frac{ \left( \mathbf{r} - \mathbf{r}' \right) \, \left( \mathbf{r} - \mathbf{r}' \right) } {\left| \mathbf{r} - \mathbf{r}' \right|} \Ub_{p'} \rpp = \\
    -\oint_{S} \dS \oint_{S} \dS' \left( \Ua_{p} \rp \cdot \n \right)  {\left| \mathbf{r} - \mathbf{r}' \right|^2} \left(  \Ub_{p'} \rpp \cdot \n'  \right) + \\
    -\frac{1}{2} \int_V \dV \int_V \dV' \,   {\left| \mathbf{r} - \mathbf{r}' \right|} \Ua_p \rp  \cdot \Ub_{p'} \rpp,
    \label{eq:IntegralIdentityII}
\end{multline}
and the asymptotic expression of the transverse dyadic Green function
\begin{multline}
 \overleftrightarrow{G}^\perp \left(\rb;s \right) =  \overleftrightarrow{g}_0^\perp \left(\rb \right) -\frac{2}{12 \pi r} \left( \frac{s \, r}{c_0} \right) \\ + \frac{\left( 3 \overleftrightarrow{I} - {\mathbf{e}}_r\,{\mathbf{e}}_r\right)}{32 \pi r} \left( \frac{s \, r}{c_0} \right)^2+\mathcal{O} \left( \frac{s \, r}{c_0} \right)^3
 \label{eq:tGreenAsym}
\end{multline} 
where $\overleftrightarrow{g}_0^\perp \left(\rb \right)$ is given by \ref{eq:GreenStat}.
\\
\section{Electric field operator}
\label{sec:electricfield}

In this Appendix, we evaluate the expressions for the electric field operator in terms of the polarization field operator. The electric field operator $\hat{\mathbf{E}}$ has two contributions in $V_\infty$: the solenoidal component $\hat{\mathbf{E}}_s$ and the irrotational component $\hat{\mathbf{E}}_c$,
\begin{equation}
    \hat{\mathbf{E}} = \hat{\mathbf{E}}_{s} + \hat{\mathbf{E}}_c.
\end{equation}
\\
\subsection{Solenoidal component}
In the time domain, the solenoidal component of electric field operator is given by
\begin{equation}
    \hat{\mathbf{E}}_s=-\dot{\hat{\mathbf{A}}},
\end{equation}
hence
\begin{equation}
\label{eq:Etime}
    \hat{\mathbf{E}}_{s}(\mathbf{r};t) =- \sum_{\mu} \dot{\hat{A}}_\mu(t) \mathbf{w}_\mu \rp. \\
\end{equation}
Using Eq. \ref{eq:rad1} and Eq. \ref{eq:pola} from Eq. \ref{eq:Etime} we obtain
\begin{widetext}
\begin{equation}
\label{eq:Etime1}
    \hat{\mathbf{E}}_{s}(\mathbf{r};t) = -\frac{1}{\varepsilon_0}\sum_{\mu}\sum_{\substack{{m}, {a}}} R_{\mu m}^a w_{\mu}(t)*\dot{\hat{p}}_{m}^a(t)\mathbf{w}_\mu \rp + \hat{\mathbf{E}}_s^{free}(\mathbf{r};t)
\end{equation}
where $\hat{\mathbf{E}}_s^{free}$ is given by \ref{eq:Efree}.
In the Laplace domain, this relation becomes
\begin{equation}
\label{eq:ELapl}
    \boldsymbol{\hat{\mathcal{E}}}_{s}(\mathbf{r};s) = -\frac{1}{\varepsilon_0}\sum_{\mu}\sum_{\substack{{m}, {a}}} R_{\mu m}^a \mathscr{W}_{\mu}(s)[{s\hat{P}}_{m}^a(s)-\hat{p}_{m}^{a\,(S)}]\mathbf{w}_\mu \rp + \boldsymbol{\hat{\mathcal{E}}}_s^{free}(\mathbf{r};s),
\end{equation}
where $\mathscr{W}_{\mu}(s) = {s}/({s^2+\omega_\mu^2})$ is the Laplace transform of $w_{\mu}(t)$. By using \ref{eq:ExpS1} we obtain from \ref{eq:ELapl}
\begin{equation}
\label{eq:ELapl1}
    \boldsymbol{\hat{\mathcal{E}}}_{s}(\mathbf{r};s) =-\mu_0 \int_V \dV'\, s\overleftrightarrow{G}^\perp \left(\mathbf{r} - \mathbf{r}';s\right) [s\boldsymbol{\hat{\mathcal{P}}}(\mathbf{r}';s)-\hat{\mathbf{P}}^{(S)}(\mathbf{r}')]  + \boldsymbol{\hat{\mathcal{E}}}^{free}_s(\mathbf{r};s)
\end{equation}
where $\boldsymbol{\hat{\mathcal{P}}}(\mathbf{r}';s)$ is the Laplace transform of the polarization field operator and $\hat{\mathbf{P}}^{(S)}$ is the polarization density field operator in the Schro\"dinger picture. In time domain, this relation becomes
\begin{equation}
\label{eq:Etime2}
    \hat{\mathbf{E}}_{s}(\mathbf{r};t) =-\mu_0 \int_V\dV' \int_{0}^{\infty}d\tau \dot{\overleftrightarrow{g}}^\perp \left(\mathbf{r} - \mathbf{r}';t-\tau\right) \dot{\hat{\mathbf{P}}}(\mathbf{r}';\tau)  + \hat{\mathbf{E}}^{free}_s(\mathbf{r};t).
\end{equation}
\end{widetext}

\subsection{Irrotational component}

In the time domain, the Coulomb field operator is given by:
\begin{equation}
\label{eq:Ec0}
    \hat{\mathbf{E}}_c\rpt  =-\frac{1}{4\pi\varepsilon_0}  \nabla_\mathbf{r} \oint_{\partial V} \dS'\,\frac{ \hat{\mathbf{P}}(\mathbf{r}';t)\cdot\hat{\mathbf{n}}(\rb ')}{\left| {\bf r} - {\bf r}' \right|} .
\end{equation}
It is convenient to move the gradient operator under the integral sign in \ref{eq:Ec0}. For $\mathbf{r}\notin \partial V$ we obtain 
\begin{equation}
\label{eq:Ec}
    \hat{\mathbf{E}}_c (\mathbf{r};t)  =\frac{1}{4\pi\varepsilon_0} \oint_{\partial V} \dS'\, \frac{ ({\bf r} - {\bf r}')}{\left| {\bf r} - {\bf r}' \right|^3} \hat{\mathbf{P}}(\mathbf{r}';t)\cdot\hat{\mathbf{n}}(\rb ') .
\end{equation}
When $\mathbf{r} \in \partial V$ an additional additive term for the normal component is needed, it is given by
\begin{eqnarray}
  \left. \delta \hat{\mathbf{E}}_{c} (\mathbf{r};t) \cdot \hat{\mathbf{n}}(\mathbf{r}) \right|_{\partial V^\pm} = \mp \frac{1}{2\epsilon_0} \left. \hat{\mathbf{P}}(\mathbf{r};t)  \cdot \hat{\mathbf{n}}(\mathbf{r})\right|_{\partial V}
\end{eqnarray}
where $\partial V^\pm$ are the external and internal pages of the surface $\partial V$. Nevertheless, inside the dielectric, using Eq. \ref{eq:EQS}, we obtain
\begin{equation}
 \hat{\mathbf{E}}_c(\mathbf{r};t) =- \frac{1}{\eps_0}\sum_{m=1,2,...} \frac{1}{\kappa_m^\parallel}\hat{p}_m^\parallel(t) \mathbf{U}_m^\parallel(\mathbf r) .
\end{equation}
\subsection{Total electric field operator}
The electric field operator $\hat{\mathbf{E}}$ can be expressed as function of the polarization field operator $\hat{\mathbf{P}}$ by using the dyadic Green function for the vector potential in the temporal gauge and in the free space
\begin{multline}
    \overleftrightarrow{g}\left(\rb;t \right)=\frac{1}{4\pi r}\left\{(\overleftrightarrow{I}-{\mathbf{e}}_r\,{\mathbf{e}}_r)\delta(t')\right.+ \\
    +\left. \frac{c_0^2}{r}(\overleftrightarrow{I}-3{\mathbf{e}}_r\,{\mathbf{e}}_r)\,u(t')\left(\frac{1}{c_0}+\frac{t'}{r}\right)\right \},
\end{multline}
where $t' = t-r/c_0$. To show this, we first need to express the Coulomb field operator as a volume integral of the polarization field operator. By using the Gauss theorem, from Eq. \ref{eq:Ec0} we obtain
\begin{equation}
\label{eq:philong}
 \hat{\mathbf{E}}_c (\mathbf{r};t)=\frac{1}{4\pi\varepsilon_0} \nabla_{{\bf r}} \int_{V} \dV'\, \nabla_{{\bf r}}\left(\frac{ 1}{\left| {\bf r} - {\bf r}' \right|}\right)\cdot\hat{\mathbf{P}}(\mathbf{r}';t).
\end{equation}
For $\mathbf{r}\notin V$ we can move the gradient operator on the left hand side under the integral operator
\begin{equation}
 \hat{\mathbf{E}}_c (\mathbf{r};t)= \frac{1}{4\pi\epsilon_0 }\int_{V} \dV'\, \nabla_{{\bf r}}\nabla_{{\bf r}} \frac{1}{|{\bf r} - {\bf r}'|} \hat{\mathbf{P}}(\mathbf{r}';t).
\end{equation}
For $\mathbf{r}\in V$, the singularity in $1/| {\bf r} - {\bf r}'|$ yields an additional term (e.g., \cite{van_bladel_singular_1996}),
\begin{multline}
\label{eq:Ecoul2}
 \hat{\mathbf{E}}_c (\mathbf{r};t)=\frac{1}{4\pi\epsilon_0 }\left[-\frac{4\pi}{3}\hat{\mathbf{P}}(\mathbf{r};t) \right. + \\ \left.+ \lim_{\delta\to 0}\int_{V-V_\delta} \dV'\,\nabla_{{\bf r}}\nabla_{{\bf r}}\frac{1}{|{\bf r} - {\bf r}'|} \hat{\mathbf{P}}(\mathbf{r}';t)\right],
\end{multline}
where $V_\delta$ is a sphere of radius $\delta$ centered at $\mathbf{r}$, and
\begin{equation}
\nabla \nabla \frac{1}{r}=\frac{1}{r^3} (3{\mathbf{e}}_r\,{\mathbf{e}}_r-\overleftrightarrow{I})\qquad \mathbf{r}\in V-V_\delta.
\end{equation}
The expression between the square brackets on the right hand side of Eq.\ref{eq:Ecoul2} is the irrotational component of the polarization density field in $V_\infty$. Therefore, the total electric field operator is given by
\begin{widetext}
\begin{equation}
\label{eq:ELapl4}
    \boldsymbol{\hat{{E}}}(\mathbf{r};t) =-\frac{\boldsymbol{\hat{{P}}}(\mathbf{r};t)}{3\epsilon_0}-\mu_0\lim_{\delta\to 0}\int_{V-V_\delta} \dV' \dot{\overleftrightarrow{g}} \left(\mathbf{r} - \mathbf{r}';t\right)* \dot{\hat{\mathbf{P}}}(\mathbf{r}';\tau)  + \hat{\mathbf{E}}^{free}(\mathbf{r};t),
\end{equation}
\end{widetext}
where
\begin{multline}
\label{eq:gtotdin}
    \dot{\overleftrightarrow{g}} \left(\mathbf{r} ;t\right)=\frac{1}{4\pi r}\left\{(\overleftrightarrow{I}-{\mathbf{e}}_r\,{\mathbf{e}}_r)\dot{\delta}(t')\right.+ \\
    +\left. \frac{c_0^2}{r}(\overleftrightarrow{I}-3{\mathbf{e}}_r\,{\mathbf{e}}_r)\left[\frac{1}{c_0}\delta(t')+\frac{1}{r}\,u(t')\right]\right\},
\end{multline}
\begin{equation}
    \hat{\mathbf{E}}^{free} = \hat{\mathbf{E}}^{free}_s +\hat{\mathbf{E}}^{free}_c,
\end{equation}
\begin{equation}
    \hat{\mathbf{E}}^{free}_c(\mathbf{r};t) = \frac{u(t)}{\varepsilon_0}\lim_{\delta\to 0}\int_{V-V_\delta} \dV' {\overleftrightarrow{g}}_0 (\mathbf{r} - \mathbf{r}') {\hat{\mathbf{P}}}^{(S)}(\mathbf{r}'),
\end{equation}
and
\begin{equation}
\label{eq:gtotstat}
    \overleftrightarrow{g}_0(\rb)=\frac{1}{4\pi r^3}\left(3{\mathbf{e}}_r\,{\mathbf{e}}_r-\overleftrightarrow{I}\right).
\end{equation}

\end{document}